\documentclass[hyper,a4paper]{JHEP3}

\usepackage{epsfig}
\usepackage{latexsym}
\usepackage{graphicx}
\usepackage{amsmath}
\usepackage{amsfonts}   
\usepackage{amssymb}    
\usepackage{float}
\usepackage{bm}
\usepackage{cite}
\usepackage{url}

\def\k{\kappa}

\def\lsim{\raise0.3ex\hbox{$\;<$\kern-0.75em\raise-1.1ex\hbox{$\sim\;$}}}
\def\gsim{\raise0.3ex\hbox{$\;>$\kern-0.75em\raise-1.1ex\hbox{$\sim\;$}}}

\def    \beq            {\begin{equation}}
\def    \eeq            {\end{equation}}
\def    \bea           {\begin{eqnarray}}
\def    \eea           {\end{eqnarray}}

\def \mn{\mu\nu{\rm SSM}}
\def\n{\widetilde\chi^0}

\def\S{S^0}
\def\P{P^0}
\def\nc{\nu^c}
\def\ak{A_\kappa}
\def\al{A_\lambda}

\def\g2{{\rm GeV}^2}

\def\SSSd{S^0_4\to S^0_i S^0_j}

\def\SPPd{S^0_4\to P^0_i P^0_j}

\def\SNNd{S^0_4\to \widetilde\chi^0_{i+3} \widetilde\chi^0_{j+3}}
\def\mgg{\mu_{\gamma\gamma}(S^0_4)}
\def\mbb{\mu_{b\bar{b}}(S^0_4)}
\def\mtata{\mu_{\tau^+\tau^-}(S^0_4)}
\def\mWW{\mu_{WW^*}(S^0_4)}
\def\mZZ{\mu_{ZZ^*}(S^0_4)}
\def\GNP{\Gamma^{\rm NP}_{\rm tot}}
\def\GNPSM{\Gamma^{\rm SM'}_{\rm tot}}
\def\RSd{\left|R^{S^0}_{41}\right|^2}
\def\RSu{\left|R^{S^0}_{42}\right|^2}
\def\RSds{|R^{S^0}_{41}|^2}
\def\RSus{|R^{S^0}_{42}|^2}
\def\hsm{h^0_{\rm SM}}
\def\Us{\mathcal{U}_{i1} S^0_M + \mathcal{U}_{i2} S^0_{U_1} + \mathcal{U}_{i3} S^0_{U_2}}
\def\Up{\mathcal{U}_{i1} P^0_M + \mathcal{U}_{i2} P^0_{U_1} + \mathcal{U}_{i3} P^0_{U_2}}
\def\Un{\mathcal{U}_{i1} \widetilde\chi^0_M + \mathcal{U}_{i2} \widetilde\chi^0_{U_1} + \mathcal{U}_{i3} \widetilde\chi^0_{U_2}}
\def\tb{{\rm tan}\beta}
\def\sb{{\rm sin}\beta}
\def\cb{{\rm cos}\beta}
\def\mht{m^{\rm tree}_h}

\def\n{\widetilde\chi^0}
\def\nc{\nu^c}

\def\sw2{sin^2 \theta_w}

\def\a^tau{\alpha_{\tau}}

\def\beq{\begin{equation}}
\def\eeq{\end{equation}}
\def\beqa{\begin{eqnarray}}
\def\eeqa{\end{eqnarray}}

\newcommand{\newc}{\newcommand}
\newc\BR{BR}
\newc{\akappa}{A_{\kappa} }
\newc\deltagmtwo{\delta (g-2)_{\mu}} 
\newc\deltaamu{\Delta a_{\mu}}

\def\anti{\overline}

\def\la{\lambda}
\def\bla{\bm \lambda}
\def\ka{\kappa}
\def\Ala{A_\lambda}
\def\Aka{A_\kappa}
\def\rpv{{R}_{p} \hspace{-0.4cm}\slash\hspace{0.2cm}}
\def\MET{{E}_{\rm T} \hspace{-0.43cm}\slash\hspace{0.2cm}}
\def\MPT{{P}_{\rm T} \hspace{-0.43cm}\slash\hspace{0.2cm}}
\newc{\haa}{BR\(h_1\to a_1 a_1\)}
\newc{\abb}{BR\(a_1\to b\anti{b}\)}
\newc{\hbb}{BR\(h_1\to b\anti{b}\)}
\newc{\abund}{\Omega h^2}
\newc\bsgamma{b\rightarrow s \gamma }
\newc\bxsgamma{\overline{B}\rightarrow X_{s}\gamma}
\newc\brbsgamma{\BR(\overline{B}\rightarrow X_s\gamma)}

\title
{Probing the $\bm{\mu\nu}$SSM with light scalars, pseudoscalars
and neutralinos from the decay of a SM-like Higgs boson
at the LHC}

\author{Pradipta Ghosh,$^{a,b}$ 
~Daniel~E.~L\'opez-Fogliani,$^c$ ~Vasiliki~A.~Mitsou$^d$,
~Carlos~Mu\~noz$^{a,b}$ and Roberto~Ruiz~de~Austri$^d$\\

$^a$Departamento de F\'{\i}sica Te\'{o}rica, Universidad Aut\'{o}noma de Madrid,
Cantoblanco, 28049 Madrid, Spain\\
$^b$Instituto de F\'{\i}sica Te\'{o}rica UAM--CSIC, Campus de Cantoblanco, 
28049 Madrid, Spain

$^c$Departamento de F\'{\i}sica, Universidad de Buenos Aires 
\& IFIBA-CONICET, \\ 1428 Buenos Aires, Argentina

$^d$Instituto de F\'{\i}sica Corpuscular CSIC--UV, c/ Catedr\'atico 
Jos\'e Beltr\'an 2, \\
46980 Paterna (Valencia), Spain

E-mails:~\email{pradipta.ghosh@uam.es, daniel.lopez@df.uba.ar,
 vasiliki.mitsou@ific.uv.es, carlos.munnoz@uam.es, rruiz@ific.uv.es}}

\abstract{The ``$\mu$ from $\nu$'' supersymmetric 
standard model ($\mu\nu$SSM) can accommodate the newly discovered Higgs-like 
scalar boson with a mass around $125$ GeV.
This model provides a solution to the $\mu$-problem and simultaneously 
reproduces correct neutrino physics by the simple use of right-handed neutrino superfields.
These new superfields together with the introduced $R$-parity violation can 
produce novel and characteristic signatures of the $\mu\nu$SSM at the LHC.
We explore the signatures produced through two-body Higgs decays into the new states,
provided that these states lie below in the mass spectrum.
For example, a pair produced light neutralinos
depending on the associated decay length can give rise to 
displaced multi-leptons/taus/jets/photons 
with small/moderate missing transverse energy. In the same spirit,
a Higgs-like scalar decaying to a pair of scalars/pseudoscalars
can produce final states with prompt multi-leptons/taus/jets/photons.}

\preprint{FTUAM-14-31,~~~IFT-UAM/CSIC-14-074,~~~IFIC-14-21}
\date{\today}
\begin{document}
\section{Introduction}
\label{Introduction}

The ATLAS and CMS collaborations have finally discovered a 
new scalar boson ~\cite{:2012gk,:2012gu} of mass about 
$125$ GeV at the LHC~\cite{:2012gk,:2012gu,
ATLAS:2012oga,Chatrchyan:2013lba,Chatrchyan:2014vua}. This new scalar 
has properties \cite{Chatrchyan:2013lba,Aad:2013wqa,Aad:2013xqa,CMS:yva,
Chatrchyan:2013vaa,Chatrchyan:2013zna,
Chatrchyan:2013iaa,ATLAS:couplings,Chatrchyan:2013mxa,Chatrchyan:2014nva,
Chatrchyan:2014vua,Aad:2014aba,Aad:2014xva,CMS:2014ega}
similar to that of the much awaited standard model (SM) Higgs boson. 
However, issues like missing precise experimental measurements 
over all the SM decay modes 
(e.g., $b\bar{b}$), hitherto existing \textit{mild} excess in the di-photon channel 
\cite{ATLAS:couplings,CMS:2014ega,Khachatryan:2014ira,Aad:2014eha}, etc., 
keep the possibility of having a beyond SM 
origin alive to date. Among a plethora of candidate beyond the SM theories, 
weak scale supersymmetry (SUSY) has extensively been 
analysed  over a long period of time. Missing experimental
evidence of SUSY to date \cite{ATLAS-susy13,CMS-susy13}, 
especially when the experimental observations are interpreted
with the simplified models, together with a class of theoretical issues, motivates one to 
consider models beyond the minimal structure.

The ``$\mu$ from $\nu$'' supersymmetric 
standard model ($\mu\nu$SSM) \cite{LopezFogliani:2005yw, Escudero:2008jg} solves  
the $\mu$-problem \cite{Kim:1983dt} of the 
minimal supersymmetric standard model (MSSM) (see \cite{Nilles:1983ge,Haber:1984rc,
Simonsen:1995cf,Martin:1997ns} for reviews)
and simultaneously accommodates the correct neutrino 
physics \cite{LopezFogliani:2005yw,Escudero:2008jg,
Ghosh:2008yh,Bartl:2009an,Fidalgo:2009dm,Ghosh:2010zi,LopezFogliani:2010bf,
Ghosh:2010ig}, as guided by the three flavour global neutrino data 
\cite{Tortola:2012te,GonzalezGarcia:2012sz,Capozzi:2013csa}. A set of three right-handed 
neutrino superfields has been utilised to address both purposes, 
relating the origin of the $\mu$-term to the origin of neutrino masses and mixing.
As a consequence of the construction, the $R$-parity \cite{Fayet:1974pd,Fayet:1977yc,
Farrar:1978xj,Weinberg:1981wj,Aulakh:1982yn,Hall:1983id,Lee:1984kr,Lee:1984tn,
Ross:1984yg,Ellis:1984gi} is explicitly broken $(\rpv)$ in the $\mu\nu$SSM. 
Non-zero neutrino masses in this model appear through a dynamically generated 
electroweak-scale seesaw \cite{LopezFogliani:2005yw,
Escudero:2008jg,Ghosh:2008yh,Bartl:2009an,Fidalgo:2009dm,
LopezFogliani:2010bf,Ghosh:2010zi,Ghosh:2010ig}. Thus, the only scale associated
with the $\mu\nu$SSM is the scale of electroweak symmetry breaking (EWSB) or in 
other words the scale of the soft SUSY-breaking terms, which is in the 
ballpark of a TeV. This nice feature can produce realistic signatures
of this model at colliders \cite{Ghosh:2008yh,Bartl:2009an,
Bandyopadhyay:2010cu,Fidalgo:2011ky,Ghosh:2012pq,Ghosh:2014rha}, 
well verifiable at the LHC or at upcoming accelerator 
experiments \cite{Erler:2001ja,Abe:2001wn,Abe:2001swa,
Heinemeyer:2010yp,Gomez-Ceballos:2013zzn}.  
As a consequence of $\rpv$, the lightest supersymmetric particle 
(LSP) is no longer a valid candidate for cold dark matter. Nevertheless,
embedding the model in the context of supergravity 
(see ref. \cite{Nilles:1983ge} for a review) 
one can accommodate the gravitino \cite{Choi:2009ng}
as an eligible decaying dark matter candidate with a life-time greater
than the age of the Universe. 
Its detection is also possible in principle through the observation 
of a gamma-ray line in the Fermi satellite \cite{Choi:2009ng,GomezVargas:2011ph,
Munoz:2012zea,Albert:2014hwa}. In ref.~\cite{Chung:2010cd}, the generation
of the baryon asymmetry of the universe was analysed in the
$\mu\nu$SSM, with the interesting result that electroweak baryogenesis 
can be realised.

In the $\mu\nu$SSM, the bilinear $\mu\hat H_d \hat H_u$ term of the MSSM 
superpotential is replaced by the trilinear terms $\la_i \hat \nu^c_i \hat H_d\hat H_u$. 
Here $\hat \nu^c_i$  are the right-handed neutrino superfields, 
singlets under the SM gauge group. New trilinear terms 
like $Y_{\nu_{ij}} \hat H_u\hat L_i\hat \nu^c_j$, where 
$Y_{\nu_{ij}}$ are the neutrino Yukawa couplings, are also introduced.
An effective $\mu$ term with $\mu_{\text{eff}}\equiv\la_i \nu^c_i$ is generated 
after the successful EWSB, where $\nu^c_i$ denotes the vacuum expectation value (VEV) acquired 
by the scalar component of the $i$-th right-handed neutrino superfield. 
In the same spirit, after the EWSB, $Y_{\nu_{ij}} \hat H_u\hat L_i\hat \nu^c_j$ terms
generate effective bilinear $\rpv$ parameters as $Y_{\nu_{ij}} \nu^c_j$.
Following the trend, effective Majorana masses
for right-handed neutrinos, $2\kappa_{ijk}\nu^c_k$,
are produced from $\kappa_{ijk}\hat\nu^c_i \hat\nu^c_j \hat\nu^c_k$ terms.
The explicit breaking of $R_p$ is apparent in all the three above mentioned 
trilinear terms.
The order of magnitude for $\nu^c_i$ is determined from
the soft SUSY-breaking terms. Thus, as emphasised before, along with 
the aforementioned features, the EWSB scale, the origin of the $\mu$-term
and the scale of the right-handed neutrino Majorana 
masses (instrumental in the generation
of neutrino mass through a seesaw mechanism) in the $\mu\nu$SSM are connected
to the one and only scale of the model, namely the scale of the 
soft SUSY-breaking terms.

It is worthy to discuss here the number of right-handed neutrino superfields
in the $\mu\nu$SSM. Although it is possible to accommodate the correct neutrino
data \cite{GonzalezGarcia:2012sz,Tortola:2012te} at the tree level 
\cite{LopezFogliani:2005yw,Escudero:2008jg,Fidalgo:2009dm,Ghosh:2008yh,Bartl:2009an}, 
provided one works with at least two $\hat \nu^c_i$, we stick to 
three $\hat \nu^c$ scenario which appears natural from the SM family symmetry.
Nevertheless, the $\mu\nu$SSM with arbitrary number of right-handed neutrino superfields
has also been discussed in the literature \cite{Bartl:2009an}.

It is well evident that the presence of a set of new couplings 
in the $\mu\nu$SSM will trigger
a few new decay modes for a SM Higgs-like scalar
provided that the new states are lighter than it. Some 
of these modes, for example Higgs decay into a new scalar/pseudoscalar pair, 
are well known for extended models (with or without SUSY) with 
a singlet \cite{Gunion:1996fb,Ellwanger:2001iw,Ellwanger:2003jt,Ellwanger:2004gz,
Dermisek:2005ar,Dermisek:2005gg,Schuster:2005py,Dermisek:2006wr,
Zhu:2006zv,Dermisek:2007yt,Carena:2007jk,Belyaev:2008gj,Dermisek:2008uu,
Lisanti:2009uy,Belyaev:2010ka,Almarashi:2010jm,Almarashi:2011hj,
Almarashi:2011bf,Ellwanger:2011sk,Almarashi:2011qq,Rathsman:2012dp,Dermisek:2012cn,Kang:2013rj,
Cerdeno:2013cz,Bhattacherjee:2013vga,Cerdeno:2013qta,Cao:2013gba,
Dermisek:2013cxa,Grinstein:2013fia,Ellwanger:2014hia,King:2014xwa,Das:2014fha,Bomark:2014gya}. 
The singlet nature\footnote{Throughout this article,
a singlet-like state implies a state with singlet 
composition larger than about $90\%$.} 
of these states is useful to evade a class of 
LEP constrains  \cite{Abbiendi:2002qp,
Abbiendi:2002in,Barate:2003sz, Abbiendi:2004ww,
Abdallah:2004wy,Schael:2006cr,Beacham:2010hf} as well as constraints
from hadron colliders \cite{Abazov:2009yi,Klemetti:2011sda,ATLAS:2011cea,
ATLAS:2012soa,ATLAS:2012dsy,Chatrchyan:2012cg,CMS:2013lea}. 
Light states are also constrained from a group of 
low-energy observables\cite{Love:2008aa,Domingo:2008rr,Aubert:2009cp,
Aubert:2009cka,Dermisek:2010mg,Andreas:2010ms,delAmoSanchez:2010ac,
Lees:2011wb,Ablikim:2011es,Lees:2012iw,Lees:2012te,Lees:2013vuj}
where the presence of these states can 
yield enhanced contribution to some processes,
often in an experimentally unacceptable way. These
issues will be addressed later with further detail.

In the case of SUSY models, an additional decay mode for a Higgs-like scalar  
into a pair of light neutralinos \cite{Accomando:2006ga,Draper:2010ew,
Huang:2013ima,Curtin:2013fra,Huang:2014cla} is also a viable 
option\footnote{If allowed kinematically
a Higgs-like scalar can also decay into a pair of heavier neutralinos. 
For example, into a pair of next-to lightest 
neutralinos. This scenario is
constrained from the measured $Z$ decay width for neutralino mass 
$\lsim M_{Z}/2$.}. In the case of a pair of the lightest neutralinos, 
this mode contributes to the invisible
Higgs decay since the lightest neutralino is usually the 
LSP for a large region of the parameter space. The latter being 
neutral \cite{Goldberg:1983nd,Ellis:1983ew} 
and stable, leaves only missing transverse momentum ($\MPT$) signature
at colliders. In the $\mu\nu$SSM, however, with
$\rpv$ this mode can lead to displaced leptons/taus/jets(hadronic)/photons at colliders
depending on the associated decay length \cite{Bartl:2009an,Bandyopadhyay:2010cu,
Fidalgo:2011ky,Ghosh:2012pq}.
In addition to the displaced objects, signals of the $\mu\nu$SSM
are accompanied by a small or moderate missing transverse energy $(\MET)$, the origin
of which relies on the light neutrinos and/or possible mis-measurements.
This is an apparent contradiction to $R_p$ conserving SUSY scenarios 
where the stable, neutral and hence undetected LSPs leave their collider imprint 
in the form of large $\MPT$. 
Nevertheless, a pure $\MPT$/$\MET$ signature is also possible for $\rpv$ scenario
when a neutralino LSP, being lighter than 
40 GeV, decays beyond the detector coverage~\cite{Bartl:2009an} 
or decays to three neutrino final states.

The rich collider phenomenology of the $\mu\nu$SSM with $\rpv$
and extra superfields makes it 
absolutely legitimate to ask two of the most 
appealing possibilities, namely:
\begin{enumerate}
 \item 
How much room do we have for non-standard (non SM-like) decays of 
the newly discovered Higgs-like scalar boson with a mass about $125$ GeV?

It is well known that so far ATLAS and CMS collaborations have not
observed any significant deviation from the SM expectations
while analysing this $125$ GeV scalar \cite{Aad:2013xqa,Chatrchyan:2013lba,CMS:yva}. 
The window of \textit{non-standardness}, however, is not closed to date, e.g.
the mild excess in the di-photon decay mode remains in the ATLAS measurements 
\cite{ATLAS:couplings,Aad:2014eha} and now is also supported by 
the CMS results \cite{Khachatryan:2014ira,CMS:2014ega}.
At the same time a precise estimation of the total decay width of
this scalar is still missing \cite{Khachatryan:2014iha,Aad:2014aba,ATLAShwidth,CMS:2014ala}.
Furthermore, missing precision information about all the SM
decay modes (e.g., $b\bar{b}$ \cite{Chatrchyan:2013lba,bb:2013lia,CMS:2013sea} and also 
$\tau^+\tau^-$ \cite{Chatrchyan:2013lba,CMS:2013sea,CMS:2013hja,Chatrchyan:2014nva} 
to some extent) allows a big open window for the branching fraction of 
the non-standard decay modes to date \cite{ATLAS:couplings,
Desai:2012qy,Espinosa:2012vu,Ghosh:2012ep,
ATLAS:2013pma,CMS:2013bfa,CMS:2013yda,CMS:13-28,Belanger:2013kya,Ellis:2013lra,
Ananthanarayan:2013fga,Belanger:2013xza,Aad:2013oja,Aad:2014iia,
Chatrchyan:2014tja,Pandita:2014nsa}. 
Thus, it is rather crucial
to investigate these new modes systematically even
before developing a linear collider.

\item
Experimentally allowed singlet-like light scalars, pseudoscalars and 
neutralinos are well affordable in the $\mu\nu$SSM \cite{Escudero:2008jg,
Bartl:2009an,Ghosh:2012pq}. So, what will be 
the consequences of these light states at colliders? For example,
how these states can affect the decay phenomenology of other
heavier SM/SUSY particles? See ref. \cite{Ghosh:2014rha} for example. 

\end{enumerate}

The enriched spectrum of the $\mu\nu$SSM, as introduced 
in refs. \cite{LopezFogliani:2005yw,Escudero:2008jg}, 
admits the aforesaid novel Higgs decays which have already been addressed in
refs. \cite{Bartl:2009an,Bandyopadhyay:2010cu,Fidalgo:2011ky,
Ghosh:2012pq}. Further, detail collider analyses
for a Higgs-like scalar decaying into a pair of neutralinos have also been discussed
in refs. \cite{Bandyopadhyay:2010cu,Ghosh:2012pq}. 
However, a concise yet complete description of the
resultant phenomenology involving those light states is missing
to date and this is exactly what we aim to address in the current
article in the light of a Higgs-like scalar discovery. 
Note that, as stated above in point 2, those light states
can also modify final state particle multiplicity/signal topology
when appear in the decay cascades of SUSY particles.
Such analyses are beyond the theme of the current paper
and we hope to address them elsewhere.

The paper is organised as follows. We start with a brief description of the 
model in section \ref{The-model}. A complete overview
of all the possible final states at colliders together with 
the identification of crucial backgrounds, when the SM-like Higgs boson in the $\mn$
decays into a pair of light scalars/pseudoscalars/neutralinos, is discussed 
in section \ref{lightest}. In section \ref{Higgs-sector}
we present a discussion about the tree-level SM-like Higgs 
boson mass followed by the effect and relevance of loop corrections 
in the light of a Higgs-like scalar with a mass around 125 GeV.
Additionally, we also identify the crucial set of parameters. 
Following this discussion, in section \ref{spsn} 
with approximate analytical formulae we 
identify the set of most relevant parameters and discuss 
how they determine the masses of those light states.
In section \ref{decay}, we investigate the relevance
of these parameters in controlling the decays of the SM-like Higgs 
boson into a pair of light states,
covering all possible new two-body decays.
We also derive the expressions of the decay widths
for the new decay modes and also evaluate the same for the SM modes,
in the presence of new physics. Finally, we also estimate 
the various reduced signals strengths in the presence of new decays
and compare them with the experimentally measured values.
We elaborate our analysis over relevant regions
of the parameter space also in the same section. 
Our concluding remarks are summarised and presented in 
section~\ref{Summary-conclusion}.


\section{The Model}
\label{The-model}

The $\mu\nu$SSM superpotential following the line of 
refs.\cite{LopezFogliani:2005yw, Escudero:2008jg} is given by
\bea 
W &=& \epsilon_{ab}(Y_{u_{ij}}\hat H^b_u\hat Q^a_i\hat u^c_j +
Y_{d_{ij}}\hat H^a_d \hat Q^b_i\hat d^c_j + Y_{e_{ij}}\hat H^a_d\hat
L^b_i\hat e^c_j + Y_{\nu_{ij}}
\hat H^b_u\hat L^a_i\hat \nu^c_j)\nonumber \\
&-&\epsilon_{ab} \lambda_i\hat \nu^c_i\hat H^a_d\hat H^b_u +
\frac{1}{3}\kappa_{ijk}\hat \nu^c_i\hat \nu^c_j\hat \nu^c_k,
\label{superpotential}
\eea
where $i,j,k$ are family indices and $\epsilon_{12}=1$. Here $\rpv$
is the combined effect of the $4^{\rm th},\,5^{\rm th}$ and $6^{\rm th}$ terms.
It is worthy to note in this connection that 
in the limit $Y_\nu\to 0$, $\hat \nu^c$ can be identified as
a pure singlet superfield without lepton 
number, similar to the next-to minimal supersymmetric standard 
model (NMSSM, see ref. \cite{Ellwanger:2009dp} for a review), where $R_p$ is not broken.
Thus $\rpv$ is small since the electroweak-scale seesaw implies 
small values for the neutrino Yukawa couplings,
 $Y_\nu$ $\sim 10^{-6} - 10^{-7}$ \cite{LopezFogliani:2005yw,
Escudero:2008jg,Ghosh:2008yh,Bartl:2009an,Fidalgo:2009dm,Ghosh:2010zi}.
This \textit{minimal} superpotential of eq.~(\ref{superpotential})
serves both the purposes of solving the $\mu$-problem
and generating non-zero neutrino masses and mixing, 
as already mentioned in the introduction. Although conventional
trilinear $\rpv$ terms are absent from the superpotential, 
the leptonic ones can, however, appear through
loop processes as shown in ref. \cite{Escudero:2008jg}. 

Working in the framework of supergravity, the Lagrangian 
$\mathcal{L}_{\text{soft}}$ containing the soft supersymmetry breaking 
terms is given by \cite{LopezFogliani:2005yw,Escudero:2008jg}:
\bea
-\mathcal{L}_{\text{soft}} &=&
(m_{\widetilde{Q}}^2)_{ij} {\widetilde Q^{a^*}_i} \widetilde{Q^a_j}
+(m_{\widetilde u^c}^{2})_{ij}
{\widetilde u^{c^*}_i} \widetilde u^c_j
+(m_{\widetilde d^c}^2)_{ij}{\widetilde d^{c^*}_i}\widetilde d^c_j
+(m_{\widetilde{L}}^2)_{ij} {\widetilde L^{a^*}_i}\widetilde{L^a_j} \nonumber \\
&+&(m_{\widetilde e^c}^2)_{ij}{\widetilde e^{c^*}_i}\widetilde e^c_j 
+ m_{H_d}^2 {H^{a^*}_d} H^a_d + m_{H_u}^2 {H^{a^*}_u} H^a_u +
(m_{\widetilde{\nu}^c}^2)_{ij}  {\widetilde{\nu}^{c^*}_i} \widetilde\nu^c_j \nonumber \\
&+& \epsilon_{ab} \left[
(A_uY_u)_{ij} H_u^b\widetilde Q^a_i \widetilde u_j^c +
(A_dY_d)_{ij} H_d^a \widetilde Q^b_i \widetilde d_j^c +
(A_eY_e)_{ij} H_d^a \widetilde L^b_i \widetilde e_j^c + \text{H.c.}  \right] 
\nonumber \\
&+&\left[\epsilon_{ab}(A_{\nu}Y_{\nu})_{ij} H_u^b \widetilde L^a_i \widetilde 
\nu^c_j-\epsilon_{ab} (A_{\lambda}\lambda)_{i} \widetilde \nu^c_i H_d^a  H_u^b+
\frac{1}{3} (A_{\kappa}\kappa)_{ijk} \widetilde \nu^c_i \widetilde \nu^c_j \widetilde 
\nu^c_k\ + \text{H.c.} \right] \nonumber \\
&-& \frac{1}{2} \left(M_3 \widetilde{\lambda}_3 \widetilde{\lambda}_3
+ M_2 \widetilde{\lambda}_2 \widetilde{\lambda}_2 + M_1 \widetilde{\lambda}_1 
\widetilde{\lambda}_1 + \text{H.c.} \right),
\label{Lsoft}
\eea
where the last term of the $2^{\rm nd}$ line and all terms appearing in the $4^{\rm th}$ line 
are generic to the $\mu\nu$SSM.  Remaining soft terms are the same as those of the MSSM, 
but without the $\mu B_\mu \hat H_u \hat H_d$ term. 

With the choice of CP-conservation,\footnote{$\mu\nu$SSM with spontaneous 
CP-violation has been studied in ref.\cite{Fidalgo:2009dm}.} VEVs acquired by 
neutral scalars are given by
\bea
\langle H_d^0 \rangle = v_d \, , \quad \langle H_u^0 \rangle = v_u \,
, \quad \langle \widetilde \nu_i \rangle = \nu_i \, , \quad
\langle \widetilde \nu_i^c \rangle = \nu^c_i.
\label{vevs}
\eea
%
As already stated, it is apparent that after the EWSB from $4^{\rm th}$ and $5^{\rm th}$ terms
of eq. (\ref{superpotential}) one can extract the effective 
$\rpv$ terms ($\varepsilon_i$), like in the bilinear $R_p$ 
violating (BRpV) model (see ref.~\cite{Barbier:2004ez} for a review) and the $\mu$ term. 
They are given by $\sum Y_{\nu_{ij}}\nu^c_j$ and $\sum \la_i \nu^c_i$, respectively.

A dedicated analysis of the model parameter space with minimisation conditions 
has been addressed in ref. \cite{Escudero:2008jg}. Also the relative importance of 
various parameters in the different regions of the parameter space has been discussed there.
The enhanced mass matrices are presented in refs. \cite{Escudero:2008jg,Ghosh:2008yh,
Ghosh:2010zi}. Augmentation of the mass matrices in the $\mu\nu$SSM over the same
for the MSSM is a consequence of the additional superfield content and $\rpv$.

Being elucidate for the convenience of reading, let us mention that the enhancement 
of the neutral and the charged Higgs sectors occur through the mixing between
the neutral and the charged doublet Higgses with the three  generations of  
left- and right-handed  sneutrinos, and left- and right-handed charged 
sleptons, respectively. In a similar way, the mixing among the neutral higgsinos and 
gauginos with the three families of left- and right-handed neutrinos enlarges 
the number of neutralino  states. An analogous effect for the chargino sector appears through 
the mixing of the charged higgsino and wino with the charged leptons.

Before we address a Higgs-like scalar boson in the $\mn$ in the light of Higgs boson 
discovery and identify the key parameters to accommodate the light scalar, pseudoscalar 
and neutralino states,  it will be convenient to illustrate first their possible collider 
phenomenology. In this way, the motivation to analyse these states 
further becomes apparent and we aim to address this in the next section with 
a complete overview of all the possible collider signatures. 

\section{Phenomenology of the light neutral states}
\label{lightest}

In this section we address the collider phenomenology of all the 
neutral states lighter than the newly discovered 
scalar with SM Higgs-like properties and a mass about 125 GeV. 
Further, we also discuss how the presence of these
light states can impinge the decay kinematics of the SM-like Higgs boson 
and produces unconventional signals at colliders.
So we focus on the scenario when the decay of a 
Higgs-like scalar into a pair of light states is completely on-shell. 
Furthermore, for simplicity we assume that all the allowed
light scalar, pseudoscalar and neutralino states are closely spaced in masses,
such that an additional decay cascade \cite{Fidalgo:2011ky} among these 
states remains kinematically forbidden.
In order to continue our discussions on the
light neutral states, a prior and brief description of the mass spectrum 
would appear very relevant for the convenience of reading.

Following ref.~\cite{Escudero:2008jg}, all the eight CP-even 
neutral scalars are denoted by $S^0_\alpha$ while $P^0_\alpha$
stands for the seven CP-odd neutral scalars. 
In order to address the decay phenomenology of the SM-like Higgs
boson into non-standard modes, one 
needs states lighter than its mass. Naturally singlet-like
(i.e., right-handed sneutrino and neutrino-like) states are
the experimentally preferred possibility to meet this requirement.
These light scalar CP-even and CP-odd states are labeled by
$S^0_i$ and $P^0_i$, respectively. 
In this article the indices $i,\, j,\,k$ are used 
to represent generation indices.
With this kind of hierarchy in
the mass spectrum, $S^0_4$ represents \cite{Bandyopadhyay:2010cu,
Fidalgo:2011ky,Ghosh:2012pq} the newly discovered SM Higgs-like
scalar state. 
The seven colour-singlet charged scalar
states and the five chargino states are represented by $S^\pm_\alpha$
and $\widetilde \chi^\pm_\alpha$, respectively. 
Concerning neutralinos, we use $\widetilde \chi^0_\alpha$ 
as the generic symbol for the ten neutralino states. The three lightest
neutralinos, namely $\widetilde \chi^0_{1,\,2,\,3}$, are nothing but the
three light active neutrinos and henceforth will be denoted as
$\widetilde \chi^0_i$. Thus, for the $\mu\nu$SSM the fourth
neutralino state, namely $\widetilde \chi^0_4$, is the 
{\it lightest neutralino} in true sense. 
In the same spirit, the three lightest charginos, i.e. 
$\widetilde \chi^\pm_i$ with $i=1,2,3$, coincide with the charged
leptons, $e, \mu$ and $\tau$, respectively, with $\widetilde \chi^\pm_4$
representing the true {\it lightest chargino}.

We start our discussion with the light scalars and pseudoscalars and
successively continue with the light neutralinos. As already stated, we also address
the effect of these states in the decays of the SM-like Higgs boson. These new
decays are an important probe for new physics since they generate unusual signals
at colliders. In addition, these decays are also the leading production
sources for these lighter states, since their direct production 
is suppressed due to the singlet nature. One should
note that the direct production rate for these states can be enhanced
with the increasing doublet admixture. However, in this way 
the states may get heavier and hardly produce any unusual
decay channels. At the same time, as stated in the introduction, 
increasing doublet composition makes it harder 
for these states to evade a class of collider
constraints. Additional constraints for these light states,
especially for the light pseudoscalar,
can appear from their connection to a class
of low-energy observables \cite{Love:2008aa,Domingo:2008rr,Aubert:2009cp,
Aubert:2009cka,Dermisek:2010mg,Andreas:2010ms,
delAmoSanchez:2010ac,Lees:2011wb,Ablikim:2011es,Lees:2012iw,Lees:2012te,Lees:2013vuj}. 
Some of the constraints can be evaded with low $\tb~(=v_u/v_d)$ 
values, e.g. $\tb \lsim 10$ \cite{Dermisek:2010mg},
while a correct balance of the singlet-doublet admixing
provides an extra handle for the others. 
As an example of the latter, the branching ratio (Br)
of $B^0_s\to \mu^+\mu^-$ is sensitive to
$1/(m_{\P})^4$, where $P^0$ represents a generic pseudoscalar.
Thus, the scenario with light $P^0_i$ apparently 
enhances Br$(B^0_s\to \mu^+\mu^-)$ and thereby, seems to be excluded by 
the experimental results. In reality however, as long as the amount
of doublet mixing is small, and thus the couplings between
the SM particles and these light states are very suppressed
due to the dominant singlet nature, these scenarios can escape experimental
constraints \cite{Cerdeno:2013qta}. 
Another effect regarding $B^0_s\to \mu^+\mu^-$ must be
mentioned here, since the branching fraction of this process
possesses a high power sensitivity to $\tb$ in the numerator
while the denominator is sensitive to the high power of the pseudoscalar
mass \cite{Bobeth:2001sq,Arnowitt:2002cq}. Hence, one can either live with small $\tb$ or 
a heavy pseudoscalar to control the size of Br$(B^0_s\to \mu^+\mu^-)$
(see refs. \cite{Akeroyd:2011kd,Arbey:2011aa,Cao:2011sn,Mahmoudi:2012un,
Arbey:2012dq,Altmannshofer:2012ks,Arbey:2012ax,Arbey:2013jla} and references therein). 
In our analysis we focus on the
small $\tb$ values, the most natural option in the presence
of light $\P_i$.

\subsection{Light scalars/pseudoscalars}
\label{lightSPS}

In this subsection we discuss the 
consequences of the light scalars and/or 
pseudoscalars in the collider phenomenology of  the $\mu\nu$SSM.
Note that the masses of these states must be lighter than
the half of $\S_4$, i.e. $2m_{\S_i,\,\P_i}\lsim m_{\S_4}$
such that new two-body decays like $\S_4\to \S_i\S_j$, $\P_i\P_j$ remain 
kinematically possible. Subsequent decays of $\S_i,\,\P_i$, as will be
discussed successively, lead to multi-particle final states.
Possible final states 
strongly depend on the masses of $\S_i$ and $\P_i$, which are 
systematically addressed below. 

\vspace*{0.25cm}
\noindent
{\textit {\textbf {Decaying to leptons and taus:}}}
A light scalar/pseudoscalar decaying into a pair of leptons/taus 
or jets (will be addressed 
subsequently) occurs essentially due to a small but non vanishing admixture with 
the doublet Higgs bosons. Final states with electrons are normally suppressed since 
the couplings of the charged leptons (jets) to the doublet Higgs boson are 
proportional to their respective masses.
The decay into a pair of muons is also normally suppressed for a 
wide range of $m_{S^0_i,P^0_i}$. 
This specific mode gets sub-leading in the range $2m_c \lsim m_{S^0_i,P^0_i} \lsim 
2 m_\tau$, while it dominates in the span of $2m_\mu \lsim m_{S^0_i,P^0_i} \lsim 2 m_c$. 
The decay into a pair of $\tau$s gets dominant for 
$2m_\tau \lsim m_{S^0_i,P^0_i} \lsim 2 m_b$.

$S^0_i$ and $ P^0_i$ states with masses between $2m_\mu$ to $2m_b$ 
(i.e., $2m_\mu \lsim m_{S^0_i,P^0_i} \lsim 2 m_b$) normally lead to
multi-lepton/multi-tau final states at colliders. They are also
relatively easy to identify as the number of associated backgrounds
are lesser and differentiable. For example, $S^0_4 \to 2 S^0_i,2 P^0_i$
can easily lead to 4$\mu$, 4$\tau$ or 2$\mu2\tau$ final states in 
the $\mu\nu$SSM \cite{Fidalgo:2011ky}. The 4$\mu$ channel apparently
seems to be the most promising one as detection efficiency for muons is
rather high at the LHC. This scenario is, however, severely constrained
after the recent CMS analyses \cite{Chatrchyan:2012cg,CMS:2013lea}. 
The process $P^0_i \to \mu^+\mu^-$ itself is also experimentally
constrained from the ATLAS results \cite{ATLAS:2011cea}. 
Further, it is evident from ref.~\cite{Belyaev:2010ka} that the typical maximum 
branching fraction\footnote{This branching fraction
is $\approx~100\%$ for $2m_\mu\lsim m_{\S_i,\P_i}\lsim 3m_{\pi}$ \cite{Belyaev:2010ka}.} 
for a light pseudoscalar decaying to $\mu^+\mu^-$
is $\sim 20\%$ for $2m_\mu \lsim m_{S^0_i,P^0_i} \lsim 2 m_\tau$.
Thus, in general the 4-muon final state has only 4\% of branching
ratio available. This, despite of the large window allowed to
date for the Br of the non-standard/invisible Higgs decays 
\cite{ATLAS:couplings,Desai:2012qy,Espinosa:2012vu,Ghosh:2012ep,
ATLAS:2013pma,CMS:2013bfa,CMS:2013yda,CMS:13-28,Belanger:2013kya,Ellis:2013lra,
Ananthanarayan:2013fga,Belanger:2013xza,Aad:2013oja,Aad:2014iia,
Chatrchyan:2014tja,Pandita:2014nsa}, would yield poor statistics for 
$S^0_4 \to 4\mu$ process. This drawback, however, can be ameliorated with 
larger luminosity or moving towards $2m_\mu\lsim m_{\S_i,\P_i}\lsim 3m_{\pi}$ 
\cite{Belyaev:2010ka} region.

On the contrary, the situation is still experimentally relaxing for $\tau$s. 
Although the process pseudoscalar $\to \tau^+\tau^-$ in the MSSM is 
constrained from experimental searches \cite{ATLAS:2012dsy,Aad:2012cfr,
Khachatryan:2014wca}, in models with singlet(s) (e.g., the NMSSM or the $\mu\nu$SSM)
bypassing the experimental bounds remain possible for the two inter-related reasons:

\noindent
\vspace*{0.1cm}
(1) Additional $S^0_i, P^0_i$ states must be lighter than $ 2m_b$
to yield an enhancement for the multi-lepton/multi-tau final states at 
colliders. Thus, all the four daughter
leptons (through $S^0_4 \to 2 S^0_i,2P^0_i \to 2l^+2l^-,\,\,
l=\ell(\equiv e,\mu),\,\tau$) are usually not highly boosted and 
often not well separated from each others. In this situation
one might need to adopt modified search criteria to identify
these leptons/taus, which are somehow inadequate to date. 
With existing analysis methods a pair of leptons/taus from
such a light $S^0_i,P^0_i$ perhaps effectively appears as 
one single particle \cite{Cerdeno:2013cz}.

\noindent
\vspace*{0.1cm}
(2) A similar approach with $\tau$s is a bit more complicated
since for taus the detection efficiency strongly depends on 
their transverse momentum, $p_{\rm T}$~
\cite{Bayatian:2006zz,ATLAS:2013uma,Mahlstedt:2014vxa}
(see references in \cite{Mahlstedt:2014vxa} also). Normally for 
low $p_{\rm T}\,(\lsim 30$ GeV),
the $\tau$ detection efficiency falls very sharply \cite{Mahlstedt:2014vxa}. 
Thus, not all the four 
taus originating through $S^0_4 \to 2S^0_i,2P^0_i \to 2 \tau^+2\tau^-$
are detectable at the experiment. In addition, proper identification
of a $\tau$ as $\tau$-jet occurs only when a $\tau$ decays hadronically,
which happens only 18\% of the times with 4$\tau$. Situation with leptonic
$\tau$ decays to muon may appear favourable since muon detection
efficiency is very high at the LHC. However, this is not a realistic
analysis mode since the Br($4\tau \to 4 \mu$) is only $\sim 0.1$\%.
In the $\mu\nu$SSM, however, it is possible to generate mass splittings 
among the light $S^0_i,P^0_i$ states by tuning the relevant parameters
\cite{Fidalgo:2011ky} so that one of the lighter states
decays to di-muon while other(s) to\footnote{A similar situation is also possible in 
the NMSSM by a distribution of the different decay Brs. For the $\mu\nu$SSM 
with the three ${\hat\nu^c}$s, this emerges naturally.} $\tau^+\tau^-$. 
Concerning Br, the $2\ell2\tau$ state is intermediate to $4\tau \to 4e,4\mu, 2e2\mu$
and $4\tau$-jets with Br$(2\ell2\tau{\text{-jets}})_{\rm max}\sim 10$\% for 
$0.5~{\rm GeV} \lsim m_{S^0_i,P^0_i}\lsim 2 m_\tau$ \cite{Belyaev:2010ka} although
the problem of narrower isolation criterion, as stated already in (1) persists.

It has to be emphasised here that regarding the branching fraction, the state with 
4$\tau$-jets dominates over $2\ell2\tau$-jets. The latter, however, is advantageous
when $\tau$ detection efficiency is taken into account. Moving towards
a different aspect of the final states, the processes $S^0_i,P^0_i \to 2e,2\mu$
theoretically appear with zero $\MET$, although in reality a non-vanishing
$\MET$ may arise from the possible mis-measurements. On the contrary,
$S^0_i,P^0_i \to 2\tau$ state is always accompanied with a non-zero
$\MET$ originating from multiple neutrinos (minimum being
4 for $4\tau \to 4\tau$-jets) which appear in the $\tau$ decay.
The presence of four neutrinos, however, does not guarantee a large 
$\MET$ due to a possible collinearity among them 
\cite{Cerdeno:2013cz,Cerdeno:2013qta}.

\vspace*{0.25cm}
\noindent
{\textit {\textbf {Decaying to jets:}}}
In the same spirit, as stated earlier, the light $S^0_i,P^0_i$ states
can also decay predominantly into a pair of jets depending on $m_{S^0_i,P^0_i}$.
These decays are further classified into two groups,
(a) a pair of light jets $(m_{S^0_i,P^0_i}\lsim 2m_\mu)$
including $c\bar{c}, \, gg$ $(2m_c\lsim m_{S^0_i,P^0_i} \lsim 2m_\tau$ 
and $m_{S^0_i,P^0_i} \to 0,$ ${\rm respectively})$ and
(b) into $b\bar{b}$ $(m_{S^0_i,P^0_i} $ $\gsim$ $2m_b)$.

The first option (a) has several shortcomings. To start with, 
this scenario is not generic
in $m_{S^0_i,P^0_i}$ as in the case of leptonic 
($e$ and $\mu$) modes. Secondly, the
jets produced in this way are narrowly separated just
like the earlier discussion with leptons and taus. The third and the 
most severe issue is to disentangle these jets from the backgrounds.
These jets are naturally soft as they are originating from the
decay of the $S^0_4$, with a mass about 125 GeV. 
Thus, their information is practically lost
within the huge QCD backgrounds, associated with a hadronic collider like the LHC.

Moving towards possibility (b), the processes $S^0_i,P^0_i\to b \bar{b}$
are the most generic decay mode for $S^0_i,P^0_i$ over a 
wide range of $m_{S^0_i,P^0_i}$, 
i.e. $2m_b\lsim m_{\S_i,\P_i}\lsim m_{\S_4}/2$. 
In addition, with $m_{S^0_i,P^0_i}\gg 2m_b$,
the produced $b$-jets can be well separated in nature. Further, concerning
the backgrounds, one can use the favour of $b$-tagging to discriminate
this signature from the backgrounds. The main problem with the $b$-jets
is the same as that with the $\tau$-jets, i.e. their detection efficiency
is also $p_{\rm T}$ dependent \cite{Bayatian:2006zz,ATLASb}. 
Thus, the process $S^0_4 \to 4b$-jets 
suffers additional suppression which might lead to a poor statistics.
Actually, with a mother particle of about 125 GeV mass, 
$p_{\rm T}$ for $3^{\rm rd}$ and/or
$4^{\rm th}$ $b$-jet can be low enough to fulfil the trigger requirement.
One should note that increasing luminosity
does not assure a better statistics for this signal, 
since this also results in a potential growth
of the QCD backgrounds. It needs to be emphasised here that
one can get higher boost for these jets (leptons/taus) coming from
the $S^0_i,P^0_i$ states when cascades with heavier particles are considered. 
However, these processes normally suffer extra suppression from
Brs in longer cascades and/or in production cross-section due
to the large masses of the concerned particles. Non-zero $\MET$ can exist for the multi-jet 
final states, e.g. through semi-leptonic b-decays.

\vspace*{0.25cm}
\noindent
{\textit {\textbf {Decaying to photons:}}}
Processes like $S^0_i,P^0_i \to \gamma \gamma$ are usually
suppressed in Brs due to the singlet nature of the mother particles
on top of the loop suppression. Only in the limit of 
sufficiently light $S^0_i,P^0_i$ $(\lsim 3m_\pi)$, this mode can 
lead the race \cite{Dobrescu:2000jt}. However, with
very small $m_{S^0_i,P^0_i}$, just like two earlier
scenarios $S^0_4\to 2S^0_i/2P^0_i \to 4 \gamma$ \cite{ATLAS:2012soa} 
will appear as 2$\gamma$s at the collider \cite{Dobrescu:2000jt}.
On the contrary, with heavier $m_{S^0_i,P^0_i}$
theoretically a clean $S^0_4 \to 4 \gamma$ signal is expected.
Unfortunately, this situation suffers huge Br suppression
$(\sim 10^{-5})$ \cite{Chang:2006bw}. Hence, unless
LHC attains a very high luminosity, this unique
channel is hardly recognisable
in spite of a negligible associated backgrounds. 
Theoretical $\MET$ prediction is zero for this signal.
It is to be noted that in a scenario when $S^0_i,P^0_i$ are very pure
singlets, Br$(S^0_i,P^0_i\to 2\gamma)$ can enhance
significantly at the cost of the reduced tree-level couplings
to fermions. In this scenario Br$(S^0_4 \to 4\gamma)$ 
can rise by orders of magnitude \cite{Dermisek:2007yt}. 

\vspace*{0.25cm}
\noindent
{\textit {\textbf {Decaying to mixed final states:}}}
In the NMSSM, depending on the respective Brs and masses, 
a pair of light scalars/pseudoscalars
can decay into two different modes. For example, one
of them decays into $\tau^+\tau^-$ while the other into $\mu^+\mu^-/b\bar{b}$.
This way, depending on the mass of the mother particle,
one can get mixed final states like $2\mu2\tau,\, 2\tau2b$, 
$2\gamma\,2j~({\rm light~jets})$ etc.
Most of these novel signals are, however, suppressed due to
Br multiplication. Being precise, a scalar/pseudoscalar with
mass $\gsim 2m_b$ typically has Br$(\S/\P\to \tau^+\tau^-) \sim 0.1$
and Br$(\S/\P\to b\bar{b}) \gsim 0.9$. Thus, the resultant
$2b2\tau$ state has an effective suppressed Br $\sim 9\%$.
On the contrary, for the $\mu\nu$SSM a splitting within different
$S^0_i,P^0_i$ is naturally possible \cite{Fidalgo:2011ky}.
Hence, with the proper mass scales when one of the $S^0_i,P^0_i$
decays to $b\bar{b}$, another one can easily decay to $\tau^+\tau^-$
with Br $\sim 1$ for both of the modes. This way the $\mu\nu$SSM
can uniquely escape the problem of Br suppression as noted
in ref. \cite{Fidalgo:2011ky}. We note in passing that a similar
situation is also affordable in the NMSSM with more than one
singlet. However, this is a rather forceful construction
while in the $\mu\nu$SSM the existence of three $\hat\nu^c$s is
well motivated by the SM family symmetry. The amount of $\MET$
associated with these signatures can vary from zero to moderate
values, depending on the decay modes.

\vspace*{0.25cm}
Finally, to conclude the discussion with the light $S^0_i,P^0_i$ states,
we describe possible leading backgrounds, without which
these analyses would remain incomplete. For all the decay modes
mentioned above, the dominant SM backgrounds arise from
Drell-Yan (DY), electroweak di-boson $(WW, WZ, ZZ/\gamma)$, $b\bar{b}$,
di-leptonically decaying $t\bar{t}$ and $W/Z + {\rm jets}$.
Some other sub-leading backgrounds can appear from electroweak 
tri-boson $(WWW, WWZ, ZZZ/\gamma$), $t\bar{t}W/Z$, etc.,
which may yield sizable contributions with larger 
centre-of-mass energy (E$_{\rm CM}$) and higher integrated 
luminosity $(\mathcal{L})$. These backgrounds
can somehow be ameliorated by studying di-jet/di-tau or 
di-lepton ${\rm M}_{\rm T2}$ \cite{Lester:1999tx,Barr:2003rg}/invariant mass $m_{inv}$
distributions, that are expected to peak around $m_{S^0_i,P^0_i}$s. 
This same logic is also applicable for the backgrounds arising 
from the MSSM, but fails for the NMSSM backgrounds. In the NMSSM,
just like the $\mu\nu$SSM, di-lepton or di-jet/di-tau $m_{inv}$/M$_{\rm T2}$
distribution can peak around $m_{S^0_i,P^0_i}$ and thus, produces
irreducible backgrounds to these class of signals. However, if
several and non-degenerate singlets are favoured by the nature,
then the $\mu\nu$SSM can give unique collider signals \cite{Fidalgo:2011ky}
in terms of the mixed final states. As an example, one can 
observe two different peaks in the $m_{inv}$/${\rm M}_{\rm T2}$ distributions
corresponding to two different $m_{S^0_i,P^0_i}$. A similar
scenario is beyond the scope of the standard NMSSM with only one singlet.
Note that a NMSSM theory with three $\hat\nu^c$s \cite{Kitano:1999qb} 
produces an irreducible impostor to all the signals
of the $\mu\nu$SSM even with $R_p$ conserving vacua.
However, in this case $\MPT$ could be larger and the scenario is 
constrained from dark matter searches.
%

\subsection{Light neutralinos}
\label{lightN}
In this subsection we moved to the study of 
light neutralinos and their phenomenological
consequences in $S^0_4$ decays. Considering only the on-shell
$S^0_4$ decay, as stated earlier, one concludes 
$2m_{\n}$ $\lsim$ $m_{S^0_4}$.
Clearly, from the lighter chargino mass bound \cite{Beringer:1900zz} the
possible leading composition for such light neutralinos is either bino- or
singlino-like (i.e., right-handed neutrino-like) or a bino-singlino mixed state. 
The chargino mass bound also implies that the minimum of $(\mu,M_2)$
(the parameters that control $\widetilde\chi^\pm$ mass with
$M_2$ as the $SU(2)$ gaugino soft-mass) must be $\gsim 100$ GeV.
Further, a bino- \cite{Heinemeyer:2007bw,Dreiner:2009ic} or 
singlino-like \cite{Ghosh:2014rha} nature is also necessary for 
a light\footnote{It has been reported in ref.\cite{Adhikari:1994wh} that a very
light neutralino can receive constraints from $B$-physics. However,
ref. \cite{Dreiner:2009er} has argued the absence of any such effects
for the MSSM with minimal flavour violation.} $\n$ 
to survive the constraints of measured $Z$-decay
width \cite{Beringer:1900zz}.
In this article we stick to a situation where $\widetilde \chi^0_{4,5,6}$
are singlino-like while $\widetilde \chi^0_{7}$ is bino-like. In addition,
we choose $2m_{\widetilde \chi^0_{7}}\gsim m_{S^0_4}$ 
(will be explained subsequently) and thus, concentrate
on singlino-like light neutralinos with\footnote{For simplicity, we consider 
singlinos that are quasi-degenerate in masses.} 
$2m_{\widetilde\chi^0_{4,5,6}}\lsim m_{S^0_4}$.

\FIGURE{\epsfig{file=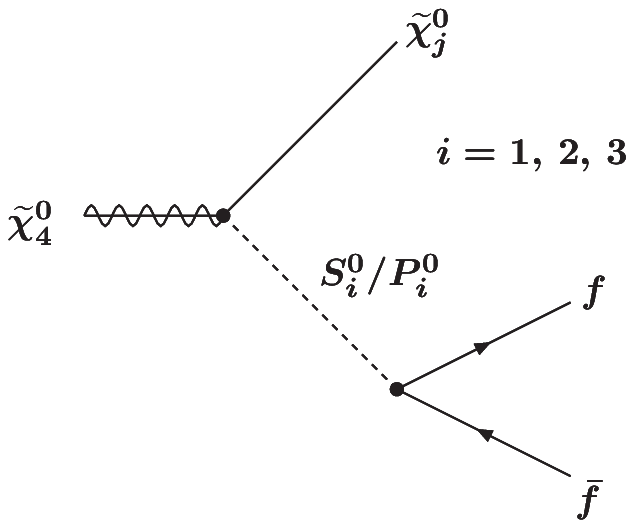,height=4.5cm,width=8.7cm,angle=0}
\caption{Figure showing $\n_{4} \to \n_j + \S_i/\P_i$
decay processes, followed by $\S_i/\P_i \to f \bar{f}$ decays.
with $f$ denoting a possible final state particle,
e.g. a lepton/tau/jet/photon etc.} 
\label{fig:chidecay}}

We begin our discussion with the novel aspect
of the $\mn$ to accommodate displaced and yet detectable
leptons/taus/jets/photons at colliders 
\cite{Bartl:2009an,Fidalgo:2011ky,Ghosh:2012pq}. The 
normal decay modes for the lightest
neutralino, $\n_4$, is primarily through
an electroweak SM gauge boson. However, when $m_{\n_4}<M_W$,
the associated decay lengths are often beyond
the charge tracker of the LHC, i.e. larger than $1$ m,
due to the presence of an off-shell intermediate $W^\pm,\,Z$. Particularly,
for $m_{\n_4}\lsim 30$ GeV the decays occur
outside the detector coverage \cite{Bartl:2009an}. 
Hence, $\S_4 \to \n_4\n_4$ process
yields a pure $\MPT$ signal, just like the SUSY models
with conserved $R_p$.
In the $\mn$ with extended field content
one can, however, get lighter $\S_i,\P_i$ states
below $m_{\n_4}$ for suitable parameter choices 
\cite{Bartl:2009an,Fidalgo:2011ky}.
Hence, the presence of a new two-body $\n_4$ decay
like $\n_4 \to \S_i/\P_i + \n_j$ can reduce
the $\n_4$ decay length drastically \cite{Bartl:2009an} even when it is
very light\cite{Fidalgo:2011ky,Ghosh:2012pq}. These
decay modes are shown in figure \ref{fig:chidecay}. 
These decays dominate even when $\S_i/\P_i$ states are
slightly heavier than $m_{\n_4}$ \cite{Ghosh:2012pq}.
An example of this kind, when $\S_4 \to \n_4\n_4$
decay leads to the displaced but detectable multi-$\tau$
+ $\MET$ final state, has already been analysed
in ref.~\cite{Ghosh:2012pq}.
A note of caution has to be emphasised here, i.e.
reduction of the decay length in the absence
of light $\S_i,\P_i$ states makes it rather hard for 
a light $\n_4$ in other $\rpv$ models to decay within
the detector coverage. Nonetheless, for certain values 
of the concerned couplings, a very light 
$\n_4$ can decay in the range of 1 cm - 3 m
for MSSM with trilinear $\rpv$ \cite{Barbier:2004ez,Dreiner:2012ex}.

It is now important to address the composition of
a light $\n_4$. Note that with a simple choice of quasi-degenerate,
flavour diagonal $\kappa_{ijk}$, i.e. say $\kappa_{i}$
and universal $\nc$, one encounters two 
experimentally viable possibilities, (1)
$\n_{4,5,6}$ are singlino-like while $\n_7$
is bino-like and (2) a bino-like $\n_4$ lies
below singlino-like $\n_{5,6,7}$. 
The $U(1)$ gaugino soft-mass $M_1$ is the 
key parameter to control the mass scale of 
a bino-like $\n$ and thus a light ($\lsim m_{\S_4}/2$) bino-like $\n_4$ 
requires a $M_1$ lighter or around $60$ GeV. Such a small
$M_1$ value, when considered together
with the experimentally hinted scale of 
gluino mass, i.e. $m_{\widetilde g}\gsim 1.2$ TeV \cite{ATLAS-susy13,CMS-susy13},
requires breaking of the gaugino universality relation.
For a singlino, mass scale is determined by
$\ka$ and $\nu^c$ \cite{LopezFogliani:2005yw,Escudero:2008jg,Ghosh:2008yh}.
The mass scales for $\S_i,\P_i$ (see section \ref{spsn} for details)
are mainly governed by $\ka,\,\nc$ and $A_\ka$ parameters
\cite{Escudero:2008jg,Ghosh:2008yh}. Thus, simultaneous
presence of the lighter $\S_i,\P_i$ states are more
feasible with a singlino-like $\n_4$ compared
to a bino-like $\n_4$.
Note that singlino-like quasi-degenerate $\n_{5,6}$
can also decay through $\S_i/\P_i$ as shown in 
figure \ref{fig:chidecay}. With $\n_{4,5,6}$
closely spaced in masses, one also encounters
$3$-body decays like $\n_{5,6}\to \n_{4,5}+\mu^+\mu^-/$jet pair etc.
These final state particles, coming through the off-shell $\S_i/\P_i$,
normally remain experimentally undetected 
due to their soft-nature \cite{Ghosh:2012pq}, 
although the final state particle multiplicity is rather large. 
It is possible to evade these soft final states by introducing 
large splittings among $\kappa_i$s, however, at the 
cost of an enlarged set of parameters and normally reducing
the predictivity of the model.

Let us now try to justify our choice of $2m_{\n_7}$ $\gsim$ $m_{\S_4}$.
First of all, as already stated, from theoretical prejudice a 
scenario like $2m_{\n_7}\lsim m_{\S_4}$ for a
bino-like $\n_7$ requires
breaking of the gaugino universality condition
at the high scale. Secondly a light $\n_7$
naturally enters into $\S_4$ decay chains and 
yield a signal like $S^0_4 \to 2\widetilde \chi^0_7 \to 
2\widetilde \chi^0_4 + 2S^0_i,2P^0_i \to$
a combination of four leptons/taus/jets/photons 
+ $\MET$. Here $\widetilde \chi^0_4$ decays
according to figure \ref{fig:chidecay}. 
Unfortunately, with a light mother particle
like $S^0_4$, most of these jets/leptons are not well boosted as well
as most likely not well isolated. Consequently, most of these novel
multi-particle final states remain experimentally undetected. 
Thus, in this article we mainly discuss about singlet-like $\n_{4,5,6}$ 
with a bino-like $\n_7$ such that $2m_{\widetilde \chi^0_7}\gsim m_{S^0_4}$.
We note in passing that for the sake of completeness
we do discuss the scenario with a bino-like $\n_4$ while
discussing new two-body Higgs decays in section \ref{decay}.

Since we stick to $2m_{\widetilde\chi^0_4} \lsim m_{S^0_4}$,  
$\widetilde \chi^0_4 \to S^0_i/P^0_i + \widetilde\chi^0_j$ remain
the leading $\widetilde\chi^0_4$ decay modes, even when $S^0_i,P^0_i$ are slightly
heavier than $\n_4$ \cite{Ghosh:2012pq}. Now it is apparent 
that the decay products for $\widetilde \chi^0_4$
will trail the same for $S^0_i,P^0_i$ as already addressed in the previous
subsection. One should note that compared to the prompt decays, the 
amount of $\MET$ will be different with
two extra neutrinos coming form a pair of $\widetilde\chi^0_4$ decay.
A class of possible final states from $S^0_4 \to 2 \widetilde\chi^0_4$ 
are $4b+\MET,\,2b2\tau+\MET,\,2\mu2\tau+\MET,\,2\gamma2j+\MET$ etc.
However, a $\widetilde\chi^0_4$ decay has an extra advantage
over the same for $S^0_i,P^0_i$, which is the appearance of displaced vertices.
In this way the $\mu\nu$SSM can produce potentially non-standard
signals, e.g. displaced multi-photons at colliders. A displaced multi-photon
signal is normally very suppressed for minimal $\rpv$ models 
since $\widetilde \chi^0_{\rm LSP~} \to 
\widetilde \chi^0_i \gamma$ appears through the
one-loop processes \cite{Hall:1983id,DS,oai,oai:9903418}.   
The presence of displaced vertices are useful  
to reject the possible SM backgrounds efficaciously which are generically 
prompt\footnote{Normally this also includes displaced objects from
B or D meson decays, unless the boost is very high or the 
associated ${\widetilde\chi^0_4}$ decay length
is very small.}. Prompt SUSY backgrounds are also differentiable in
the same fashion. SUSY backgrounds with displaced objects can be
separated by constructing the di-lepton/di-jet/di-tau invariant mass/$\rm M_{\rm T2}$
distribution that peaks around a scalar/pseudoscalar mass with a long tail from 
possible wrong combinatorics. A possible look-alike can
appear from the NMSSM in a fine tuned corner of the parameter space 
\cite{Ellwanger:1997jj,Ellwanger:1998vi}. However, as
argued in ref. \cite{Ellwanger:1998vi}, the appearance of a mesoscopic
decay length (1~cm -- 3~m) is not possible in this scenario.
Hence, these signatures remain rather unique to 
SUSY models with singlets with or 
without $\rpv$, e.g. the $\mu\nu$SSM or the NMSSM with $3\hat\nu^c$
for a range of $m_{\widetilde\chi^0_4}$, although
the latter with $R_p$ conserving vacua produces
larger $\MPT$ and suffers additional constraints from
dark matter searches.

\vspace*{0.2cm}
To recapitulate, we have addressed the complete relevant phenomenological scenarios
that can arise from the light scalars, pseudoscalars and neutralinos. We 
have also discussed their consequences in $S^0_4$ decay modes. 
We are now in the ideal state to identify the set of
parameters which assure these light states. However, before
that it will be useful to discuss the parameter space 
in the $\mn$ that can accommodate a SM-like Higgs with a
mass about $125$ GeV. This is also rather necessary as
we aim to explore various light states in the light of the
$S^0_4$ decay that has a mass around 125 GeV.
One should note that the presence of these light states can also lead
to new signals at colliders for other heavier SM particles.
For example, consequences of the light scalars, pseudoscalars and neutralinos
in the $\mu\nu$SSM in the decays of $W^\pm$ and $Z$ bosons have already been
addressed in ref. \cite{Ghosh:2014rha}.
We note in passing that, since we aim at covering
all phenomenological consequences of the light scalars, pseudoscalars
and neutralinos in the SM-like Higgs phenomenology,
analyses with numerical examples are beyond the theme of
the current work. We will address these issues in a 
set of forthcoming publications \cite{glmmrF2}.
%

\section{The SM-like Higgs in the {$\bm \mu\nu$}SSM}
\label{Higgs-sector}
%

After the discovery of a new scalar boson \cite{:2012gk, :2012gu} 
with properties like the SM Higgs boson, the constraints on the parameter space
and mass spectrum of the SUSY models are severely tightened. It is hence absolutely
relevant to re-investigate the $\mu\nu$SSM parameter space \cite{Escudero:2008jg} 
to accommodate this new scalar and to analyse its general 
phenomenological consequences respecting various experimental results.

We start with a note on the tree-level analysis of Higgs mass 
and discuss the effect and relevance of the loop
corrections in succession. Further, we also highlight the possible differences
of the concerned mass spectrum with that of the MSSM. We want to 
emphasis here that the analysis presented in this section has 
notable similarity with that of the NMSSM Higgs sector. However, 
$\rpv$ and an enhanced particle content offer a novel and unconventional
phenomenology for the $\mu\nu$SSM \cite{Ghosh:2008yh,Bartl:2009an,
Bandyopadhyay:2010cu,Fidalgo:2011ky,Ghosh:2012pq,Ghosh:2014rha} which deserves 
a systematic analysis.

At this juncture it is relevant to mention the value of $m_{\S_4}$ that will be
used to estimate some other relevant quantities
in this section. The latest ATLAS result
gives $m_{\S_4} = 125.36 \pm 0.41$ GeV,
after combining the measured values from $\S_4\to ZZ^* \to 4$ leptons
and $\S_4\to \gamma\gamma$ decay modes \cite{Aad:2014aba}.
For the CMS the latest number, after combining the measurements over the same
two decay modes, gives $m_{\S_4}=$ $125.03^{+0.29}_{-0.31}$ GeV 
\cite{CMS:2014ega}. In this article we choose to work with $m_{\S_4} = 125$ GeV
which will be used henceforth. 
This value of $m_{\S_4}$ is within the $1\sigma$ 
range of the ATLAS and CMS observations.


In the $\mu\nu$SSM, as already stated in section \ref{The-model},
the doublet-like Higgses mix with the three families of the left- and the right-handed
sneutrinos. Through the mixing with the right-handed sneutrinos, 
the lightest doublet-like Higgs mass at the tree-level 
receives an extra contribution\footnote{
A similar feature exists for the NMSSM \cite{Drees:1988fc,
Ellis:1988er,Binetruy:1991mk, Espinosa:1991gr,Espinosa:1992hp}, however, with
only one $\la$.} in the $\mu\nu$SSM in such a way
that the upper bound is now given by \cite{Escudero:2008jg}
\bea
(m^{tree}_{h})^2 \leq M^2_Z \left ( \cos^2 2\beta + \frac{2 {\bm \lambda}^2 
\cos^2 \theta_W}{g^2_2} \sin^2 2\beta \right ) \approx  M^2_Z \left 
( \cos^2 2\beta + 3.62 \, {\bm \lambda}^2  \sin^2 2\beta \right ),
\label{boundHiggs}
\eea
where $M_Z$ denotes the mass of $Z$ boson, $g_2$ is the $SU(2)$
gauge coupling, $\tb=\frac{v_u}{v_d}$ 
and $\theta_W$ is Weinberg mixing angle. The extra piece
of contribution grows with small tan$\beta$ and large 
${\bm \la} (\equiv |\sqrt{\sum \la^2_i}|=\sqrt{3}\la$ assuming
universal $\la_i$, which will be used henceforth throughout the 
text). Equation~(\ref{boundHiggs}) can be written in a more
elucidate form as
\bea
(m^{tree}_{h})^2 \leq  \frac{M^2_Z}{(1 + {\rm tan}^2\beta)^2} \left[
(1 - {\rm tan}^2\beta)^2  + 14.48 \, {\bm \lambda}^2 \, {\rm tan}^2\beta \right].
\label{boundHiggs1}
\eea
%
In the case of the MSSM the $2^{\rm nd}$ term of eq.~(\ref{boundHiggs1}) is absent.
Hence, the maximum possible tree-level mass is about $M_Z$ as
$\tb \gg 1$ and consequently a contribution as large as $0.38$ 
times of the tree-level mass from other sources (for example through the loops) 
is essential to reach the target of 
$125$ GeV. The necessity of a 
larger contribution over the tree-level mass
to reach the target of 125 GeV grows as $\tb$ takes moderate 
to small values. For example, with $\tb=2$, eq.~(\ref{boundHiggs1})
predicts the upper bound of $\mht$ about $55$ GeV. Hence,
to reach $125$ GeV one needs a contribution which is 
at least $\approx 1.3$ times larger compared to the $\mht$. 

On the contrary, as has already been mentioned in
ref. \cite{Escudero:2008jg}, in the $\mu\nu$SSM one can reach 125 GeV solely 
with the tree-level contribution. One can observe from eq.~(\ref{boundHiggs1}) 
that $\mht$ enhances with an increase in $\bla$. Thus, even at the limit
$\tb\to 1$, $\mht \neq 0$. The variation of the tree-level mass, $\mht$ as 
calculated using eq.~(\ref{boundHiggs1}), with a change in $\tb$
values for the different fixed values of $\bla$ is shown in figure~\ref{fig:lamHigg}.
Here $\tb$ values greater than $25$ have been intentionally truncated
since $\mht$ practically saturates around $M_Z$ for $\tb> 25$ with a mild
exception for $\bla=2$. It is also worth noticing that a 
class of flavour observables (e.g., $B_s^0\to \mu^+\mu^-$),
as already stated, depending on the other parameters 
posses high power sensitivity to tan$\beta$
which in turn can put strong constraints on the large $\tb$ values.
\FIGURE{\epsfig{file=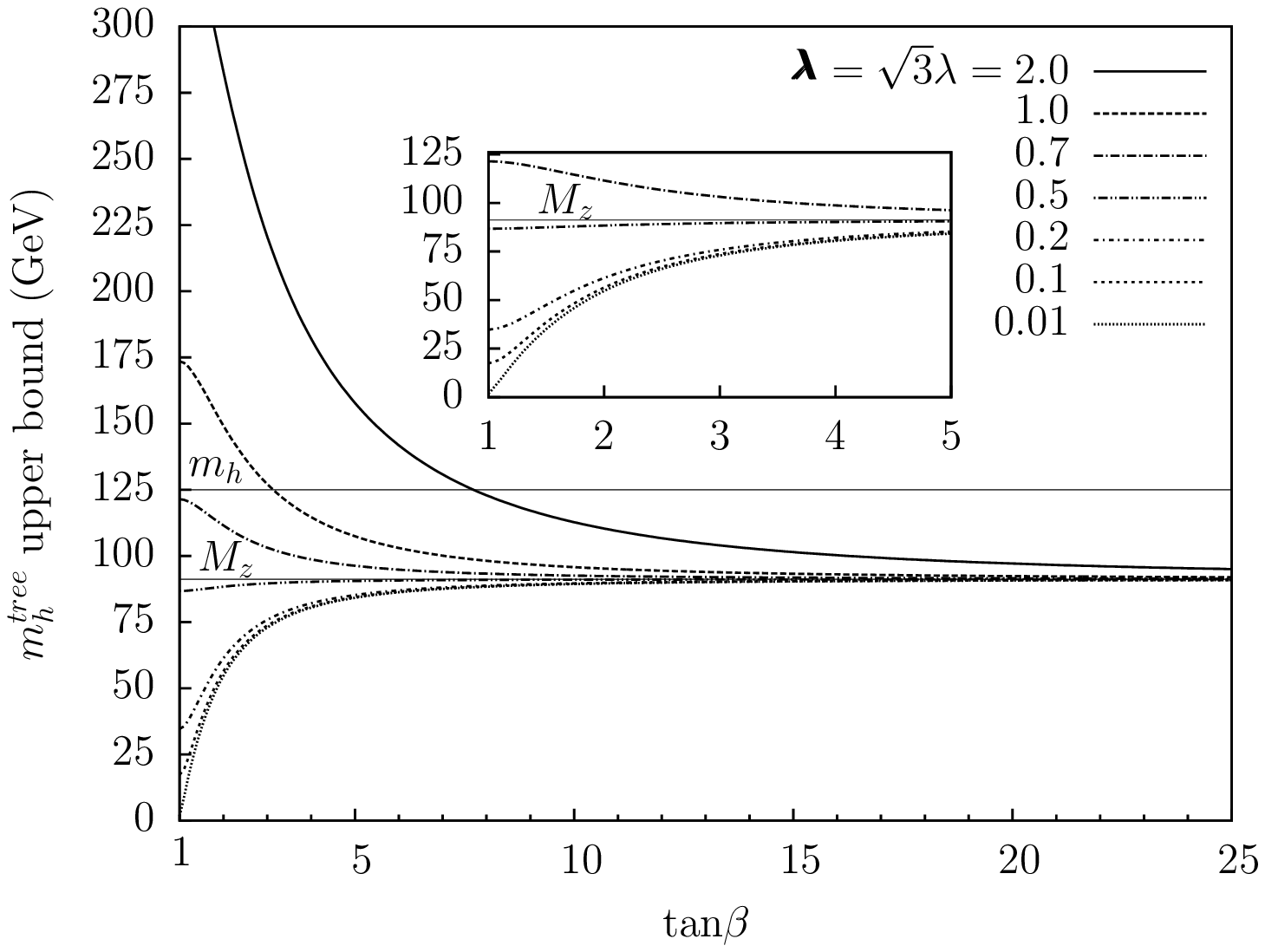,height=8cm,width=12cm,angle=0} 
\caption{Plot showing the variation of $\mht$ upper bound with $\tb$ 
for different $\bla~(\equiv \sqrt{3}\la)$ values as 
calculated using eq.~(\ref{boundHiggs1}). Here $m_h$ and $M_Z$ 
represent the experimentally measured mass of the Higgs and $Z$ boson, i.e.
125 GeV and 91.187 GeV, respectively. The region of $\tb\leq5$ has been \textit{zoomed}
for $\bla \leq 0.7$ values for the convenience of representation.} 
\label{fig:lamHigg}}
%

In order to discuss figure~\ref{fig:lamHigg}, let us choose three
regions in $\bla$ values, namely 
(a) {\textit{small to moderate, i.e. $\bla \lsim 0.01\,
{~\rm to~} \bla \leq 0.1$}}, (b) {\textit{moderate to large, i.e. $\bla > 0.1
{~\rm to~} \bla \leq 0.7$}}, which is the maximum
possible value of $\bla$ maintaining its perturbative
nature up to the scale of a grand unified theory (GUT) ($\sim 10^{16}$ GeV), 
and finally (c) {\textit{dominant, $\bla > 0.7$}}. These ranges will also be
useful later when we continue our discussion in 
section \ref{spsn} and section \ref{decay}.

\vspace*{0.25cm}
\noindent
(a) {\bf Small to moderate $\bla$:} In this range of $\bla$ values
the quantity $\bla^2$ takes values from $\sim 10^{-4}$ to $10^{-2}$.
Hence the maximum value of $\mht$ using eq.~(\ref{boundHiggs1}) with
$\bla=0.1$ goes as $\approx 56.5$~GeV for $\tb=2$ which is $\approx 1$ GeV
more compared to a similar situation in the MSSM. Keeping 
all the other parameters fixed, $\mht$ is estimated as
$\approx 55$ GeV for $\bla=0.01$ as well as for $\bla=0$. 
The real difference is only $\sim 0.02$ GeV when $\bla$
changes from $0$ to $0.01$. Since this change is rather 
insignificant, we do not explicitly show $\bla=0$ (MSSM-like) scenario
in figure~\ref{fig:lamHigg}. With larger $\tb$
values (say $10$ or more) this extra contribution diminishes
and $\mht \to M_Z$ as $\tb \gg1$. This feature is also apparent
from figure \ref{fig:lamHigg}. It is thus essential to have 
additional contributions to raise $\mht$ up to 125 GeV, as 
has been measured experimentally.

A possible source of extra tree-level mass can also arise
through the mixing of doublet-like states with other 
states like the left- and the right-handed sneutrinos. 
The mixing between the doublet-like states with the left-handed sneutrinos,
however, has negligible effect on the tree-level Higgs
mass as the concerned terms are suppressed 
through very small $Y_{\nu_{ij}}$ and $\nu_i$ \cite{Escudero:2008jg,Ghosh:2008yh}.
On the other hand, the mixing between the doublet-like
and the right-handed sneutrino-like states
appears through $\la_i$, which are usually
several orders of magnitude larger compared to $Y_{\nu_{ij}}$. These mixing can raise
the tree-level lightest doublet-like Higgs mass in the case when 
the right-handed sneutrino-like 
states are lighter compared to the lightest
doublet-like Higgs. 
Note that the parameters $\ka$ and $A_\ka$ are the key ingredients to determine
the mass scale of these right-handed sneutrino states \cite{Escudero:2008jg,Ghosh:2008yh}.
In this situation, the lightest 
doublet-like state feels a push away effect from
the lighter singlet-like states which can contribute
to push $\mht$ (as estimated using eq.~(\ref{boundHiggs1}))
a bit further towards 125 GeV. Unfortunately, for this range
of $\bla$ values the push-up effect is normally small 
owing to the small singlet-doublet mixing which is driven by 
$\bla$ \cite{Escudero:2008jg,Fidalgo:2011ky}. 

One can also get heavy singlet-like states with the other 
choices of $\ka, \Aka$. This scenario, however,
has the opposite effect on the doublet-like lightest
state, namely to lower the mass.

Necessity for an additional contribution is now apparent for this
corner of the parameter space to accommodate a 125 GeV 
doublet-like Higgs. This time the contribution is coming from a 
well known source, namely the loop effects \cite{Haber:1990aw,Altarelli:1990zd,
Hempfling:1993qq,Casas:1994us,Carena:1995bx,Carena:1995wu,Haber:1996fp,
Zhang:1998bm,Heinemeyer:1998np,Heinemeyer:1999be,Espinosa:1999zm,
Carena:2000dp,Espinosa:2000df,Ambrosanio:2001xb,Brignole:2001jy,
Carena:2002es,Martin:2002wn,Degrassi:2002fi,Frank:2006yh}. For 
$\tb\lsim 5$, a loop contribution as large as the 
tree-level mass (e.g., $\mht\approx 56$~GeV for $\tb=2$ and $\bla=0.1$) is required.
Thus, in this region of the parameter space the issue of accommodating
a 125 GeV Higgs is practically similar to that of the MSSM, where
large masses for the third-generation squarks and/or large trilinear soft-SUSY
breaking terms are essential \cite{Hall:2011aa,Heinemeyer:2011aa,Draper:2011aa}, 
without which a 125 GeV Higgs mass is hardly attainable.
The smallest A-terms and the average squark masses can 
be (with $\tb>20$) around $1000$ GeV and $500$ GeV, respectively. 
A small A-terms is possible only by decoupling the scalars to at least 
5 TeV \cite{Draper:2011aa}. 
A light third generation squark, especially a stop, 
on the other hand, is \textit{natural} in the 
so-called maximal mixing scenario \cite{Haber:1996fp}.
These issues indicate that the novel signatures
from the SUSY particles (e.g., from a light stop or sbottom)
are less generic in this region of $\bla$. 
Nevertheless, novel differences are feasible for Higgs decay phenomenology,
especially in the presence of singlet-like lighter states
which has already been discussed in section \ref{lightest}.

Let us finally note that
the effects of loop contributions are normally negligible for the
singlet states, however, when $\ka\sim 0.1$ or larger, the singlet states
can receive a large loop correction $\propto \ka^2$. 
This happens when the singlet-like states are heavier compared to the lightest
doublet-like state.

\vspace*{0.25cm}
\noindent
(b) {\bf Moderate to large $\bla$:} For this range of $\bla$
values $(>0.1 {~\rm to~} \leq 0.7)$ the maximum of 
$\mht$ can go beyond $M_Z$, especially
for $\tb\lsim 5$ and $\bla\sim 0.7$. This is also clear
from figure~\ref{fig:lamHigg}. In fact, depending on $\bla$,
in this region the maximum of $\mht$ can remain close to the 125 GeV
target. For example with $\bla=0.7$, $\tb=2~(5)$ gives
$\mht\sim 112~(96)$~GeV using eq.~(\ref{boundHiggs1}). This 
is $\approx 100\%~(14\%)$ enhancement compared to the MSSM scenario
with the same $\tb$. Thus, one needs $\sim 12\%~(30\%)$ contribution
from other sources to reach the 125 GeV milestone. The necessity
of larger (compared to the given numbers) additional contribution
emerges as $\bla$ picks up smaller values, say $0.2$. 
In this case with $\tb=2~(5)$, $\mht$ is estimated as
$\sim 61~(85)$~GeV and one needs rather large, $\sim 100\%~(47\%)$
contribution over the tree-level mass to achieve 125 GeV.
With an intermediate value, say $\bla=0.5$, $\tb=2~(5)$
gives $\mht$ as $\sim 88~(90)$~GeV and thus, $\sim 40\%$
extra contribution over the $\mht$ is needed. It is interesting
to see from the last calculation and also from figure \ref{fig:lamHigg}
that the $\bla=0.5$ line is almost overlapping to the $M_Z$ line
and consequently magnitude of $\mht$ or the amount of extra
contribution to reach 125 GeV remains practically
the same for all $\tb$ values. Being quantitative, as
$\tb$ changes from 2 to 10, the requirement of an extra
contribution over the $\mht$ to reach the goal of 125 GeV
changes by an amount of $\sim 4\%$. One 
should note that $\bla=0.2$ and $0.7$ are translated as
$\la\sim 0.12$ and $\sim 0.4$, assigning a universality for $\la_i$.

For this region of $\bla$, the singlet-doublet mixing is no
longer negligible, particularly as $\bla\to 0.7$. Thus,
a state lighter than 125 GeV with the leading singlet composition
appears rather difficult without a certain degree of tuning of the other parameters,
e.g. $\ka$, $\nu^c$, $A_\ka$, $\Ala$ etc. These issues will be addressed
thoroughly later in section~\ref{spsn}. In this situation
the extra contribution to $\mht$ is favourable through a push-up
action from the singlet states compared to small to 
moderate $\bla$ scenario. However, a sizable doublet
impurity makes it rather hard for these states to escape
the collider constraints. The situation is a bit ameliorated
with smaller $\bla$, say around $0.2$ or $0.3$.

Once again a contribution from the loops is needed to reach
the 125 GeV target. However, depending on the values
of $\bla$ and $\tb$ the requirement sometime is much softer
compared to small to moderate $\bla$ scenario. Beyond 
$\tb=10$, at least $\sim 35\%$ of the tree-level contribution
from the other sources
is required to reach 125 GeV even with $\bla=0.7$, which
is $\sim 5\%$ small compared to a similar scenario
with $\bla=0.1$. Considering the same analysis for 
$\tb=3$ one gets $\sim 48\%$ difference between $\bla=0.7$
and $\bla=0.1$ scenarios. Hence, depending on $\tb$ and $\bla$, the
necessity of heavy third-generation squarks and/or
large trilinear soft-SUSY breaking term may or may not appear
essential for this region \cite{Hall:2011aa}. For example, 
for the scenario studied in ref. \cite{Ghosh:2012pq},
where $\tb=3.7$ and $\la_i=0.11$ (i.e., $\bla \approx 0.2$), one needs 
$A_t=2.4$ TeV and stop masses about $1$ TeV.
Moving towards $\bla\sim 0.7$, on the contrary, room for 
the third-generation squarks lighter than 1 TeV
is possible. 
For example, with $\tb=2$ and $\bla=0.7$, stop masses and 
A-terms of about $300$ GeV are sufficient to raise the 
Higgs mass to 125 GeV \cite{Hall:2011aa}. 
It is also worth noticing that the naturalness is therefore improved 
with respect to the MSSM or smaller values of $\bla$.

Lighter singlet states, as already stated, are also feasible here with some degree
of parameter tuning. Although they lead to unusual
signatures at the LHC, however, a sizable doublet component
makes it hard for these states to escape a group of the experimental
constraints, as mentioned in the introduction.

\vspace*{0.25cm}
\noindent
(c) {\bf Dominant $\bla$:} If one relaxes the idea of 
perturbativity up to the GUT scale, large values 
$(>0.7)$ for $\bla$ emerge naturally\footnote{A similar scenario 
in the context of the NMSSM has been popularised as $\la$-SUSY 
\cite{Barbieri:2006bg}.}. Assuming a scale
of new physics around $10^{11}$ GeV, the perturbative
limit on $\bla$ gives $\bla \sim 1.0$ (i.e., $\la\sim 0.58$)\cite{Escudero:2008jg}. 
Pushing the scale of new physics further below to 10~TeV,
this limit gives $\bla\sim 2$ (i.e., $\la\sim 1.1$).
In this region, as also shown in figure \ref{fig:lamHigg},
the maximum of $\mht$ as evaluated from eq.~(\ref{boundHiggs1})
can remain well above 125 GeV even up to $\tb\sim 8$ for
$\bla\sim 2$. For $\bla=1$, a similar analysis gives $\tb\sim 2$
as the upper limit. Here with $\bla=1$, the 
maximum of $\mht$ for $\tb=2,\,5$ and $10$
is estimated as $\sim 150$~GeV, $108$~GeV and $\sim 96$~GeV,
respectively. With $\bla=2$ these numbers increase further,
for example, $\sim 113$ GeV when $\tb=10$. The requirement of
an extra contribution to reach the target of 125 GeV is thus, rather
small and even negative in this corner of the parameter space unless $\tb$
goes beyond 10 or 15 depending on the values of $\bla$.

A singlet-like state lighter than 125 
GeV is rather difficult in this corner of
the parameter space due to the large singlet-doublet mixing.
In fact even if one manages to get a scalar lighter
than 125 GeV with drastic parameter tuning, a push-up action 
can produce a sizable effect to push the mass of the lightest
doublet-like state beyond 125 GeV, especially for 
$\tb \lsim 10$ taking $\bla=2$. Moreover, a huge doublet
component makes these light states hardly experimentally acceptable.
In this region of the parameter space a heavy singlet-like 
sector is more favourable which can push $\mht$ down towards
125 GeV. A set of very heavy singlet-like states, even with
non-negligible doublet composition is 
also experimentally less constrained.

It is needless to mention that the amount of the loop correction
is much smaller in this region compared to the two previous
scenarios. For example, with $\tb=10$, one needs a loop
effect $\sim 11\%$ and $30\%$ with $\bla=2$ and $1$,
respectively. One should compare this with the 
maximum value of $\bla$ keeping perturbative nature up to
the GUT scale, i.e. $0.7$, where one needs $\sim 35\%$ contribution 
over the tree-level mass for $\tb=10$. 
Following the above discussion for large values of $\bla$, this 
region of the parameter space also favours third-generation 
squarks lighter than $1$ TeV, which can be produced
with enhanced cross sections and can lead to novel signatures
of the model at the LHC. 
Note that the light third generation of squarks
is still allowed by the LHC results, see e.g. 
refs. \cite{ATLAS-susy13,CMS-susy13}.
This feature can produce new signatures at colliders
with $\rpv$ for this region $\bla$ values, even when the singlet-like states remain
heavier, as stated earlier. One should note that 
for such a large $\bla$ value, new loop effects
from the right-handed sneutrinos with contributions $\propto \bla^2$
can generate an additional enhancement \cite{Degrassi:2009yq}.

We end our discussion on the dominant $\bla$ scenario
with a note of caution. It apparently seems that 
pushing the scale of perturbativity as low as possible
is useful to yield larger and larger $\bla~(>2$ for instance) values. 
However, $\bla \sim 3$ indicates the scale of new physics 
around $1$ TeV which appears to be an extinct possibility 
from the experimental observations since no definite excess over 
the SM predictions has been observed to date.

\vspace*{0.2cm}
The discussion presented so far favours, in order 
to obtain the light singlet-like states, small to moderate $\bla$
region where the singlet-doublet mixing is small.
Hence in this corner of the parameter space 
one can easily get the light singlet-like states with suitable choices
of $\ka,\,\Aka$ and $\nu^c$ \cite{Bartl:2009an,Fidalgo:2011ky,Ghosh:2012pq}. 
Although a large loop contribution is essential
for this region of the parameter space to reach the 125 GeV target, the associated
lighter states have notable consequences in the collider phenomenology of 
the scalar sector, as already stated in section \ref{lightest}. It is 
now absolutely essential to investigate the behaviour of 
$S^0_i,P^0_i$ and $\n_{i+3}$ masses for these three regions of 
$\bla$ values, which is what we plan for the next section. 

\section{Masses of the singlet-like states in the {$\bm \mu\nu$}SSM}
\label{spsn}

In this section we first aim to identify the relevant set of
parameters which controls the mass scale
of the singlet-like scalars, pseudoscalars and neutralinos in
the $\mu\nu$SSM. Subsequently, we present a set of general
expressions for the mass terms of the singlet-like $\S_i,\,\P_i$
and $\n_{4,5,6}$ states. We further extend our analyses
over the three regions of $\bla$ values, as of the last section,
accompanied by a discussion regarding the scale of the other crucial
parameters.  
In this section and from henceforth
we use $\widetilde \chi^0_{i+3},~i=1,2,3$, to denote 
the three lightest neutralinos in lieu of $\widetilde \chi^0_{4,5,6}$.

In order to proceed systematically it is crucial to identify 
first the set of most relevant parameters which controls the tree-level 
masses and mixing of the electroweak scalar and fermion sectors in the $\mu\nu$SSM. 
Considering universal $\nu^c_i~(\equiv \nu^c)$, 
flavour-diagonal but quasi-degenerate $\kappa_{ijk}~(\equiv \ka_i)$ together with the
universal and flavour-diagonal $A_\la$ and $A_\ka$, the  
parameters that control the electroweak fermions are 
\bea 
M_1,\,M_2,\,\bla,\, \kappa_{i}, \,\nu^c, \,{\rm tan}\beta \,.
\label{EWF-param1}
\eea
In the same spirit, the relevant parameters for the scalars (CP-even and odd)
are 
\bea 
\bla,\,\kappa_{i},\, \nu^c, \,{\rm tan}\beta,\, A_\la,\, A_\kappa.
\label{EWF-param2}
\eea
Note that with our choice of quasi-degenerate $\ka_i$
and universal $\Aka$, each of the three $S^0_i,\,P^0_i $ and $\n_{i+3}$ states are
closely spaced in masses. Assumptions for $Y_{\nu_{ij}}$ (chosen
to be flavour diagonal), $\nu_i$ and $A_\nu$ (chosen
to be flavour diagonal and universal) are not explicitly mentioned  
in eqs. (\ref{EWF-param1}) and (\ref{EWF-param2}). The 
left-handed neutrinos and sneutrinos, as already stated, couple to the remaining 
states through $Y_{\nu_{ij}}$ or $\nu_i$ \cite{LopezFogliani:2005yw,
Escudero:2008jg,Ghosh:2008yh}. 
Both of these $(Y_{\nu_{ij}},\,\nu_i)$ 
are constrained to be small $({\cal{O}}(10^{-6}-10^{-7}),\,{\cal{O}}(10^{-4}-10^{-5}),$
respectively \cite{LopezFogliani:2005yw,
Escudero:2008jg,Fidalgo:2009dm,Ghosh:2008yh,Bartl:2009an,Ghosh:2010zi}), 
in order to accommodate the
measured neutrino data \cite{Tortola:2012te,GonzalezGarcia:2012sz,Capozzi:2013csa}
with a electroweak scale seesaw mechanism \cite{LopezFogliani:2005yw,
Escudero:2008jg,Ghosh:2008yh,LopezFogliani:2010bf,Ghosh:2010ig,Bartl:2009an}.
Hence, the admixture of these states does not produce any significant
changes in the phenomenological analyses considered here and
thus, are not shown in eqs.~(\ref{EWF-param1}) and~(\ref{EWF-param2}).

It has already been emphasised that we are looking for the hints
of new physics with $S^0_4 \to XX$ decay modes with $X$ as the
light singlet-like $S^0_i,\,P^0_i,\,\widetilde \chi^0_{i+3}$.
The mass scales of these states, as shown in refs. \cite{Escudero:2008jg,Ghosh:2008yh}
depend on the set of parameters shown in eqs.~(\ref{EWF-param1}) and 
(\ref{EWF-param2}). We work in the region of low tan$\beta$ to avoid a
class of flavour physics constraints, e.g. $B^0_s \to \mu^+\mu^-$. 
Further, we assume  a higgsino-like $\widetilde\chi^\pm_4$ and 
$\mu \gsim 100$ GeV, consistent with the LEP lighter chargino 
mass bound \cite{Beringer:1900zz}. One advantage of 
this choice is that one can push $M_{2}$ to
proper values such that $m_{\widetilde g}\gsim 1.2$ TeV \cite{ATLAS-susy13,CMS-susy13}
appears naturally without spoiling the gaugino universality at the GUT scale. 
On the dark side, depending on the value of $\bla~(\sqrt{3}\la)$, a singlino-like
neutralino with mass $\lsim M_{Z}/2$ may posses sizable
higgsino admixture (remember $5^{th}$ term of eq.
(\ref{superpotential})) and thereby gets severely constrained from
the measured $Z$ decay width \cite{Beringer:1900zz}. Of course, one can live with 
a light $(\sim {\mathcal{O}}(100$ GeV)) gaugino-like $\widetilde \chi^\pm_4$ without 
the gaugino universality relation for $M_3$ yet maintaining\footnote{If one 
considers a heavy gaugino-like $\widetilde \chi^\pm_4$, for example with
$M_2 \sim 400$ GeV, the gaugino universality appears 
naturally with $m_{\widetilde g}\gsim 1.2$ TeV. The scenario 
with a heavy higgsino-like $\widetilde \chi^\pm_4$ is 
somewhat inconsistent with the idea of naturalness. The breaking of
universality relation between $M_{1},M_2$ will also increase the number of free 
parameters further.} $M_2 =2M_1$. In this case $\widetilde \chi^0_7$ is 
bino-like and can coexist with the measured $Z$ decay width \cite{Beringer:1900zz}
even being lighter than ${M_Z}/2$, since a tree-level \textit{$Z-$bino-bino}
coupling does not exist. We, however, do not consider 
this possibility in order to work with a minimal number of
the free parameters.

Summarising, the parameters relevant for this analysis are 
%
\bea
\bla,\,\kappa_{i}, \nu^c,\, tan\beta, \,M_1,\,A_\la \, {\rm and} \, A_\kappa.
\label{EWF-param3A}
\eea
%
It is clear from the mass matrices \cite{Escudero:2008jg,Ghosh:2008yh} that
$\kappa_i$ and $A_\kappa$ are the two crucial parameters to determine the
masses of the singlet states, originating from 
the \textit{self-interactions}. The remaining parameters $\bla$ (through
$\la$) and $A_\la$ not only 
appear in the said interactions, but also control the mixing between
the singlet and the doublet states and hence, contribute
in determining the mass scale.
In the limit of a vanishingly small $\bla$, the singlet
states are completely decoupled from the doublets\footnote{One should
simultaneously consider a very large $\nu^c$ such that $\mu$ ($\sqrt{3}\bla\nc$)
remains $\gsim 100$ GeV, as required by the LEP lighter chargino mass bound.}. It 
is thus apparent that $\bla$ is undoubtedly the most relevant parameter for this analysis.
Another aspect of the parameter $\bla$, i.e.
to yield additional contribution to the
tree-level lightest doublet-like Higgs mass has already been 
discussed in the previous section.

In order to proceed further, we continue with the three regions of
$\bla$ values as already introduced in the last section. 
Similar ranges of $\bla$ values, but in the context of SUSY 
signatures for the NMSSM has been mentioned in ref. \cite{Dreiner:2012ec}.
For each of these three $\bla$-zones, 
we will address in section \ref{decay} the phenomenological signatures
from the new $S^0_4$ decays, including effects coming from the variations
of $\ka_{i},\,A_\kappa $, tan$\beta$ and $A_\la$ parameters. 

In order to give a better interpretation of these scenarios,
we start with the approximate analytical formulae for $m^2_{S^0_i},\,
m^2_{P^0_i}$ and $m_{\widetilde\chi^0_{i+3}}$. 
A set of expressions for these masses with three families of the right-handed 
neutrino superfields using a simplified parameter choice 
(see eqs.~(\ref{EWF-param1}) and (\ref{EWF-param2})),
even in the region of small to moderate $\bla$, appears rather complicated 
due to the index structure of the parameters $\ka_i$s.
The expressions for the mass terms are relatively simpler for $P^0_i$ and $\chi^0_{i+3}$
in the limit of a complete degeneracy 
in all the relevant parameters, i.e. when eq.~(\ref{EWF-param3A}) is rewritten as
\bea
\bla~(\equiv \sqrt{3}\la),\,\kappa,\, \mu, 
\,{\rm tan}\beta,\,M_1,\, A_\la,\, A_\kappa,
\label{EWF-param3}
\eea
where we have replaced the $\nu^c$ parameter of eq.~(\ref{EWF-param3A}) with the $\mu$
parameter $\equiv \sqrt{3}\bla\nu^c$.
Note that, even with the assumptions of eq.~(\ref{EWF-param3}), 
the expressions for the squared mass terms remain rather complicated for the scalars $S^0_i$. 
In order to investigate the mass terms for $S^0_i,\,P^0_i,\,\widetilde\chi^0_{i+3}$ states
in more detail in the light of the relevant parameters, as given by eq.~(\ref{EWF-param3}), 
we start our discussion with $S^0_i$ and $P^0_i$
and later we continue with $\widetilde\chi^0_{i+3}$.

Being illustrative, in the $\mu\nu$SSM
with the three families of $\hat\nu^c_i$, dimensions of the scalar, pseudoscalar
and neutralino mass matrices are $8\times 8$, $8\times 8$
and $10\times 10$, respectively \cite{Escudero:2008jg,Ghosh:2008yh}.
Now, as already stated in section \ref{Higgs-sector}, the left-handed
sneutrinos couple with the remaining states (i.e.,
doublet Higgses and the right-handed sneutrinos) through
$Y_{\nu_{ij}}$ and $\nu_i$. Both of these are constrained
to be tiny, as required by a electroweak-scale seesaw mechanism 
\cite{LopezFogliani:2005yw,Escudero:2008jg,Fidalgo:2009dm,Ghosh:2008yh,Bartl:2009an}.
Hence, for all practical purposes, the effect of these mixing
are negligible on the remaining $5\times5$ scalar and pseudoscalar
mass matrices. Each of these $5\times 5$ matrices contains a $2\times2$ MSSM-like block
(top-left \cite{Escudero:2008jg,Ghosh:2008yh}),
a $3\times3$ block (bottom-right 
\cite{Escudero:2008jg,Ghosh:2008yh}) with the right-handed sneutrino mass terms
and finally two $2\times3$, $3\times2$ off-diagonal blocks that contain
the mixing between the right-handed sneutrinos and the doublet Higgses.
Note that the scalar, pseudoscalar and neutralino
mass matrices in the $\mn$ are symmetric \cite{Escudero:2008jg,Ghosh:2008yh}.

Concentrating on the $5\times5$ block, as mentioned above, 
the $3\times3$ right-handed sneutrino block, both for the  scalar 
and the pseudoscalar mass matrices, in the light of eq.~(\ref{EWF-param3}) 
symbolically can be written as $A\,I_{3\times3} + B\,(\mathcal{I}-I)_{3\times3}$.
Here $I_{3\times3}$ is a $3\times3$ identity matrix while $\mathcal{I}_{3\times3}$ 
is a $3\times3$ matrix with $1$ in all the nine places, $A$ and $B$ are functions of 
$\bla,\,\kappa, {\rm tan}\beta$, $A_\la,\,A_\ka$ and ${\nu^c}$. 

At this stage it is possible to apply a $3\times3$ rotation matrix\footnote{Note 
that the actual rotation matrix must be $5\times5$ in size, however, has a $2\times2$ 
identity matrix in the top-left $2\times2$ block and zeros in the off-diagonal
$2\times3$ block.}, constructed with its eigenvectors\footnote{One needs to use 
Gram-Schmidt procedure to  obtain a proper orthonormal set of eigenvectors since 
two of the eigenvalues of $A\,I_{3\times3} + B\,(\mathcal{I}-I)_{3\times3}$ 
matrix are identical.}, to obtain a $3\times3$ {\textit{rotated}}
right-handed sneutrino mass matrix with non-zero entries, $A-B,\,A-B$ and $A+2B$
only in the diagonals.
For the pseudoscalars, one also needs to apply a $2\times2$ rotation matrix\footnote{Again 
the actual one is $5\times 5$ in size
with a $I_{3\times3}$ for the right-handed sneutrino block
and zeros in the $2\times3$ off-diagonal block.} constructed out of $\sb,\,\cb$
($\sb=\frac{v_u}{v},\cb=\frac{v_d}{v}$), to rotate away the would
be Goldstone boson.

With this simple operation, two of the entries of the
{\textit{rotated}} right-handed sneutrino mass 
matrix, both for the scalars and the pseudoscalars,
are exactly degenerate in masses and are completely
separated from the rest of the mass matrix. In other words,
after the aforementioned $3\times3$ rotation, two of the three eigenvalues of 
the right-handed sneutrino mass matrix 
get decoupled and remain as the pure singlet-like states
without any doublet contamination.
The third eigenvalue, namely the one which goes as $A+2B$,
however, mixes with the doublet-like states and 
eventually appears with a much complicated form.

In the case of the pseudoscalar, after rotating away the Goldstone boson, 
the remaining matrix is a simple $2\times2$ matrix and thus, it 
is possible to extract the exact {\textit{modified}} (i.e., after 
mixing with the doublets) formula for that $A+2B$ eigenvalue. 

The absence of Goldstone mode
for the scalars, on the other hand, leaves the resultant mass
matrix $3\times3$ in size after
separating out the two degenerate eigenvalues. Hence, it is rather difficult
to obtain a simple analytical formula for the scalar right-handed sneutrino
that mixes with the doublet Higgses. A naive attempt to extract
this eigenvalue using the idea of $x_l \approx$
${\rm det}[{\rm Mat}_{n\times n}]/{\rm det}[{\rm Mat}_{(n-1)\times (n-1)}]$
($x_l$ represents the \textit{lightest eigenvalue}
of a $n\times n$ matrix \textit{`Mat'}) fails since the 
resultant expression contains terms up to $\bla^5$ (the parameter
which controls the singlet-doublet admixing, see eqs.~(\ref{superpotential}) and 
(\ref{Lsoft})) with non-negligible coefficients in front.

A note of caution must be emphasised here, i.e. with the choice
of $\ka_{ijk}=\k \delta_{ij} \delta_{jk}$, two of the eigenvalues
of the scalar and the pseudoscalar squared mass matrices appear
degenerate in masses with no doublet impurity. These states,
when appear in the bottom of the mass spectrum,
are highly stable in nature\footnote{The stability is not 
absolute as we have neglected the tiny but non-vanishing
contributions from the terms involving $Y_\nu$ or $\nu_i$. A similar
construction of the NMSSM with multiple 
singlets will give absolute stability to the set 
of lightest degenerate states.}. This artificial stability can
be broken by introducing mild splittings 
in $\kappa_i$ values \cite{Fidalgo:2011ky,Ghosh:2012pq}.
Their composition can, nevertheless, still remain dominantly
singlet-like depending on the values of the other parameters.

We have further verified that our approximate analytical formulae agree rather 
well with a full numerical
evaluation. In the limit of mild non-degeneracy in $\ka_i$s,
all three singlet-like states adhere doublet impurity,
however, the amount of doublet component is 
small for the aforesaid two degenerate states which
are now mildly separated in masses \cite{Ghosh:2012pq}.

Turning towards the neutralinos, one can think of a similar
rotation to the $7\times7$ block that contains
a $4\times 4$ MSSM-like block (top-left 
\cite{Escudero:2008jg,Ghosh:2008yh}), a $3\times3$ block (bottom-right 
\cite{Escudero:2008jg,Ghosh:2008yh})
with the right-handed neutrino Majorana mass terms
and two off-diagonal $4\times3$ and $3\times4$ blocks
that contain the mixing terms between the MSSM-like neutralinos
and the right-handed neutrinos.
For this propose,  we  construct a set of the three new orthonormal eigenvectors 
using linear combination of the three existing trivial orthonormal eigenvectors\footnote{The
original eigenvectors are $(1,0,0),$ $(0,1,0)$ and $(0,0,1)$ while the 
modified ones are $\frac{1}{\sqrt{3}}(1,1,1),$ $\frac{1}{\sqrt{2}}(1,0,-1)$ and
$\frac{1}{\sqrt{6}}(1,-2,1)$. These new ones are also used for the rotation of the 
scalar and pseudoscalar mass matrices. From the structure of these
eigenvectors it is clear that mathematically we are rotating the initial
right-handed sneutrino/neutrino basis to a specific basis where
one of the combinations is completely symmetric (eigenvector $\frac{1}{\sqrt{3}}(1,1,1)$)
and mixes with the other states while the remaining two combinations are antisymmetric
and remain decoupled from the other states.}, arising from the diagonal $3\times3$
right-handed neutrino mass matrix.
This mathematical operation, just like the case of the $S^0_i$ and $P^0_i$,
decouples out the mass terms for the two right-handed neutrinos from the rest
of the mass matrix, while the third one mixes
with the other MSSM-like neutralinos and has 
an intricate expression for the mass term. 

Now we are in a stage to write down the analytical
expressions for the mass terms of the three singlet or right-handed
neutrino, sneutrino-like $\widetilde \chi^0_{i+3},\, P^0_i$ 
and $S^0_i$ states as
\bea
&&m_{\widetilde \chi^0_{1+3,2+3}} \equiv m_{\widetilde \chi^0_{U_{1,2}}} = 2 \kappa \nu^c, \nonumber\\
&&m_{\widetilde \chi^0_{3+3}} \equiv m_{\widetilde \chi^0_{M}} = 2\kappa \nu^c + 
\frac{1}{6} \frac{\bla^2 v^2}{\mu} 
\left(\frac{1}{f(T)}-\frac{4 \mathcal{M}\mu}{v^2}
\right) \left(1-\frac{\mathcal{M}\mu}{v^2f(T)}\right)^{-1},\nonumber\\
&&m^2_{P^0_{1,2}} \equiv m^2_{P^0_{U_{1,2}}} =  - 3 \kappa A_k \nu^c
+  \left( \frac{A_\la}{\mu} 
+ \frac{4}{\sqrt{3} } \frac{\ka}{\bla} \right)f(T)\bla^2 v^2  
- \bla^2 v^2,\nonumber\\
&&m^2_{P^0_{3}} \equiv m^2_{P^0_{M}} =  - 3 \kappa A_k \nu^c + \frac{ A_\la }{A_\la
+ \kappa \nu^c} 3\sqrt{3} f(T) \bla \, \kappa v^2 ,\nonumber\\
&&m^2_{S^0_{1,2}} \equiv m^2_{S^0_{U_{1,2}}} = \kappa A_k \nu^c + 4 \kappa^2 {\nu^c}^2 +  
\frac{A_\la}{\mu} f(T) \bla^2 v^2- \bla^2 v^2, \nonumber\\
&&m^2_{S^0_{3}} \equiv m^2_{S^0_{M}} =   \nonumber\\
&& ~~~~~~~~~ \frac{a_0 + a_1~f(T)\bla + 4a_2~{f(T)}^2 \bla^2 + 24 a_3~{f(T)} \bla^3
-24 a_4~\bla^4 + 864 a_5~{f(T)}^3 \bla^5}
{b_0 + 4 b_1 {f(T)}^3 \bla + 24 b_2~{f(T)}^2 \bla^2 -24 b_3~{f(T)}^3 \bla^3}, \nonumber\\
\label{sps-approx1}
\eea 
where $a_{0,..,5}$ and $b_{0,..,3}$, in the expression of $m^2_{\S_{M}}$ are 
complicated functions of the model parameters and are given by
\bea
a_0 &=& G_2 \ka v^4 {\nc}^3 (\al+\ka \nc)(\ak+4\ka\nc) {g(T)}^2,\nonumber\\
{\sqrt{3}}a_1&=&  - G_2  v^2 \nc\{ 4 {A^2_\la} {f(T)}^2 {v^4}  
+ \al (2{v^2_d}+{v^2})({v^2}+2{v^2_u})\ka\nc \nonumber\\
&&+4\ka\nc v^4(\ka\nc {g(T)}^2-\ak {f(T)}^2)\} f(T),\nonumber\\
{{3}}a_2&=&v^4 \{G_2 \al v^4 {f(T)}^2 +6\al (G_2 v^2 - A^2_\la){\nc}^2
+ 6\ka(\al(\ak-5\al)+2G_2 v^2){\nc}^3\nonumber\\
&& +6\ka^2(\ak-4\al){\nc}^4 \} {f(T)}^2,\nonumber\\
{3\sqrt{3}}a_3&=& v^2 \nc   \{-6G_2 v^4{\nc}^2 {f(T)}^2 
+ 2 A^2_\la (v^4{f(T)}^2 + 3 v^2{\nc}^2)\nonumber\\
&&+ \ka \nc ((5\al-\ak) v^4 {f(T)}^2 
+ 18 \al v^2{\nc}^2) +12 \ka^2 v^2{\nc}^4\} f(T),\nonumber\\
{9}a_4&=& \al v^4(v^2 {f(T)}^2+3 {\nc}^2)^2+3 \ka v^2{\nc}^3
(4 v^4 {f(T)}^2+3 v^2{\nc}^2),\,{9\sqrt{3}}a_5= v^6 {\nc}^3 {f(T)}^3,\nonumber\\
b_0&=&G_2v^4 {\nc}^2 (\al+\ka\nc) {g(T)}^2,\,
{\sqrt{3}}b_1= G_2 v^6 \nc {f(T)}^3,\nonumber\\
{3}b_2&=& v^4 {\nc}^2 (\al+\ka\nc) {f(T)}^2,\,
{3\sqrt{3}}b_3=  v^6 \nc {f(T)}^3.
\label{sps-approx1a}
\eea 
%
Here we have used $\sqrt{3}\la = \bla$, $v=\frac{v_u}{\sb}$ $=\frac{v_d}{\cb}
=$ $\sqrt{v^2_u+v^2_d}$, $G_2=g^2_1+g^2_2$,
$T=\tb$ and $\mathcal{M} = M_1 M_2/(g^2_1 M_2 + g^2_2 M_1)$ with 
$g_{1}(g_2)$ as the $U(1)(SU(2))$ gauge coupling. 
The functions $f(T) = \frac{T}{1+T^2}$ and $g(T)=\frac{1-T^2}{1+T^2}$
are derived using $v=\frac{v_u}{\sb}=\frac{v_d}{\cb}$, and finally
we use $\mu = 3\la\nu^c \equiv \sqrt{3}\bla\nu^c$. 
Subscripts `U' and `M' are used to interpret
the nature of the concerned state, i.e. whether it remains
an `U'nmixed singlet-like without a doublet contamination
or appears as a `M'ixed one with non-vanishing doublet composition.

With these formulas ready we are now in a state to investigate the behaviour 
of $m_{\n_{i+3}}$, $m^2_{\P_i}$ and $m^2_{\S_i}$ for the three different ranges of 
$\bla$ values, as already introduced in the last section, along with the necessary 
discussion about the other crucial parameters.
Before we proceed further, it will be useful to reevaluate eq.~(\ref{sps-approx1})
in the limit of $\tb \to \infty$ (i.e. $f(T)\to 0$ and 
${g(T)}^2\to 1$) when the formulae take simpler
forms as
\bea
&&m_{\widetilde \chi^0_{U_{1,2}}} = 2 \kappa \nu^c, \,
m_{\widetilde \chi^0_{M}} \approx 2\kappa \nu^c - 
\frac{1}{6} \frac{\bla^2 v^4}{\mathcal{M}\mu^2},\nonumber\\
&&m^2_{P^0_{U_{1,2}}} \approx  - 3 \kappa A_k \nu^c - \bla^2 v^2,\, 
m^2_{P^0_{M}} \approx  - 3 \kappa A_k \nu^c,\nonumber\\
&&m^2_{S^0_{U_{1,2}}} \approx \kappa A_k \nu^c + 4 \kappa^2 {\nu^c}^2 - \bla^2 v^2, \,
m^2_{S^0_{M}} \approx \kappa A_k \nu^c + 4 \kappa^2 {\nc}^2  
- \frac{8 {\mu}^2}{g^2_1+g^2_2} ~\bla^2,
\label{sps-approx3}
\eea 
where co-efficient of the $\bla^2$ term 
in the expression of $m^2_{S^0_{M}}$ is estimated 
using $\mu=\sqrt{3}\bla\nc$.

It is evident from eq.~(\ref{sps-approx3}) that unless
$\bla$ is small to moderate 
(i.e., $0.01$ $\lsim \bla$ $\leq 0.1$) it is in general hard to accommodate a 
complete non-tachyonic light spectrum (i.e. $\lsim {m_{S^0_4}}/{2}$) 
for both the scalars
and pseudoscalars in the limit of large $\tb$ without
a parameter tuning. 
A non-tachyonic $\n_M$, on the other hand, is 
possible up to $\bla \sim 0.7$ unless 
$2\ka\nc\lsim 10$ GeV or $\mathcal{M} \ll {\cal{O}}(v,\,\mu)$
in the limit of relaxing the gaugino universality condition at the 
high energy scale\footnote{Note that the minimum of $M_2$ 
is $\approx 100$ GeV from the LEP lighter
chargino mass bound \cite{Beringer:1900zz}.}.
The limit $\bla \to 0$ (with $\mu\gsim 100$ GeV as required
from the lighter chargino mass bound) together with a proper choice 
of the other relevant parameters (i.e., $\ka,\,A_\ka,\,\nc)$ assures the
light singlet-like $\widetilde\chi^0_{i+3},\,\P_i,\,\S_i$ states
in the mass spectrum with a vanishingly small doublet composition.
We emphasise here that although the expressions
for $\n_{i+3},\,\S_i$ and $\P_i$ mass terms as shown by
eq.~(\ref{sps-approx3}) are much simpler compared to
the same as given by eq.~(\ref{sps-approx1}), this region
of the parameter space with $\tb\gg1$ is severely constrained 
from diverse experimental results.
This is because the branching fractions for some low-energy
processes (e.g. $B^0_s\to \mu^+\mu^-$),
as discussed before in sections \ref{lightest}
and \ref{Higgs-sector}, depending on the other relevant parameters
are sensitive to the high powers of $\tb$ and thus, 
can produce large branching ratios for these processes in an experimentally
unacceptable way in the limit of $\tb\gg1$. 
For this reason, we will not explicitly address the behaviour
of $m_{\n_{i+3}},\,m_{\P_i},\,m_{\S_i}$ for various ranges
of $\bla$ values in this limit.

The other limit, i.e. small $\tb$, on the contrary,
is useful from the view point of raising the mass of the lightest
doublet-like scalar (see eq.~(\ref{boundHiggs1})) towards 125 GeV,
especially for moderate to large $\bla$ values as already
addressed in section \ref{Higgs-sector}. However,
as shown by eq.~(\ref{sps-approx1}), not all the mass formulas
for the light $\n_{i+3},\,\P_i,\,\S_i$ are simple structured
in this region.

In order to understand the behaviour of $m_{\n_{i+3}},\,m_{\P_i},\,m_{\S_i}$ in detail
we start once again with the small to moderate $\bla$ scenario, as of the last section, 
and will address the remaining two scenarios successively.

\subsection{Regions of the parameter space with light scalars, pseudoscalars
and neutralinos}
\label{spsn1}

\vspace*{0.15cm}
\noindent
(a) {\bf Small to moderate $\bla$:} 
For this range of $\bla$ values, as already discussed in
the previous section, the extra contribution to the lightest
doublet-like Higgs mass is small (see eq.~(\ref{boundHiggs1}))
even for small $\tb$. For example with $\tb=2$, the contribution varies
between $\sim 0.03\%$ to  $3\%$ over that of 
the MSSM contribution as $\bla$ changes from 
$0.01$ to $0.1$, respectively. 
Hence, a large stop mass and (or) a large $A$-term are much needed 
\cite{Draper:2011aa,Heinemeyer:2011aa} to produce a sizable loop correction to 
reach the target of 125 GeV, similar to the MSSM.  

It is also worthy to note that for further small $\bla$ values (i.e. $\lsim 0.01$)
or in the limit of a vanishingly small $\bla$, eq.~(\ref{sps-approx1}) 
coincides with the well-known NMSSM formulas of the same type \cite{Ellis:1988er} 
(although in the NMSSM one has only one singlet) and is given as
\bea
m^2_{S^0_i} \approx   4 \kappa^2 {\nu^c}^2 + \kappa A_k \nu^c, \quad
m^2_{P^0_i} \approx  - 3 \kappa A_k \nu^c,\quad
m_{\widetilde \chi^0_{i+3}} \approx 2\kappa \nu^c. 
\label{sps-approx2}
\eea 
So for this region of $\bla$ values, the mass scales for 
these states are solely determined
by the parameters $\ka,\,A_\ka$ (parameter $\nc$ is estimated
from $\mu = \sqrt{3}\bla\nc$ relation) and composition-wise they are completely
free from any doublet contamination. These simple
formulas can be utilised to estimate the concerned
set of parameters. Note that from eq.~(\ref{sps-approx2}) one can 
obtain the following relations between the masses:
%
\bea
m^2_{\S_i} \approx m^2_{\n_{i+3}}-m^2_{\P_i}/3,\quad 
m^2_{\P_i} \approx -3m_{\n_{i+3}} A_\kappa/2. 
\label{sps-approx2A}
\eea 
%
Thus, the simultaneous presence of non-tachyonic $\S_i$ and $\P_i$ 
implies that $A_\kappa$ and $m_{\n_{i+3}}=2\kappa \nu^c$ 
must have the opposite signs, $m_{\P_i}< \sqrt{3} |m_{\n_{i+3}}|$ 
and as a consequence $m_{\S_i}<|m_{\n_{i+3}}|$. Hence, the 
light scalar/pseudoscalar states are assured when the 
light neutralinos are present. On the other hand, 
using the expression for $m^2_{\P_i}$ in eq.~(\ref{sps-approx2A}) 
with the condition $m_{\P_i}<\sqrt{3} |m_{\n_{i+3}}|$, 
one obtains that $|A_\kappa|<2|m_{\n_{i+3}}|$. 
Hence, for the light $\n_{i+3}$ (i.e., $2|m_{\n_{i+3}}|\lsim m_{\S_4}$),
one can use this relation to estimate
$|A_\ka|\lsim 125$ GeV. If one demands $\P_i$
states comparable/lighter than $\n_{i+3}$ states, then one gets
$|A_\ka| \lsim 2|m_{\n_{i+3}}|/3 $. In this case
$2|m_{\n_{i+3}}|\lsim 125$ GeV predicts
$|A_\ka|\lsim 42$ GeV. It is thus apparent
that the existence of scalar and/or pseudoscalar states
lighter than $\n_{i+3}$ states requires 
small $A_\ka$ values. The requirement
is more stringent for lighter $\P_i$ states.

\vspace*{0.1cm}
Now before we start analysing the behaviours of $m_{\n_{i+3}}$, 
$m^2_{\P_i}$ and $m^2_{\P_i}$ in the light of
eqs. ((\ref{sps-approx1}) - (\ref{sps-approx2})), 
we want to emphasise that for the
simplicity of the analysis: (1) we estimate the scale of $\nu^c$ 
using $\frac{\mu}{\sqrt{3}\bla}$ relation with the minimum of $\mu \gtrsim 100$ GeV, 
(2) we assume $A_\la \gg \ka \nu^c$.
The last assumption emerges from the fact that we need singlinos lighter 
than $m_{\S_4}$ in order to affect the SM-like Higgs phenomenology through on-shell 
$\S_4 \to \n_{i+3} \n_{j+3}$ decay modes. 
The presence of the latter decay modes with $m_{\n_{i+3}}=2\kappa\nu^c$ 
implies $\ka\nu^c \lsim 31.5$ GeV.
Hence together with $A_\la,\,\nu^c$ in the ballpark of a TeV (as preferred by the 
scale of soft-SUSY breaking masses), $A_\la \gg \ka \nu^c$ is well justified. 
We work with $v = 174$ GeV.

\vspace*{0.2cm}
\noindent 
1. We start with the neutralinos, where the expressions for 
$m_{\n_{U_{1,2}}}$ are free from $\bla$ parameter. They are
also free from any doublet contamination. The mass
scale for these neutralinos are determined by the parameters
$\ka$ and $\nu^c$. However, through the latter, $\bla$-parameter
dependency from $\mu=\sqrt{3}\bla\nc$ relation
implicitly enters in the evaluation of $m_{\n_{U_{1,2}}}$.
One can, however, fixed the scale of $\nc$ to evade
this implicit $\bla$-dependence.
The behaviour of their mass scale remains the same 
also when $\ka_i\neq\ka_j$ with $\ka_i-\ka_j \to 0$,
although for this region of the parameter space
they adhere a small to negligible doublet admixing. 
One should note that the relative position of
$\n_{U_{1,2}}$ with respect to that of the $\n_M$
in the mass spectrum depends on the relative signs of the various parameters.
For example, from eq.~(\ref{sps-approx3}) with sign($\ka\nc)$ = sign($M_{1,2})$, 
one gets $|m_{\n_M}| \lsim |m_{\n_{U_{1,2}}}|$.

For the $\n_{M}$, from eq.~(\ref{sps-approx1}) it is clear that
the extra contribution goes from 
$\approx$ $50\bla^2 \times$ $\frac{(2.5-0.013\mathcal{M})}{(1-0.008\mathcal{M})}$
to $\approx$ $50\bla^2 
\times$ $\frac{(10.1-0.013\mathcal{M})}{(1-0.033\mathcal{M})}$ as $\tb$ 
varies from $2$ to $10$, taking the minimum of $\mu=100$ GeV. 
Hence, with $\mathcal{M}\sim {\cal{O}}(100$ GeV) and $\tb=2$
one gets a contribution like $\approx 300\bla^2$ GeV which
yields a correction of around $3$ GeV with $\bla=0.1$.
This contribution diminishes further with a larger values
of $\mu,\,\tb$ or $M_{1,2}$, i.e. larger $\mathcal{M}$. For example,
reanalysis of the last step taking $\tb=2$ and $\mathcal{M}=1$ TeV
with everything else fixed gives a correction of $\approx 0.75$ GeV.
This correction reduces to $\approx$ 0.05 GeV for $\tb=10$.
In other words, unless $2|\ka\nc|\lsim 10$ GeV, for a novel
region of the parameter space, the mass correction and the amount of doublet
admixing remain negligible for $\n_M$. Thus, for most
of the time $|m_{\n_M}|\approx |m_{\n_{U_{1,2}}}|$.

\vspace*{0.2cm}
\noindent
2. Turning towards $\P_i$, the presence of multiple terms
in the expressions of $m^2_{\P_{U_{1,2}}}$ with the same 
coefficient $\bla^2 v^2$ provides an option to 
remove the $\bla$ dependence from the mass terms for some specific
set of the parameter choice. For small to moderate $\tb$,
the $\bla$-dependent terms are given by
$\delta m^2_{\P_{U_{1,2}}} = \left[\left( \frac{A_\lambda}{\mu}   + \frac{4  
\kappa}{ \sqrt{3}\bla}\right) f(T)   - 1\right]$ $\bla^2 v^2$.
Now as we mentioned before, with $A_\la,\,\nu^c \sim {\cal{O}}(1$ TeV),
$\mu \sim {\cal{O}}(100$ GeV) one gets $\frac{A_\la}{\mu}\approx 10$.
At the same time, the light $\n_{U_{1,2}}$ in the upper limit 
(i.e., $2m_{\n_{U_{1,2}}}\approx m_{\S_4}$)
together with $\nu^c \sim 1$ TeV predict $0.7 \leq \frac{4\ka}{\sqrt{3}\bla}\lsim 7$
for $0.1\geq \bla \gsim 0.01$. It is thus apparent
that one needs at least $\ka \lsim 10^{-2}$ to use 
$\frac{A_\la}{\mu} + \frac{4\ka}{\sqrt{3}\bla} \sim \frac{A_\la}{\mu}$.
In this limit one effectively gets $\delta m^2_{\P_{U_{1,2}}} \approx 
[10 f(T)   - 1]\bla^2 v^2$ assuming the relevant signs for
the different parameters. The magnitude of this contribution is at most $\sim 0.1\bla^2 v^2$
for $9\leq\tb\leq11$ and vanishes\footnote{Since
$\frac{A_\la}{\mu}f(T)-1=0$ is a quadratic equation in $\tb$, 
one should expect another solution for $\tb$, however,
we do not consider any such solution when $\tb<1$.} around $\tb\sim 9.9$.
So in this corner of the parameter space the
light $\P_{U_{1,2}}$ are guaranteed with the proper choice
of $\ka,\,\nc$ and $A_\ka$.
Outside of this region, the lightness of $\P_{U_{1,2}}$ are possible
at the cost of a mutual cancellation between the different
components in the expressions of $m^2_{\P_{U_{1,2}}}$ (see eq. (\ref{sps-approx1})).
Note that with $\mu\gsim 100$ GeV,
$\frac{A_\la}{\mu}\sim 10 \Rightarrow A_\la \approx \nc$ when
$\bla\sim{\cal{O}}(0.1)$ while $10A_\la \approx \nc$ for
$\bla\sim{\cal{O}}(0.01)$.

With $A_\la\gg\ka\nc$, the extra piece of contribution to
$m^2_{\P_M}$ as shown in eq.~(\ref{sps-approx1})
goes as $\delta m^2_{\P_{M}} \approx 3\sqrt{3} f(T)\bla \, \kappa v^2
\approx 1.58\times 10^5 f(T)\bla\ka$ GeV$^2$. Now the scale for $\ka$
can be estimated with $2|\ka\nc|\lsim{m_{\S_4}}/{2}$ and $\nc\approx {\cal{O}}(1$
TeV), as $\lsim{\cal{O}}(10^{-2})$. On the other hand, $f(T)$
changes from $0.4$ to $\sim 0.1$ as $\tb$ varies
from $2$ to $10$, respectively. Thus, for $\tb=2$, $\delta m^2_{\P_{M}}$
goes as $\sim 632 \bla$ GeV$^2$ which is about $63$ GeV$^2$ for $\bla=0.1$.
This indicates that $\bla$-dependent contribution 
and hence the doublet admixing is non-negligible
for $\P_M$. The lightness is, however, still possible
using a possible cancellation between the two different terms
in the expression of $m^2_{\P_M}$ (see eq. (\ref{sps-approx1})). 
One can of course consider $\tb \gsim 10$ and/or
a smaller $\ka$ value to reduce $\delta m^2_{\P_{M}}$ further.

\vspace*{0.2cm}
3. Concerning  the scalars, it is clear from eqs.~(\ref{sps-approx1}) and 
(\ref{sps-approx3}) that it is in general rather hard to
estimate the correction in $\S_M$ from $\bla$ dependent terms,
in the limit of small to moderate $\tb$ (see eq.~(\ref{sps-approx1})). 
In general, one naively expects a non-negligible doublet impurity in $\S_M$ for
this range of $\bla$ values while the lightness of $\S_M$, in an experimentally viable manner, 
is still possible with a fine cancellation among various components in the expression 
of $m^2_{\S_M}$ (see eq.~(\ref{sps-approx1})).

For $\S_{U_{1,2}}$, as shown in eq.~(\ref{sps-approx1}), $\bla$-dependent
contributions are given by $\delta m^2_{\S_{U_{1,2}}} =
[\frac{A_\lambda}{\mu} f(T) - 1]\bla^2 v^2$,
which are similar to the $\delta m^2_{\P_{U_{1,2}}}$ assuming 
$\frac{A_\lambda}{\mu} \gg \frac{4 k}{\sqrt{3}\bla}$.
Hence, the analysis remains similar. 
We note in passing that another tool to reduce the contribution
from the first term of $\delta{m^2_{\S_{U_{1,2}}}}$ 
is to consider $A_\la \ll \mu$ while keeping 
$A_\la \gg \ka\nu^c$ at the limit of a very small $\ka$.
In this case, for both of $\S_{U_{1,2}}$ and $\P_{U_{1,2}}$, the
$\bla$-dependent term is given by $\bla^2v^2$ since
$4\ka$ is also $\ll \sqrt{3}\bla$ for this region of the parameter
space.

\vspace*{0.15cm}
Combining all the facts, the lightness for 
$\n_{i+3}$, $\P_{i}$ and $\S_{U}$ states are rather 
assured in this region of $\bla$ values with a negligible to small tuning of 
the other parameters. Concerning the $\S_M$, especially for $\bla \sim 0.1,$ 
a low mass is rather hard to accommodate
without a fine cancellation between the different contributors. 
A similar conclusion also holds true for the amount of doublet impurity in $\S_M$. 
The amount of the doublet admixing in the $\P_M$ and $\n_{M}$, 
on the other hand,  are rather easily 
controlled with a proper but not very fine tuned 
choice of the other model parameters.

\vspace*{0.25cm}
\noindent
(b) {\bf Moderate to large $\bla$:} Moving towards moderate to large $\bla$ 
region, as mentioned in section \ref{Higgs-sector}, the additional  
contribution to the tree-level lightest doublet-like scalar mass
over the same from the MSSM can vary from $12\%$ to $\sim 100\%$ when 
$\bla$ goes from $0.2$ to $0.7$ with $\tb=2$.
With increasing tan$\beta$, once again this extra contribution goes down, 
for example $\sim 4\%$ for $\bla=0.7$ with $\tb=10$.
Necessity of a large stop mass and/or a large $A$-term are somewhat ameliorated 
for this scenario in the region of tan$\beta\lsim 10$.
Further, in this corner of the parameter space an enhanced branching 
ratio is possible for $S^0_4\to \gamma\gamma$ compared to the SM, 
especially as $\bla$ tends to $0.7$ with a suitable choice of 
the other parameters. This enhancement is supported
by both the ATLAS \cite{Aad:2014eha}
and CMS collaborations \cite{CMS:2014ega} to date.

In this region of the parameter space, 
the lightness of $\S_{U_{1,2}}$ and $\P_{U_{1,2}}$ are 
not assured without a moderate tuning of the relevant parameters (e.g., $\bla$, 
$|\ka|$, $|A_\ka|$ etc.). Their purities, however, remain unaffected by the virtue 
of the construction.

Before beginning the discussion of 
$\n_{i+3},\, \P_i$ and $\S_i$ masses, note that for this range of $\bla$ 
values, the estimation
of $\nu^c$ from $\mu/\sqrt{3}\bla$ relation with $\mu$ around $100$ GeV
is somewhat inconsistent with the TeV-scale soft masses. 
For example, with this assumption
one would get $\nc\approx 80$ GeV for $\bla=0.7$ and so we
fixed $\nu^c$ at $1$ TeV for this region of the parameter space.

\vspace*{0.2cm}
\noindent
1. Discussion for $m_{\n_{U_{1,2}}}$ remain the same as before while 
interpretation for the $m_{\n_M}$ is more involved than the one 
for the  small to moderate $\bla$ region with $\nc$ fixed
at $1$ TeV.

\vspace*{0.1cm}
\noindent
(i) In the region of the parameter space with large $\mu$ or $\mathcal{M}$ 
(such that $\frac{\mathcal{M}\mu}{v^2}\gg\tb$), the extra term,  $\delta m_{\n_{M}}$ 
is written as $\frac{2\bla v^2}{3\sqrt{3}\nc}\times\frac{T}{(1+T^2)}$.
This term changes from $\sim 1$ GeV  to $\sim 3$ GeV when $\bla$ changes from $0.2$ 
to $0.7$ for $\nc=1$ TeV and $\tb=2$.
Thus, unless $2\ka\nc$ is very small $(\lsim 10$ GeV), effect from this new term in 
the lightness of $m_{\n_{M}}$ is small to negligible and its effect in spoiling  the singlet 
purity of $\n_{M}$ is also moderate to suppressed. 
A Larger value of $\tb$ is another way to reduce this extra contribution.

\vspace*{0.1cm}
\noindent
(ii) Now we investigate another corner of the parameter space with 
$\mathcal{M}\approx \mu \approx v$ where the new contribution,  $\delta m_{\n_{M}}$
goes as  $\approx -\frac{\bla v^2}{6\sqrt{3}\nc}$ $\frac{(1+T^2-4T)}{(1+T^2-T)}$, or 
numerically $\sim 3 \bla$ with $\tb=2$ and $\nc=1$ TeV. Thus, as $\bla$ changes from 
$0.2$ to $0.7$, this varies from $\sim 0.6$ to $\sim 2$ GeV and decreases further 
for larger $\tb$ values. 
Consequently its contribution to the $\n_{M}$ mass term as well to the composition 
from the doublet-like states remains negligible unless $|2\ka\nu^c|\lsim 10$ GeV.
Note that the sign of this contribution changes for $\tb\geq4$ when
it appears as a negative one.

\vspace*{0.2cm}
\noindent
2. Concerning the $\P_i$, the situation remains flexible in order to secure light 
the $\P_i$ states or to minimise the doublet composition in $\P_M$.

In this range of $\bla$ values $4\ka/\sqrt{3}\bla$ decreases
further compared to the last region 
(i.e., about $0.1\leq \frac{4\ka}{\sqrt{3}\bla}\lsim 0.4$
for $0.7 \geq \bla \gsim 0.2$) and hence, just like the 
small to moderate $\bla$ scenario, the $\bla$-dependent contribution
are given by $\delta m^2_{\P_{U_{1,2}}} \approx [\frac{A_\la}{\mu} f(T) - 1]\bla^2 v^2$.
However, now with $A_\la,\,\nc \approx {\cal{O}}(1$ TeV), one gets
$350$ GeV $\lsim \mu$ $\lsim 1200$ GeV as $\bla$ moves
from $0.2$ to $0.7$. 
With our choice of $A_\la,\,\nu^c$, the quantity $A_\la/\mu$ varies between
$\sim 0.8$ to $2.8$ and hence depending on the value of $\tb$
this $\bla$-dependent contribution may appear negligible.
For example, with $\bla=0.2$, magnitude of this extra contribution is
about $0.1\times\bla^2 v^2$ for $2\lsim\tb\lsim2.9$ and vanishes
around $\tb\approx 2.38$. With $\bla=0.7$, keeping
$\nc$ fixed at $1$ TeV, a similar
phenomenon remains missing for any real values of $\tb$.
Nevertheless, depending on the relative signs of the different
terms with the proper choice of parameters, e.g. $\ka,\nc$ and $A_\ka$,
the light $\P_{U_{1,2}}$ are well affordable in this corner of
the parameter space, at the cost of a partial cancellation
between the different components.

Moving towards $\P_M$, with $A_\la/\nu^c \approx 1$, 
the $\bla$ dependent correction, $\delta m^2_{\P_{M}}$ 
is $\approx$ $3\sqrt{3} f(T)\times$ $\bla \kappa v^2$
$\approx 6.3\times 10^4 \bla\ka$ GeV$^2$ with $\tb=2$.
Larger $\tb$ values of course provide an additional
tool to reduce this contribution. The quantity $\delta m^2_{\P_M}$
 varies from $\sim 1.3\times10^4\ka~\g2$ to $\sim 
4.4\times10^4\ka~\g2$ as $\bla$ goes from $0.2$ to $0.7$. 
It is clear now that if we stick to $\ka\sim {\cal{O}}(10^{-2})$
(as guided by $|2\ka\nc|\lsim m_{\S_4}/2$), these
corrections are significant, e.g. about $130~\g2$
for $\bla=0.2$. So we need to move to the region of
$\ka \lsim 10^{-3}$ to reduce this extra contribution
so that the lightness and the singlet purity for $\P_M$
remain assured for this region of the parameter space
with a proper choice of the other relevant parameters (e.g., $A_\ka$).
The choice of $\ka\sim {\cal{O}}(10^{-3})$ also makes
the assumption $\frac{A_\la}{\mu}f(T)+\frac{4\ka}{\sqrt{3}\bla} 
\approx \frac{A_\la}{\mu}$ more reliable. One would, however,
need a higher $\nc$ value to get $2\ka\nu^c\gsim 10$ GeV.

We note in passing that it is still possible to accommodate
a light $\P_M$ with $\ka\sim 10^{-2}$ at the cost of
a cancellation between the different parts in the 
expression of $m^2_{\P_M}$.

\vspace*{0.2cm}
\noindent
3. Once again like the earlier $\bla$ scenario, an interpretation
about the singlet purity and the lightness for $\S_M$ is highly complicated.
The situation gets worse as one moves to higher $\bla$ values.
The only possibility to assure a light $\S_M$ for this corner of the
parameter space appears through a cancellation among the various
terms in the expression of $m^2_{\S_M}$, at the cost of a large fine tuning
of the different relevant parameters.

Moving towards $\S_{U_{1,2}}$, the $\bla$-dependent contributions are the same as
that of the pair of $\P_{U_{1,2}}$ with the assumption
$\frac{A_\mu}{\mu}\gg\frac{4\ka}{\sqrt{3}\bla}$. Hence,
the discussion remains the same as that of the $\P_{U_{1,2}}$.

\vspace*{0.15cm}
Summarising the discussion, we conclude that the light singlet-like $\n_{i+3}$ states are 
well feasible in this range of $\bla$ values without a large parameter tuning or a 
strange cancellation. A light singlet-like $\P_{M}$ appears with a bit of parameter tuning, 
especially for $|\ka|$, however, more easily compared to the light $\P_{U_{1,2}}$ or $\S_{U_{1,2}}$. 
The presence of the light $\P_{U_{1,2}}$ and $\S_{U_{1,2}}$ 
even for $\bla=0.2$, requires certain degree of 
parameter tuning which grows with increasing $\bla$.
A singlet-like $\S_M$ is rather hard in this region without a high amount of 
parameter tuning. A similar argument holds true for the lightness of $m_{\S_M}$.
Note that in this region of the $\bla$ values or higher, the
existence of a light $\S_M$ for $\bla > 0.1$ and large $\tb$ excludes the 
possibility of having the light $\P_i,\,\n_{M}$ in the mass spectrum. 
In a similar fashion, for the same corner of the parameter space, the presence of 
the light $\P_i,\,\n_{i+3}$ states discards the existence of a light $\S_M$.

\vspace*{0.25cm}
\noindent
(c) {\bf Dominant $\bla$:} In this region of the parameter space,  as already addressed 
in section \ref{Higgs-sector}, one can really accommodate a 125 GeV Higgs boson even 
with only tree-level contribution, at the cost of relaxing perturbativity of $\bla$ 
up to the GUT scale. 
Hence, a small $A$-term and(or) small stop masses are well affordable in this scenario. 
Assuming perturbativity up to 10 TeV, $\bla$  as large as $1.7$ is possible, as can be 
interpreted from figure \ref{fig:lamHigg}. 
With this scenario an enhancement over the MSSM tree-level Higgs mass as large as $\approx7\%$ 
is possible with tan$\beta=10$ and $\bla=1$. At lower $\tb$, say $5$, this contribution 
gives $\approx 28\%$ enhancement over the  MSSM limit.
A larger contribution for a higher $\bla$ values is possible with the 
larger tan$\beta$ values 
as shown in figure \ref{fig:lamHigg}, although constraints from flavour observables can 
restrict this scenario.

Similar to the moderate to large $\bla$ region, we 
do not determine the scale of  $\nu^c$ using $\mu=\sqrt{3}\bla\nc$ relation for this 
corner of the parameter space as then $\mu|_{\min}\sim 100$~GeV would 
predict $|\nu^c|\lsim 60$ GeV for $\bla\gsim 1$. Note 
that, however, here keeping $|\nc|$ fixed at 1 TeV would give
$\mu\approx 2$ TeV for $\bla=1.2$. Hence, here we keep
$\mu$ fixed at 1 TeV.

\vspace*{0.2cm}
\noindent
1. Concentrating  on the neutralinos, behaviour of the two $m_{\n_{U_{1,2}}}$ remain
the same as of the last two $\bla$ regions. For $m_{\n_M} $ the discussion 
is also similar to the last $\bla$ region,
although the amount of $\bla$-dependent correction increases in
the magnitude due to the larger $\bla$ values.

\vspace*{0.2cm}
\noindent
(i) In the corner of the parameter space with the large $\mu$ or $\mathcal{M}$ (such that 
$\frac{\mathcal{M}\mu}{v^2}\gg\tb$), this extra term, i.e. $\frac{2\bla^2 v^2}{3\mu}
\times \frac{T}{(1+T^2)}$ for $\tb=2$, $|\nc|=1$ TeV and 
$\bla=1$ is estimated as $\sim 8$ GeV. 
Hence,  unless $|2\ka\nc|\lsim 10$ GeV, effect from this new term in 
the lightness of $m_{\n_{M}}$ and also in determining the amount of the doublet 
impurity in $\n_{M}$ remains moderate to small.
Larger values for $\tb$ is another tool to reduce the doublet contamination in $\n_{M}$ apart 
from ensuring its lightness with the suitable $|\ka|$ and $|\nc|$ values.

\vspace*{0.1cm}
\noindent
(ii) In the region of the parameter space 
with $\mathcal{M}\approx \mu \approx v$  the new contribution, 
$-\frac{\bla^2 v^2}{6\mu}\times \frac{(1+T^2-4T)}{(1+T^2-T)}$ goes 
as $\sim 5$ GeV with $\bla=1,\,\tb=2$  and $\nc=1$ TeV. 
This is again a small contribution unless $|2\ka\nu^c|\lsim 10$ GeV. 
So the effect of this term in the lightness of $\n_{M}$ and  in the doublet composition 
of $\n_{M}$ remains moderate to small, especially for larger $\tb$ values.  

\vspace*{0.2cm}
\noindent
2. The discussion with $\P_i$ states are very similar to that of
the last $\bla$ region. With $|\ka|\sim {\cal{O}}(10^{-3})$
and $\bla\gsim 1$, $\frac{4\ka}{\sqrt{3}\bla}\lsim 0.0023$.
Thus, with $A_\la,\mu\approx {\cal{O}}(1$ TeV),
$\frac{A_\la}{\mu}\gg\frac{4\ka}{\sqrt{3}\bla}$ is well
justified and we end up with $\delta m^2_{\P_{U_{1,2}}} \approx 
[f(T)   - 1]\bla^2 v^2=3[f(T)   - 1]\times 10^4~\g2$ for $\bla=1$.
This contribution does not vanish for any real values of $\tb$,
however, $\lsim \bla^2 v^2$ for small $\tb$ values. 
So the lightness of $\P_{U_{1,2}}$ are difficult for this 
corner of the parameter space without a large cancellation
between the different components in the expressions
of $m^2_{\P_{U_{1,2}}}$ (see eq.~(\ref{sps-approx1})) with a proper
choice of the other relevant parameters.

Regarding $\P_M$, with $A_\la/(A_\la+\ka\nc)\approx 1$,
the $\bla$-dependent contribution in $\delta m^2_{{\P_{M}}}$
is given by $\approx 3\sqrt{3} f(T)\bla \, \kappa v^2$
$\approx 1.57\times 10^2 f(T)$ GeV$^2$ with $\bla=1$ and $\ka\sim 10^{-3}$.
This is clearly a non-negligible correction unless
one moves to large $\tb$ values $(\gsim 10)$.
Hence, the lightness and the singlet purity are not generic
to $\P_M$ for this region of $\bla$ values. However,
with a suitable sign choice of the relevant parameters
it remains possible to yield a light $\P_M$ using a moderate
to large cancellation between the different components in
the expression of $m^2_{\P_M}$.

\vspace*{0.2cm}
\noindent
3. Concerning $\S_i$, it is in general hard to accommodate 
a light $\S_M$ for this region of the parameter space,
especially for small to moderate $\tb$, without a
severe cancellation among the various components in the expression
of $m^2_{\S_M}$ (see eq.~(\ref{sps-approx1})). This large
cancellation also indicates a very high doublet contamination in $\S_M$. 

Regarding the two $\S_{U_{1,2}}$ the analysis is the same as that of the two
$\P_{U_{1,2}}$ with the valid assumption $\frac{A_\la}{\mu}\gg\frac{4\ka}{\sqrt{3}\bla}$.
The lightness of $m^2_{\S_{U_{1,2}}}$ for this region of $\bla$ values, 
appears mainly with a large possible cancellation between 
the different terms in the expressions of $m^2_{\S_{U_{1,2}}}$.

\vspace*{0.15cm}
In summary, the simultaneous presence of light $\n_{i+3},\,\S_i$ and $\P_i$ states are 
hardly possible in the dominant $\bla$ region. 
Concerning the lightness of all the states and singlet purity of the mixed states, 
the neutralinos appear as the most favoured ones in terms of the 
amount of fine tuning of the parameters. 
The pseudoscalars $\P_i$ as well as $\S_{U_{1,2}}$ are second on the list with a 
moderate to large fine tuning, for small to moderate $\tb$ values.
A pure $\S_M$ is hardly possible for this range
of $\bla$ values although the lightness can be
achieved with a large to severe tuning of the relevant parameters.

\vspace*{0.3cm}
\textit{In a nutshell}, so far we have given a complete overview of the relevant 
parameters, not only to accommodate a 125 GeV 
SM Higgs-like scalar boson, but at the same time to 
investigate the possibility of having the light singlet-like 
scalars, pseudoscalars and neutralinos in the mass spectrum. Thus, 
it remains to address the only remaining part of our analysis, namely 
the effects of the aforesaid light states in the decay phenomenology
of the SM Higgs-like $\S_4$. We aim to address these issues in the next section, 
once again giving special emphasis on the three different $\bla$ regions.

\section{New decays of the SM-like Higgs in the {$\bm\mu\nu$}SSM}
\label{decay}

In this section we present analytical estimates 
of the decays of the SM Higgs-like $S^0_4$ into
a pair of $\S_i,\,P^0_i$ and $\widetilde \chi^0_{i+3}$ states.
Note that we consider only new two-body
decays of $\S_4$ and thus, more complex or longer decay cascades like 
the ones addressed in ref. \cite{Fidalgo:2011ky} will be skipped. 
It will be useful to compute first the complete expressions 
of the decay widths for these processes as:

\bea
\Gamma_{\S_4\to \S_i\S_j} &=&
\left|\widetilde g O^{SSS}_{4ij}\right|^2 \times \frac{ 
\mathcal{F}\left(m^2_{\S_4},m^2_{S^0_i},m^2_{S^0_j}\right)}
{16\pi m_{\S_4}(1+\delta_{ij}) }, \nonumber\\
\Gamma_{\S_4\to \P_i\P_j} &=&
\left|\widetilde g O^{SPP}_{4ij}\right|^2 \times \frac{
\mathcal{F}\left(m^2_{\S_4},m^2_{P^0_i},m^2_{P^0_j}\right)}
{16\pi m_{\S_4}(1+\delta_{ij}) }, \nonumber\\
\Gamma_{\S_4\to \n_{i+3}\n_{j+3}} &=&
\left[\left(\frac{1}{2}\left|\widetilde g O^{nnh}_{L(i+3)(j+3)4}\right|^2 
+ \frac{1}{2}\left|\widetilde g O^{nnh}_{R(i+3)(j+3)4}\right|^2\right)
\left(m^2_{\S_4}-m^2_{\n_{i+3}} -m^2_{\n_{j+3}}\right)
\right.\nonumber\\
&-&\left. 
2 \Re\left(|\widetilde g|^2 O^{nnh}_{L(i+3)(j+3)4} 
O^{{nnh}^*}_{R(i+3)(j+3)4}\right)m_{\n_{i+3}} m_{\n_{j+3}}\right]\nonumber\\
&&
\times \frac{\mathcal{F}\left(m^2_{\S_4},
m^2_{\n_{i+3}},m^2_{\n_{j+3}}\right)}{16\pi m_{\S_4} (1+\delta_{ij})}.\,\,
\label{decay-width}
\eea
Here $m^{2}_{\S_4}\,\mathcal{F}(m^2_{\S_4},m^2_{X^0_i},m^2_{X^0_j})=
\sqrt{\left(m^2_{\S_4}-{m^2_{X^0_i}}-{m^2_{X^0_j}}\right)^2
-4 {m^2_{X^0_i}} {m^2_{X^0_j}}}$ with $X^0_i=\S_i,\,\P_i$ and
$\n_{i+3}$, the couplings
$\widetilde g O^{SSS}_{mnp}$ and $\widetilde g O^{SPP}_{mnp}$
are complicated functions which are given in the appendix B of
ref.~\cite{Fidalgo:2011ky}, with a notation
$h_\delta h_\epsilon h_\eta \equiv \widetilde g O^{SSS}_{\delta\epsilon\eta}$
and $h_\delta P_\epsilon P_\eta \equiv \widetilde g O^{SPP}_{\delta\epsilon\eta}$, 
and the couplings $\widetilde g O^{nnh}_{Lijk}$ and $\widetilde g O^{nnh}_{Rijk}$ 
are given in the appendix E of ref.~\cite{Ghosh:2010zi}. Note that
the kinematic factor $\mathcal{F}(m^2_{\S_4},m^2_{X^0_i},m^2_{X^0_j}) \approx 1$  
for $m_{\S_i},\,m_{\P_i},\,m_{\n_{i+3}}$ $\ll m_{\S_4}$.
This holds roughly true for higher values of $m_{X^0_i}$ also, 
e.g. $\mathcal{F}(m^2_{\S_4},m^2_{X^0_i},m^2_{X^0_j})$
$\sim 0.83$ for $m_{X^0_i}= 35$ GeV.

At this point we want to stress that since our goal is to describe
a complete picture of the possible new two-body $S^0_4$ 
decay phenomenology with the $\mu\nu$SSM,
our analyses are confined up to the level of analytical
estimates. Note that a full numerical
analysis using eq.~(\ref{decay-width}), as anticipated 
in a set of forthcoming publications \cite{glmmrF2},
should satisfy a class of existing experimental
observations \cite{ATLAS-susy13,CMS-susy13,CMS:yva,Aad:2013wqa,
TheATLAScollaboration:2013hia,Chatrchyan:2013vaa,Chatrchyan:2013zna,
Chatrchyan:2013iaa,Chatrchyan:2014nva,TheATLAScollaboration:2013wia,ATLAS:2012dsy,
CMS:2013hja,Arbey:2012bp,CMS:2013lea,ATLAS:couplings,
Aad:2014eva,Aad:2014eha,CMS:2014ega,CMS:2014ala,Khachatryan:2014iha,ATLAShwidth,
Chatrchyan:2014tja}.
%

%
\FIGURE{\epsfig{file=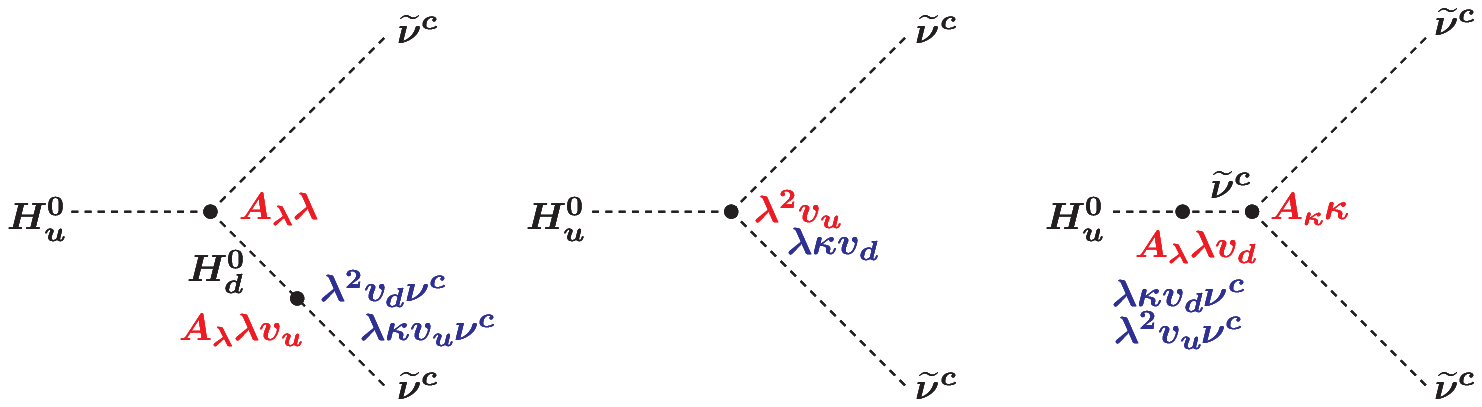,height=3.5cm,width=12.5cm,angle=0} 
\epsfig{file=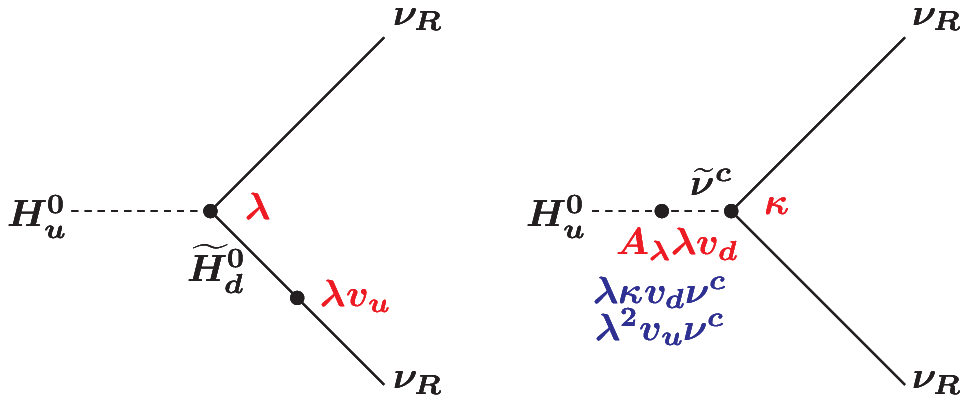,height=3.5cm,width=9.5cm,angle=0} 
%
\caption{Diagrams showing SM-like $S^0_4$ decays into a pair of singlet-like 
CP-even scalars, CP-odd scalars and neutralinos in the flavour basis with
the leading contributions. Symbol $\nu_R$ 
has been used to represent a right-handed neutrino.
Red (blue) colour has been
used to represent couplings of certain (alternate) kind. 
An extra factor that appears
when a complex scalar field $\bm \Phi$ is decomposed
as $\bm\Phi = v_\Phi + \frac{\Re \bm\Phi + i \Im \bm\Phi}{\sqrt{2}}$ with
$v_\Phi$ as the acquired VEV, is not explicitly shown here.
Diagrams with $Y_{\nu_{ij}}$ or $\nu_i$ in the couplings are not shown
since they give rise to negligible contributions.} 
\label{fig:mixing}}

Assuming $S^0_i,\,P^0_i,\,\widetilde\chi^0_{i+3}$ with 
a leading\footnote{In figure \ref{fig:mixing}, we label 
a state as X-like when the composition of X in that
state dominates $(\gsim 90\%)$ over the others. For example, a SM-like
$\S_4$ requires leading $H^0_u$ composition
although certain amount of $H^0_d$ component is essential
so that it can couple to the down-type fermions, e.g. $b\bar{b},\,
\tau^+\tau^-$, etc.}
singlet composition and the SM-like $S^0_4$, we present diagrams
giving leading contributions to $S^0_4 \to\S_i\S_j,\,\P_i\P_j,
\,\widetilde\chi^0_{i+3} \widetilde\chi^0_{j+3}$ processes in
figure \ref{fig:mixing}. We adopt the flavour basis for the convenience of 
analysis.
We emphasise here that these simple analytical analyses are purely qualitative
although agreed rather well with the full numerical results.
However, when the amount of doublet impurity is high in 
$\S_i,\,\P_i,\,\n_{i+3}$ (e.g., for larger $\bla$ values), 
these estimations differ significantly.

Figure \ref{fig:mixing} in the mass or physical basis represents $\S_4\to \S_i\S_j$,
$\P_i\P_j$ and $\n_{i+3}\n_{j+3}$ processes. Following
our discussion of section \ref{spsn}, especially
for the chosen set of parameters (see eq.~(\ref{EWF-param3})), it is clear that 
two of these $\S_i,\,\P_i,\,\n_{i+3}$ states are $\S_U,\,\P_U$
and $\n_U$, respectively, while the remaining $\S_i,\,\P_i,\,\n_{i+3}$
states represent $\S_M,\,\P_M,\,\n_M$. It is thus, important
to emphasise here that all of these states do not 
couple to $\S_4$ with identical strengths. To start with, it is
convenient first to write down all the relevant terms used
to draw figure \ref{fig:mixing}. Following ref. \cite{Escudero:2008jg}
they are\footnote{Our ($\widetilde H^0_d) \equiv 
(\widetilde H_d$) of ref. \cite{Escudero:2008jg}.}:

\bea
\mathcal{L} &=& \left[-(A_\la\la)_i\widetilde \nu^c_i H^0_u H^0_d
+ \frac{1}{3} (A_\ka\ka)_{ijk} \widetilde \nu^c_i \widetilde \nu^c_j \widetilde \nu^c_k
+ \ka_{ijk}\la^*_j H^{0^*}_u H^{0^*}_d \widetilde\nu^c_i \widetilde\nu^c_k
+ {\rm H.c.}\right]\nonumber\\
&+& \la_i\la^*_j (H^0_u H^{0^*}_u+H^0_d H^{0^*}_d)\widetilde\nu^c_i \widetilde\nu^{c^*}_j 
+ \ka_{ijk}\ka_{ljm} \widetilde \nu^c_i \widetilde \nu^{c^*}_l 
\widetilde \nu^c_k \widetilde \nu^{c^*}_m \nonumber\\
&+&\frac{1}{2}\left[\la_i H^0_u \widetilde H^0_d \nu_{R_i}
- 2 \kappa_{ijk} \widetilde\nu^c_i \nu_{R_j} \nu_{R_k} +  {\rm H.c.}\right].
\label{lagrangian-rel1}
\eea
Assuming real parameters, eq.~(\ref{lagrangian-rel1}) in the light of
eq.~(\ref{EWF-param3}) can be rewritten as
\bea
\mathcal{L} &=& \left[-\frac{A_\la\bla}{\sqrt{3}} H^0_u H^0_d \sum^3_{i=1}\widetilde \nu^c_i
+\frac{1}{3} A_\ka\ka \sum^3_{i=1}\widetilde \nu^{c^3}_i
+ \frac{\ka\bla}{\sqrt{3}} H^{0^*}_u H^{0^*}_d \sum^3_{i=1} \widetilde\nu^{c^2}_i 
+ {\rm H.c.}\right]\nonumber\\
&+& \frac{\bla^2}{3} (H^0_u H^{0^*}_u+H^0_d H^{0^*}_d) 
\sum^3_{i=1} \widetilde\nu^c_i  \sum^3_{j=1} \widetilde\nu^{c^*}_j
+ \ka^2 \sum^3_{i=1} |\widetilde \nu^c_i|^4 \nonumber\\
&+&\frac{1}{2}\left[\frac{\bla}{\sqrt{3}} H^0_u \widetilde H^0_d 
\sum^3_{i=1} \nu_{R_i}
- 2\ka \sum^3_{i=1} \widetilde \nu^c_i \nu^2_{R_i}+ {\rm H.c.}\right].
\label{lagrangian-rel2}
\eea
Following the footnote 17, it is possible to relate 
$\widetilde\nu^c_i$ and $\nu_{R_i}$ states with $\S_M,\S_{U_{1,2}}$,
$\P_M,\P_{U_{1,2}}$ and $\n_M,\n_{U_{1,2}}$ states, respectively, as:
\bea
\Re\widetilde\nu^c=\mathcal{U}\, S^{0'},\quad
\Im\widetilde\nu^c=\mathcal{U}\, P^{0'}, \quad
\nu_R=\mathcal{U} \, \widetilde\chi^{0'},
\label{matrix-reln}
\eea
where $\Re\widetilde\nu^c=(\Re\widetilde\nu^c_1,\,
\Re\widetilde\nu^c_2,\,\Re\widetilde\nu^c_3)$, $\Im\widetilde\nu^c
=(\Im\widetilde\nu^c_1,\,\Im\widetilde\nu^c_2,\,\Im\widetilde\nu^c_3)$,
$\nu_R=(\nu_{R_1},\,\nu_{R_2},\,\nu_{R_3})$, $S^{0'}=(\S_M,\,\S_{U_1},\,\S_{U_2})$, 
$P^{0'}=(\P_M,\,\P_{U_1},\,\P_{U_2})$, 
$\widetilde\chi^{0'}=(\n_M,\,\n_{U_1},\,\n_{U_2})$
are all $3\times1$ matrices. The $3\times3$ matrix
$\mathcal{U}$, following the footnote 17, is given by
\bea
\mathcal{U}=\left(\begin{array}{ccc}
\frac{1}{\sqrt{3}} & \frac{1}{\sqrt{2}} & \frac{1}{\sqrt{6}} \\ 
\frac{1}{\sqrt{3}} & 0 & -\sqrt{\frac{2}{3}} \\ 
\frac{1}{\sqrt{3}} & -\frac{1}{\sqrt{2}} & \frac{1}{\sqrt{6}} 
\end{array}\right).
\label{matrix}
\eea
Note that these transformations give $\sum \widetilde\nu^c_i$
$=\sum \nu^c_i$ $+\,(\Re \widetilde\nu^c_i$  
$+\,i \Im \widetilde\nu^c_i)/\sqrt{2}$ $= 3\nu^c + \sqrt{3}(\S_M+i\P_M)/\sqrt{2}$.
Now using eqs.~(\ref{matrix-reln}), (\ref{matrix}) and the field decomposition 
for $\widetilde\nu^c_i$ (mentioned in figure \ref{fig:mixing}),
it is possible to extract the relevant (concerning figure \ref{fig:mixing}) 
parts of eq.~(\ref{lagrangian-rel2}) as:
\bea
\mathcal{L'} &=& \left[-A_\la\bla H^0_u H^0_d 
\frac{(\S_M + i \P_M)}{\sqrt{2}} 
+ \frac{A_\ka\ka}{6\sqrt{2}}\sum^3_{i=1}
\left\{\left(\Us\right)^3 \right.\right.\nonumber\\
&-&\left.\left.
i\left(\Up\right)^3 \right.\right.\nonumber\\
&+& \left.\left.3i\left(\Us\right)^2\left(\Up\right)\right.\right.\nonumber\\
&-& \left.\left.3\left(\Us\right)\left(\Up\right)^2\right\} \right.\nonumber\\
&+&\left. \frac{\ka\bla}{\sqrt{3}} H^{0^*}_u H^{0^*}_d
\left(\sqrt{6} \nu^c (\S_M+i\P_M)+\frac{1}{2}\left\{S^{0^2}_M+S^{0^2}_{U_1}+S^{0^2}_{U_2}
-P^{0^2}_M-P^{0^2}_{U_1}-P^{0^2}_{U_2}\right\}\right)+ {\rm H.c.}\right]\nonumber\\
&+& {\bla^2} (|H^0_u|^2 + |H^0_d|^2 ) 
\left(\sqrt{6}\nu^c\S_M + \frac{S^{0^2}_M + P^{0^2}_M}{2} 
\right) + \sqrt{2}\ka^2\nu^c \left(\Us\right)^3 \nonumber\\
&+&\sqrt{2}\ka^2\nu^c \left(\Us\right) \left(\Up\right)^2\nonumber\\
&-&\frac{1}{2}\left[\sqrt{2}\ka\nu^c\left(\Us\right)\left(\Un\right)^2
+ {\rm H.c.} \right]\nonumber\\
&-&\frac{1}{2}\left[\sqrt{2}i\ka\nu^c\left(\Up\right)\left(\Un\right)^2
+ {\rm H.c.} \right]\nonumber\\
&+&\frac{1}{2}\left[{\bla} H^0_u \widetilde H^0_d \n_M + {\rm H.c.}\right].
\label{lagrangian-rel3}
\eea
Here $\mathcal{U}_{ij}$ represent the elements of the matrix $\mathcal{U}$,
shown in eq.~(\ref{matrix}). It is apparent\footnote{In the last line
of eq.~(\ref{lagrangian-rel3}) we have kept
things at the level of two-component spinors.} from eq.~(\ref{lagrangian-rel3}) 
that processes like $\S_4 \to \S_M\S_{U_{1,2}}$,
$\P_M\P_{U_{1,2}}$, $\n_M\n_{U_{1,2}}$ or
$\S_4\to \S_{U_1}\S_{U_1}$, $\S_4\to \n_{U_2}\n_{U_2}$, 
etc., are suppressed compared to $\S_4 \to \S_M\S_M,\,
\P_M\P_M$ and $\S_4\to \n_M\n_M$ processes. This is due to the presence
of smaller couplings for the former, e.g. powers of $\kappa$,
that are estimated to be around ${\cal{O}}(10^{-2})$ or smaller in the last section.
This conclusion weakens in the limit of $\bla \gsim 0.7$, when
a coupling like $\bla\ka$ appears to be ${\cal{O}}(10^{-1})$.
Thus, for small to moderate $\bla$ values
as well as for some regions of moderate to large 
$\bla$ values, one can use $\sum$ Br($\S_4\to \S_i\S_j$) $\approx$ 
Br($\S_4\to \S_M\S_M$). This statement 
holds true for $\P$ and $\n$ also and has been verified numerically.
Further, we have checked numerically
that our observations remain valid even when small splittings
exist within $\kappa_i$ (see eq.~(\ref{EWF-param3A})) values.
Furthermore, the relative sign difference for $\S$ and $\P$
normally predicts Br$(\S_4\to \S_M\S_M)>$ Br$(\S_4\to \P_M\P_M)$.

Now, looking at figure \ref{fig:mixing},
one can estimate the effective couplings
that control the decays of the SM Higgs-like $\S_4$
into $\widetilde \nu^c$-like $\S_i,\,\P_i$ states,
which are given as: 
$\frac{A^2_\la\bla^2 v_u}{3m^2_s}$, 
$\frac{A_\la\bla\kappa v_u\mu}{3\sqrt{3}m^2_s}$,  
$\frac{A_\la\bla^2 v_d\mu}{9m^2_s}$, 
$\frac{\bla^2v_u}{3}$, $\frac{\bla\kappa v_d}{\sqrt{3}}$,
$\frac{A_\la A_\ka\bla \ka v_d}{\sqrt{3}m^2_s}$,
$\frac{A_\ka\ka^2 \mu v_d}{3m^2_s}$ and $\frac{A_\ka\bla\ka \mu v_u}{3\sqrt{3}m^2_s}$. 
Here $m^2_s$ represents the scale of $H^0_{u,d}$
and $\widetilde\nu^c$ soft squared 
masses, and we have used $\mu=\sqrt{3}\bla\nu^c$
with $\la_i =\la =\frac{\bla}{\sqrt{3}}$. 
Thus, following our discussion of the last section,
if one considers (1) $A_\la \approx \mu \sim 1$ TeV\footnote{ 
Note that $\mu\approx 100$ GeV, that holds
true for small to moderate $\bla$ region, reduces
the $2^{nd}$ and $3^{rd}$ terms further.}, (2) $\ka\sim 10^{-2}$,
(3) $v_u > v_d$ for $\tb>1$,
(4) maximum of $A_\ka \sim 125$ GeV (see eq.~(\ref{sps-approx2A})) 
and (5) fixed $m_s$ at $1$ TeV, then
the leading coupling goes as $\frac{\bla^2v_u}{3}$ GeV.
In the same fashion, out of $\frac{\bla^2 v_u}{3\mu}$, 
$\frac{A_\la\bla \ka v_d}{\sqrt{3}m^2_s}$,
$\frac{\ka^2 \mu v_d}{3m^2_s}$ and $\frac{\bla\ka \mu v_u}{3\sqrt{3}m^2_s}$,
the leading coupling that controls the SM Higgs-like $\S_4$
decays into right-handed neutrino-like $\n_{i+3}\n_{j+3}$
is given by $\frac{\bla^2v_u}{3\mu}$. Here
the parameter $\mu$ has been used to represent the
$\widetilde H^0_d$ mass scale.

With these couplings, assuming 
$\mathcal{F}(m^2_{\S_4},\,m^2_{X^0_i},\,m^2_{X^0_j})\approx 1$ 
in eq.~(\ref{decay-width}), the
\textit{maximum}\footnote{With heavier
$\S_i,\,\P_i$ and $\n_{i+3}$ states, these $\mathcal{F}$
functions reduce further and thereby 
justify the \textit{maximum} estimate 
of these decay widths as shown in eq.~(\ref{decay-width3}).} 
approximate leading decay widths for $\S_4\to \S_i\S_j$
and $\P_i\P_j$, $\n_{i+3}\n_{j+3}$ 
processes are then given as
%
\bea
\Gamma_{\S_4\to \S_i\S_j, \P_i \P_j} 
&\approx& \frac{\bla^4 v^2 {\rm tan^2}\beta}{9(1+{\rm tan^2}\beta)}
\times \frac{1}{16 \pi m_{\S_4}(1+\delta_{ij})}
\approx \frac{0.5\bla^4  
{\rm tan^2}\beta}{(1+{\rm tan^2}\beta)(1+\delta_{ij})}~{\rm GeV},\nonumber\\
\Gamma_{\S_4\to \n_{i+3}\n_{j+3}} 
&\approx& \frac{\bla^4 v^2 {\rm tan^2}\beta}{9\mu^2(1+{\rm tan^2}\beta)}
\left[m^2_{\S_4}-\left(m_{\n_{i+3}}+ m_{\n_{j+3}}\right)^2\right]
\times \frac{1}{16\pi m_{\S_4} (1+\delta_{ij})}\nonumber\\
&\approx& \frac{8366 \bla^4 {\rm tan^2}\beta}{(1+{\rm tan^2}\beta)(1+\delta_{ij})}
\left[1-\left(\frac{m_{\n_{i+3}}}{m_{\S_4}}+
\frac{m_{\n_{j+3}}}{m_{\S_4}}\right)^2\right]~\frac{{\rm GeV}^3}{\mu^2}. 
\label{decay-width3}
\eea
Here we have used $v=174$ GeV and $m_{\S_4}=125$ GeV.
From eq.~(\ref{decay-width3}) note that at the limit
of $\tb\gg1$, $\S_4\to$ $\S_i\S_j$, $\P_i\P_j$, $\n_{i+3}\n_{j+3}$
decays become independent of $\tb$.

With the formulae as shown in eq.~(\ref{decay-width3}),
one can estimate the relative importance of the new decays, namely
$S^0_4\to S^0_i S^0_j$, $P^0_i P^0_j$, $\widetilde\chi^0_{i+3}\widetilde\chi^0_{j+3}$
with respect to the known 
and five well-measured SM decay modes, namely
$S^0_4 \to b \bar{b}$ , $\tau^+\tau^-,\,
\gamma\gamma$, $W^\pm {W^\mp}^*$ and $ZZ^*$  
\cite{ATLAS:couplings,Aad:2014eva,Aad:2014eha,CMS:2014ega}. 
The branching ratios into these modes and the total decay widths for the SM Higgs boson,
from theoretical analyses, are given in refs.~\cite{HiggsBr,Heinemeyer:2013tqa}
assuming a huge variation in Higgs mass, $80~{\rm GeV} - 1000$ GeV.
The total decay width for the SM Higgs boson with a mass of $125$ GeV is 
$\Gamma^{\rm SM}_{\rm tot}=4.07^{+0.162}_{-0.160}$ MeV  \cite{HiggsBr}.

It is now essential to discuss the various 
measured experimental constraints on the SM Higgs-like $\S_4$
that are relevant for the discussion of this section.
The stringent set of constraints are coming
from the measured reduced signal strengths over 
the five aforesaid SM decay modes. 
The reduced signal strength, when an \textit{on-shell}  
$\S_4$ decays into a pair of $X$
particles, $\mu_{XX}(\S_4)$ is given by
%
\bea
\mu_{XX}(\S_4)&=&\frac{\sigma_{\rm prod}(\S_4)\times Br(\S_4\to XX)}{
\sigma_{\rm prod} (\hsm)\times Br(\hsm\to XX)}\nonumber\\
&=& \frac{\sigma_{\rm prod}(\S_4)}{\sigma_{\rm prod} (\hsm)}
\times \frac{\Gamma_{\S_4\to XX}}{\Gamma_{\hsm\to XX}}
\times \frac{\Gamma^{\rm SM}_{\rm tot}}{\Gamma^{\rm SM'}_{\rm tot}
+\Gamma^{\rm NP}_{\rm tot}}.
\label{reduced}
\eea
Here $\hsm$ denotes the SM Higgs boson, 
$\sigma_{\rm prod}(\S_4)$ and $\sigma_{\rm prod}(\hsm)$
represent the production cross-section of the $\S_4$ and $\hsm$, respectively.
We use $\Gamma_{\S_4\to XX}$ and
$\Gamma_{\hsm\to XX}$ to represent Higgs $\to XX$ decay width
in the new physics (NP) theory (in this case the $\mu\nu$SSM)
and in the SM, respectively. The total decay width for the NP
is written as a sum of the pure NP decay width ($\Gamma^{\rm NP}_{\rm tot}$)
and that of the SM modes in NP theory ($\Gamma^{\rm SM'}_{\rm tot}$). 
The quantity $\Gamma^{\rm NP}_{\rm tot}$, following
eq.~(\ref{decay-width3}) is written as
\bea
\Gamma^{\rm NP}_{\rm tot} = \sum^{3}_{i,j=1} \frac{1+\delta_{ij}}{2}
\left(\Gamma_{\S_4\to\S_i\S_j}
+ \Gamma_{\S_4\to\P_i\P_j} + \Gamma_{\S_4\to\n_{i+3}\n_{j+3}}\right).
\label{29NPtot}
\eea
The latest measured $\mu_{XX}(\S_4)$ values 
for $X=b,\,\tau,\,\gamma,\,W^\pm$ and $Z$ are given in table \ref{tabcross}.

\TABLE[ht]
{\begin{tabular}{cc|c|c|l}
\cline{3-4}
& & Measured value & $m_{\S_4}$ (GeV) & \\ \cline{1-4}
\multicolumn{1}{|c|}{} &
\multicolumn{1}{|c|}{$\mu_{b\bar{b}}(\S_4)$} 
& $0.2^{+0.7}_{-0.6}$ \cite{ATLAS:couplings} & 125.5 &   \\ \cline{2-4}
\multicolumn{1}{|c|}{}                        &
$\mu_{\tau^+\tau^-}(\S_4)$ 
& $1.4^{+0.5}_{-0.4}$ \cite{ATLAS:couplings} & 125.5 &\\ \cline{2-4}
\multicolumn{1}{|c|}{ATLAS}                        &
$\mu_{\gamma\gamma}(\S_4)$& $1.17 \pm 0.27$ \cite{Aad:2014eha} & 125.4 &\\ \cline{2-4}
\multicolumn{1}{|c|}{}                        &
$\mu_{WW^*}(\S_4)$& $1.0^{+0.32}_{-0.29}$ \cite{ATLAS:couplings}& 125.5 &\\ \cline{2-4}
\multicolumn{1}{|c|}{}                        &
$\mu_{ZZ^*}(\S_4)$&$1.44^{+0.40}_{-0.33}$ \cite{Aad:2014eva} & 125.36 &\\ \cline{2-4}
\multicolumn{1}{|c|}{}                        &
Combined&$1.30^{+0.18}_{-0.17}$ \cite{ATLAS:couplings} & 125.5 &\\ \cline{1-4}
\multicolumn{1}{|c|}{} &
\multicolumn{1}{|c|}{$\mu_{b\bar{b}}(\S_4)$} 
& $0.93 \pm 0.49$\cite{CMS:2014ega} & 125 &   \\ \cline{2-4}
\multicolumn{1}{|c|}{}                        &
$\mu_{\tau^+\tau^-}(\S_4)$& $0.91 \pm 0.27$ \cite{CMS:2014ega} & 125 &\\ \cline{2-4}
\multicolumn{1}{|c|}{CMS}                        &
$\mu_{\gamma\gamma}(\S_4)$&$1.13 \pm 0.24$ \cite{CMS:2014ega} & 125 &\\ \cline{2-4}
\multicolumn{1}{|c|}{}                        &
$\mu_{WW^*}(\S_4)$&$0.83 \pm 0.21$ \cite{CMS:2014ega} & 125 &\\ \cline{2-4}
\multicolumn{1}{|c|}{}                        &
$\mu_{ZZ^*}(\S_4)$&$1.00 \pm 0.29$ \cite{CMS:2014ega} & 125 &\\ \cline{2-4}
\multicolumn{1}{|c|}{}                        &
Combined&$1.00 \pm 0.13$ \cite{CMS:2014ega} & 125 &\\ \cline{1-4}
\end{tabular}
\caption{The measured signal strengths up to 
the $1\sigma$ errors from the ATLAS and
CMS measurements, with the concerned values of Higgs mass.
For the ATLAS measurements, $\mbb,\,\mWW$ correspond to
a combined data set of E$_{\rm CM}=7$ TeV
with $\mathcal{L}=4.6-4.8$ fb$^{-1}$
and E$_{\rm CM}=8$ TeV with $\mathcal{L}=20.3$ fb$^{-1}$ \cite{ATLAS:couplings}. 
Only $8$ TeV data set has been used \cite{ATLAS:couplings}
to evaluate $\mtata$. In the measurements of $\mgg,\,\mZZ$, the ATLAS collaboration 
has used a data set corresponds to E$_{\rm CM}=7$ with $\mathcal{L}=4.5$ fb$^{-1}$
combined with E$_{\rm CM}=8$ TeV with $\mathcal{L}=20.3$ fb$^{-1}$ \cite{Aad:2014eha,Aad:2014eva}.
A data set corresponds to E$_{\rm CM}=7$ with $\mathcal{L}=5.1$ fb$^{-1}$
and E$_{\rm CM}=8$ TeV with $\mathcal{L}=19.7$ fb$^{-1}$ has been
used for the CMS analyses \cite{CMS:2014ega}.
\label{tabcross}}}

Additional constraints can appear 
from the other measurements, e.g. the total decay width 
$\Gamma_{\rm tot}=\Gamma^{\rm NP}_{\rm tot}+\Gamma^{\rm SM'}_{\rm tot}$,
room for the invisible/non-standard branching fractions, etc. 
For the former, the concerned
CMS limit is $\Gamma_{\rm tot}< 22$ MeV \cite{CMS:2014ala,Khachatryan:2014iha} 
assuming $m_{\S_4}=125.6$ GeV. 
The other constraint, i.e. the experimentally allowed
window for the invisible/non-standard $\S_4$ decay branching
fraction at $95\%$ C. L. is $< 0.41$
from the ATLAS \cite{ATLAS:couplings} while 
$<0.58$ from the CMS \cite{Chatrchyan:2014tja} observation.

At the LHC, $gg\to \S_4$ is the leading source of Higgs production.
Assuming stops above $1$ TeV, $gg\to \S_4$ process in the NP
occurs mainly through the top loop, just like the SM. The only
difference appears from the concerned coupling, through
which a SM-like $\S_4$ couples to $t\bar{t}$ in NP.
In one line, the ratio of the decay widths for a Higgs-like scalar
decaying into $XX$ final state, in the NP and in the SM,
is proportional to the ratio of the respective squared couplings.
Hence, one gets 
\bea
\frac{\Gamma_{\S_4\to XX}}{\Gamma_{\hsm\to XX}}
=\frac{\mathcal{G}^2_{\S_4 XX}}{\mathcal{G}^2_{\hsm XX}},
\label{29coupratio}
\eea
where $\mathcal{G}$ represents the relevant coupling.
The ratios of the relevant squared couplings are given as
\bea
&&\hspace*{1.5cm}\frac{\mathcal{G}^2_{\S_4 t\bar{t}}}{\mathcal{G}^2_{\hsm t\bar{t}}}
= \frac{(1+{\rm tan}^2\beta)}{{\rm tan}^2\beta}\left|R^{\S}_{42}\right|^2,\quad
\frac{\mathcal{G}^2_{\S_4 b\bar{b}}}{\mathcal{G}^2_{\hsm b\bar{b}}}
= \frac{\mathcal{G}^2_{\S_4 \tau^+\tau^-}}{\mathcal{G}^2_{\hsm \tau^+\tau^-}}
= {(1+{\rm tan}^2\beta)}\left|R^{\S}_{41}\right|^2,\nonumber\\
&&\frac{\mathcal{G}^2_{\S_4 WW}}{\mathcal{G}^2_{\hsm WW}}
=\frac{\mathcal{G}^2_{\S_4 ZZ}}{\mathcal{G}^2_{\hsm ZZ}}
= \left|\cb\,R^{\S}_{41}+\sb\, R^{\S}_{42} +\frac{\nu_i}{v}\, R^{\S}_{4,i+5}\right|^2 
\approx \frac{{\rm tan}^2\beta}{(1+{\rm tan}^2\beta)}
\left|\frac{R^{\S}_{41}}{{\rm tan}\beta}
+  R^{\S}_{42}\right|^2,\nonumber\\
&&\hspace*{4.5cm}\frac{\mathcal{G}^2_{\S_4 \gamma\gamma}}{\mathcal{G}^2_{\hsm \gamma\gamma}}
\approx \frac{(1+{\rm tan}^2\beta)}{{\rm tan}^2\beta}\left|R^{\S}_{42}\right|^2.
\label{reduced-2}
\eea
Here we have used $v_u=v\,\sb$, $v_d=v\,\cb$ and 
$\nu_i/v\sim \mathcal{O}(10^{-6})\ll 1$. The couplings
$R^{\S}_{41},\,R^{\S}_{42},\,R^{\S}_{4,i+5}$,
following ref. \cite{Escudero:2008jg}, are given
in ref. \cite{Ghosh:2010zi}. These are related
to the composition of $H^0_d,\,H^0_u$ and left-handed
sneutrinos in $\S_4$ with the maximum possible squared value equal to $1$.
Thus, neglecting $(\nu_i/v)R^{\S}_{4,i+5}$ in the last
line of eq.~(\ref{reduced-2}) is well justified.
In the derivation of $\mathcal{G}^2_{\S_4 \gamma\gamma}/
\mathcal{G}^2_{\hsm \gamma\gamma}$, we have assumed
that the primary contribution
to the SM-like $\S_4\to \gamma\gamma$ emerges through
the top loop, similar to the SM. The latter is well motivated
in the absence of light charged SUSY particles.

It is now possible to use eqs.~(\ref{29coupratio})
and (\ref{reduced-2}) to write $\Gamma^{\rm SM'}_{\rm tot}$
as 
\bea
\Gamma^{\rm SM'}_{\rm tot}
&&= \sum_m \left(\frac{\mathcal{G}^2_{\S_4mm}}{\mathcal{G}^2_{\hsm mm}}\right) 
\times \Gamma_{\hsm\to mm},\nonumber\\
&&\approx
\frac{(1+{\rm tan}^2\beta)}{{\rm tan}^2\beta}\RSu 
(\Gamma_{\hsm\to \gamma\gamma}+\Gamma_{\hsm\to Z\gamma}
+\Gamma_{\hsm\to gg}
+\Gamma_{\hsm\to c\bar{c}})\nonumber\\
&&+ {(1+{\rm tan}^2\beta)} \RSd
(\Gamma_{\hsm\to b\bar{b}}+\Gamma_{\hsm\to s\bar{s}}
+\Gamma_{\hsm\to\tau^+\tau^-}
+\Gamma_{\hsm\to\mu^+\mu^-})\nonumber\\
&&+ \frac{{\rm tan}^2\beta}{(1+{\rm tan}^2\beta)}\left|\frac{R^{\S}_{41}}{\tb}
+ R^{\S}_{42}\right|^2(\Gamma_{\hsm\to WW^*}+\Gamma_{\hsm\to ZZ^*}),
\label{reduced-4}
\eea
where the sum exists over all the known SM modes.
Here we have used the fact
that $c\bar{c}$ and $s\bar{s},\,\mu^+\mu^-$
couples to the $\S_4$ like $t\bar{t}$ and $b\bar{b}$, respectively.
We also assume that the leading source of 
$\S_4\to Z\gamma$ process is the top loop.
%
\TABLE[ht]
{\small
\begin{tabular}{|c|c|c|c|c|c|c|c|c|c|c|}
\hline
$\hsm\to mm$ &$\gamma\gamma$&$Z\gamma$&$gg$&$c\bar{c}$&$b\bar{b}$&
$s\bar{s}$&$\tau^+\tau^-$&$\mu^+\mu^-$&$WW^*$&$ZZ^*$   \\ \hline
$\Gamma_{\hsm\to mm}$ &&&&&&&&&&  \\ 
(MeV)  &0.009&0.006&0.349&0.118
&2.348&0.001&0.257&0.001&0.875& 0.107 \\ \hline
\end{tabular}
\caption{Theoretical decay widths for a 125 GeV $\hsm$
with $\Gamma^{\rm SM}_{\rm tot}=4.07$ MeV, 
as given in ref. \cite{HiggsBr}. The corresponding
errors are not shown. \label{tabwidth}}}
%
One can rewrite eq.~(\ref{reduced-4}) using the
decay widths for a 125 GeV $\hsm$ into different 
modes as given in table \ref{tabwidth}. The result 
is given by 
\bea
\Gamma^{\rm SM'}_{\rm tot}~({\rm MeV})
&&\approx
\frac{(1+{\rm tan}^2\beta)}{{\rm tan}^2\beta}\RSu 
\times 0.48 + {(1+{\rm tan}^2\beta)} \RSd
\times 2.61\nonumber\\
&&{\hspace*{0.3cm}}+ \frac{{\rm tan}^2\beta}{(1
+{\rm tan}^2\beta)}\left|\frac{R^{\S}_{41}}{\tb}
+ R^{\S}_{42}\right|^2 \times 0.98.
\label{reduced-5}
\eea
 
In the light of these discussions,
together with eqs.~(\ref{29coupratio})
and (\ref{reduced-2}), one can re-interpret eq.~(\ref{reduced}) as
\bea
\mu_{XX}(\S_4)&\approx&
\frac{\mathcal{G}^2_{\S_4 t\bar{t}}}{\mathcal{G}^2_{\hsm t\bar{t}}}
\times \frac{\mathcal{G}^2_{\S_4 XX}}{\mathcal{G}^2_{\hsm XX}}
\times \frac{\Gamma^{\rm SM}_{\rm tot}}{\left(\Gamma^{\rm SM'}_{\rm tot}
+ \,\Gamma^{\rm NP}_{\rm tot}\right)},
\label{reduced-1}
\eea
and consequently, 
\bea
\mu_{\gamma\gamma}(\S_4)&\approx&
\frac{(1+{\rm tan}^2\beta)^2}{{\rm tan}^4\beta}\left|R^{\S}_{42}\right|^4
\times \frac{4.07~\rm MeV}
{\left(\Gamma^{\rm SM'}_{\rm tot}
+ \,\Gamma^{\rm NP}_{\rm tot}\right)}
,\nonumber\\
\mu_{b\bar{b}}(\S_4),\,\mu_{\tau^+\tau^-}(\S_4) &\approx&
\frac{(1+{\rm tan}^2\beta)^2 }{{\rm tan}^2\beta}
\left|R^{\S}_{41}\right|^2 \left|R^{\S}_{42}\right|^2 
\times \frac{4.07~\rm MeV}{\left(\Gamma^{\rm SM'}_{\rm tot}
+ \,\Gamma^{\rm NP}_{\rm tot}\right)}
,\nonumber\\
\mu_{WW^*}(\S_4),\, \mu_{ZZ^*}(\S_4) &\approx&
\left|\frac{R^{\S}_{41}R^{\S}_{42}}{\tb}
+ R^{{\S}^2}_{42}\right|^2
\times \frac{4.07~\rm MeV}{\left(\Gamma^{\rm SM'}_{\rm tot}
+ \,\Gamma^{\rm NP}_{\rm tot}\right)}.
\label{reduced-all}
\eea
Here we have used $\Gamma^{\rm SM}_{\rm tot}=4.07$ MeV for $m_{\hsm}=125$ GeV
and the units of $\Gamma^{\rm SM'}_{\rm tot}$ and $\Gamma^{\rm NP}_{\rm tot}$ are
given in MeV. Expressions for $\Gamma^{\rm NP}_{\rm tot}$ 
and $\Gamma^{\rm SM'}_{\rm tot}$ are given in eqs.~(\ref{decay-width3}), (\ref{29NPtot})
and eq.~(\ref{reduced-5}), respectively. 

Let us consider now, as an example\footnote{The set of numbers used here,
i.e. $\RSds\cong 0.1$ and $\RSus\cong 0.9$, is valid mainly
for small to moderate $\bla$ region. For larger $\bla$ values (e.g., $\bla\gsim 0.3$),
singlet-doublet mixing is enhanced and consequently
a non-negligible singlet composition $\sum |R^{\S}_{4,i+3}|^2$
appears in $\S_4$. The concerned formulas ((\ref{reduced-5}) and (\ref{reduced-all})),
however, remain still valid but with a smaller values of $\RSds$ and $\RSus$.},
$|R^{\S}_{42}|^2 \cong 0.9$, $|R^{\S}_{41}|^2 \cong 0.1$ and
negligible $\widetilde \nu^c$ composition in $\S_4$. These numbers
will be used henceforth. 
Note that $(1+{\rm tan}^2\beta)/{\rm tan}^2\beta$ varies
from $1.25$ to $1$ as $\tb$ changes from $2$ to very large values.
On the other hand, $(1+{\rm tan}^2\beta)$ grows very
fast with $\tb$. Hence, a small $\RSds$ is essential 
to accommodate $\Gamma^{\rm SM'}_{\rm tot}$
as well as $\mbb,\,\mtata$ in an
experimentally allowed way \cite{CMS:2014ala,Khachatryan:2014iha,
ATLAS:couplings,Aad:2014eha,Aad:2014eva,CMS:2014ega}. An alternate way to reduce
$|R^{\S}_{41}|^2$ and hence 
$\Gamma^{\rm SM'}_{\rm tot}$, and consequently
$\mbb,\,\mtata$, with respect to the SM, 
is to introduce more singlet component in $\S_4$. 
This procedure, however, is valid for small
$\tb$ values, unless one considers $\RSds \to 0$. In this way
one can also increase $\mu_{\gamma\gamma}(\S_4)$ \cite{Ellwanger:2011aa,
SchmidtHoberg:2012yy}. For the latter, the
existence of light charged SUSY particles (e.g., stau, chargino) in the 
spectrum \cite{Carena:2012gp,Casas:2013pta,Batell:2013bka,Hemeda:2013hha}
is another possibility.

The choice of $\RSds=0.1$ and\footnote{Note that the limit on 
$\tb$ attains smaller values with increasing $\RSds$,
e.g. $\tb$ $\lsim 1.9$ for $\RSds=0.2$.} $\tb$ $\lsim 3$,
using eq.~(\ref{reduced-5})
implies $\Gamma^{\rm SM'}_{\rm tot}\lsim \Gamma^{\rm SM}_{\rm tot}=4.07$ MeV,
and hence $\Gamma^{\rm SM}_{\rm tot}/(\Gamma^{\rm SM'}_{\rm tot}
+\Gamma^{\rm NP}_{\rm tot})$ $\gsim 1$, given that 
$\Gamma^{\rm NP}_{\rm tot}\ll \Gamma^{\rm SM'}_{\rm tot}$, 
as expected for small to moderate $\bla$ values (see eqs.~(\ref{decay-width3})
and (\ref{29NPtot})).
This behaviour is well expected since with small $\RSds$ and small $\tb$
(see eq.~(\ref{reduced-5})), the decay 
widths for all the down-type fermions reduce below their SM values.
This reduction, especially for $b\bar{b}$ which is the leading decay
mode for the SM-like $\S_4$, diminishes $\Gamma^{\rm SM'}_{\rm tot}$
mainly through a reduction in $\Gamma_{\S_4\to b\bar{b}}$.
Using $\RSus=0.9$ and $\RSds=0.1$,
in the limit of a negligible $\GNP$, 
one evaluates from eq.~(\ref{reduced-all}) that
$\mgg$, $\mWW$, $\mZZ \gsim 1$ for $\tb\lsim 3$ 
while $\mbb$, $\mtata$ $< 1$.
This observation has an important consequence, i.e. 
departure of all the five reduced signal strengths from the value $1$ remains
possible even when $\Gamma^{\rm NP}_{\rm tot}=0$, i.e. when
no additional decay modes exist for $\S_4$. This phenomenon,
in the context of the $\mu\nu$SSM can occur when $m_{\S_i,\,\P_i}$ and $m_{\n_{i+3}}$
are larger or comparable to $m_{\S_4}$. 

Following now the trend of our past analyses, as of the last two sections,
we again start with the discussion of the $0.01\lsim\bla\leq 0.1$ scenario.

\vspace*{0.25cm}
\noindent
(a) {\bf Small to moderate $\bla$:} 
In this region of the parameter space, the maximum value of
$\bla$ is $0.1$. It is thus evident from eq.~(\ref{decay-width3})
that $\SSSd,\,\SPPd,\,\SNNd$ decays are naturally suppressed
in this corner of the parameter space due to the smallness of
the $\bla$ parameter. 
Following the discussion of the last paragraph,
one gets for example with $\tb=2$ and $\mu=100$ GeV,
$\GNP\approx 0.036$ MeV while $\GNPSM \approx 2.81$ MeV, 
as evaluated from eqs. (\ref{29NPtot}) and (\ref{reduced-5}).
Hence, numerically one gets $\mgg \approx 1.81$, $\mbb,\,\mtata \approx 0.81$
and $\mWW,\,\mZZ \approx 1.58$. 
Clearly, from table \ref{tabcross}, for $\gamma\gamma$
one needs to consider at least $3\sigma$ variation
to accommodate this scenario experimentally. The conclusion is also
very similar for $WW^*$ and $ZZ^*$ while for 
$b\bar{b}$ the number is within the $1\sigma$
range of the experimental measurements. 
For $\tau^+\tau^-$, the number falls
within the $1\sigma$ and $3\sigma$ range
of the CMS and the ATLAS measurements, respectively.
For the evaluation of $\GNP$, we have considered the proper 
numerical factors as mentioned in the caption of figure \ref{fig:mixing},
took $i=j$ and $m_{\n_{i+3}}=10$ GeV. Furthermore, we
have used the fact, as already stated, that for this  
corner of the parameter space 
$\GNP\approx \Gamma_{\S_4\to \S_M\S_M}+\Gamma_{\S_4\to \P_M\P_M}
+\Gamma_{\S_4\to \n_M\n_M}$.

Taking $\tb=3$ and keeping everything else the same,
one gets $\GNP\approx0.041$ MeV, $\GNPSM\approx 4.07$ MeV,
$\mgg$, $\mbb,\,\mtata$, $\mWW$, $\mZZ$ $\approx 0.99$. 
These numbers, as evident
from table \ref{tabcross}, are within the $1\sigma$ ranges
of the CMS measurements while for some of the cases fall
within the respective $2\sigma$ ranges of the ATLAS
measurements. 
It is thus important to note that even in the presence
of a non-vanishing new physics effect, all the five reduced
signal strengths can remain very close to 1, the expected
SM value.
With larger $\tb$, $\mbb,\,\mtata$
enhance while $\mgg$, $\mWW$, $\mZZ$ decrease further.
The $\mgg$ goes beyond the respective $3\sigma$ range around $\tb\gsim 5$. 
In this corner of the parameter space, using the chosen 
values of $\RSus=0.9$ and $\RSds=0.1$, all of the three
$\mgg,\,\mWW,\,\mZZ$s remain larger than 1,
but within the respective $2\sigma$ CMS ranges, for $2.5\lsim\tb\lsim 2.95$. 
For $\tb > 2.98$, $\mgg,\,\mWW$ and $\mZZ$ reduce
below $1$. In this region of the parameter space,
concerning the CMS measurements (see table \ref{tabcross}),
all the five reduced signal strengths 
remain within their respective $2\sigma$ 
ranges for $2.5\lsim \tb\lsim 3.9$. 
With a different choice of $\RSus$ and $\RSds$,
one gets a shift in the range of $\tb$
towards smaller values. For example, with $\RSus=0.75$ and $\RSds=0.25$,
one evaluates $1.45\lsim \tb\lsim 2.15$ as the preferred
range of $\tb$, where
all the five reduced signal strengths lie within
their respective CMS $2\sigma$ ranges (see table \ref{tabcross}).

In this connection, note that a bino-like 
lightest neutralino (defined as $\n_b$) of the same mass,
i.e. $m_{\n_{i+3}}\approx m_{\n_b}$, can also
contribute to the $\S_4$ decay phenomenology. In this case
the coupling goes as $\frac{g_1 v_u}{\mu}$. Hence,
one needs to multiply $\Gamma_{\S_4 \to \n_{i+3}\n_{i+3}}$
with ${9g^4_1}/{\bla^4}$ to get $\Gamma_{\S_4 \to \n_{b}\n_{b}}$.
With $g_1\approx 0.352$ one evaluates 
$\Gamma_{\S_4 \to \n_{b}\n_{b}}\approx 25.34$ MeV for $\bla=0.1$
and $\tb=3$, with the proper numerical factor as mentioned
in figure \ref{fig:mixing}. This gives $\mgg,\,\mWW,\,\mZZ$
$\approx 0.14$ which are beyond/at the boundary of 
the respective $3\sigma$ ranges as measured from 
the CMS (see table \ref{tabcross}).
Further, $\Gamma_{\S_4 \to \n_{b}\n_{b}} > 22$ MeV 
is also excluded by the CMS decay
width measurement \cite{CMS:2014ala,Khachatryan:2014iha}.
It is nonetheless possible to accommodate
a light $\n_b$ with larger $\mu$ values,
e.g. $\mu=500$ GeV (with $\bla=0.1$ and $\tb=3$), that gives 
$\GNP=\Gamma_{\S_4 \to \n_{b}\n_{b}}\approx 1.01$ MeV,
$\mgg$, $\mbb,\,\mtata$ and $\mWW,\,\mZZ \approx 0.80$. All these
numbers\footnote{For ATLAS $\tau^+\tau^-$, one 
needs to consider a variation in the $3\sigma$ range.} are within the $2\sigma$ measured
values as shown in table \ref{tabcross}.

\vspace*{0.15cm}
To summarise, depending on $\tb$ values, 
experimentally allowed new decay channels for the SM Higgs-like $\S_4$ 
into scalars, pseudoscalars and neutralinos are possible with 
respect to known SM decay modes. Taking into account the 
results of section \ref{lightest}, we note that:
prompt leptons/taus/jets/photons from a singlet-like scalar/pseudoscalar 
are favourable in the small to moderate $\bla$ scenario. 
A similar conclusion holds for the displaced objects from a singlino-like 
neutralino. On the contrary, with a bino-like 
lightest neutralino displaced leptons/taus/jets/photons are
difficult in this region of the parameter space unless
one considers a large $\mu$-value.
The associated decay length for a $\n_b$, with a mass smaller than $M_W$, is normally
larger compared to a right-handed neutrino-like $\n_{i+3}$ of the same mass.
The reason, as already mentioned in subsection \ref{lightN}, is related
to the \textit{natural} feasibility of having lighter right-handed sneutrino-like 
$\S_i,\,\P_i$ states for the latter.

\vspace*{0.25cm}
\noindent
(b) {\bf Moderate to large $\bla$:} 
In this region of the parameter space
we consider two limiting representative scenarios:
(1) $\bla=0.2$ and $\mu=350$ GeV,
and (2) $\bla=0.7$ and $\mu=1200$ GeV.

For scenario 1, $\mgg,\,\mWW,\,\mWW$ remain
larger than $1$ for $\tb\lsim 2.75$. The quantities
$\mbb$ and $\mtata$ go above $1$ from $\tb\gsim 3.60$
when $\mgg$, $\mWW$, $\mZZ$ diminish to $\approx 0.70$, $0.72$ and
$0.72$, respectively.
In this region of $\tb$ values, $\Gamma^{\rm SM'}_{\rm tot}$
wins over $\Gamma^{\rm SM}_{\rm tot}$. Concerning
the measured values from the CMS (the ATLAS limits
are more flexible), as given in table \ref{tabcross},
$\mgg$ goes beyond respective $2\sigma$ range
from $\tb\gsim 3.75$ while $\mWW,\,\mZZ$ attain
the same from $\tb\gsim 5.15$. 
The reduced signal strengths for all the five measured modes,
with our choice of $\RSus=0.9$ and $\RSds=0.1$,
remain within their respective $2\sigma$ CMS 
ranges for $2.3$ $\lsim$ $\tb$ $\lsim 3.75$.
The total decay width $\GNPSM+\GNP$ remains
below the CMS limit of $22$ MeV 
\cite{CMS:2014ala,Khachatryan:2014iha} 
unless $\tb\gsim 8.8$. This region
of $\tb$ value, however, is also excluded
from $\mgg$, $\mWW$ and $\mZZ$ measurements. 
In this region, especially as $\bla\to 0.7$, additional constraints 
can appear through the possible $Z\to S^0_i P^0_j,\,
\n_{i+3}\n_{j+3}$ decay modes \cite{Ghosh:2014rha}.

Scenario 2, on the contrary, even with $\tb\to 1$
predicts $\mgg$, $\mWW$, $\mWW$ and $\mtata$
beyond the respective $3\sigma$ ranges (see table \ref{tabcross}). 
Moreover, in this corner of the parameter space,
even with $\GNP\approx \Gamma_{\S_4\to \S_M\S_M}+\Gamma_{\S_4\to \P_M\P_M}
+\Gamma_{\S_4\to \n_M\n_M}$, one estimates
$\GNP > 22$ MeV, which is excluded by the CMS
measurement \cite{CMS:2014ala,Khachatryan:2014iha}.
Such a large $\Gamma^{\rm NP}_{\rm tot}$, for $\bla=0.7$,
is capable of giving a $\mbb$ value beyond the $2\sigma$ range, in spite of
the huge associated errors (see table \ref{tabcross}).
The large contribution appears mainly
through $\S_4\to \S_i\S_j,\,\P_i\P_j$ modes,
e.g. about $30$ MeV for $\S_4\to \S_M\S_M+\P_M\P_M$ process.
Contribution from $\n_{i+3}\n_{j+3}$ modes remain
much suppressed compared to the former, e.g.
with $\tb=2$ one gets $\Gamma_{\S_4\to\n_M\n_M}$ $\approx 0.3$ MeV
while $\S_4\to \S_M\S_M$ is $\approx 24$ MeV.
Thus, existence of new $\S_4$ decays,
especially $\S_4\to \S_i\S_j,\,\P_i\P_j$, are hardly
possible in this region. Further, as already emphasised
in section \ref{spsn}, larger singlet-doublet
admixture in $\S_i,\,\P_i$ and $\n_{i+3}$ as $\bla\to 0.7$
makes it harder to accommodate these states
in an experimentally allowed way. Nonetheless,
for this region of the $\bla$ value,  $\S_i,\,\P_i$ and $\n_{i+3}$ states
comparable or heavier than $\S_4$ remain an allowed
possibility. This scenario, as stated earlier in the last paragraph
before the discussion of small to moderate $\bla$ region,
can still predict $\mgg,\,\mWW,\,\mZZ>1$ 
depending on the scale of $\tb$. 
Note that as $\bla\to 0.7$, the severe constraint
on the presence of light $\S_i,\P_i$ states
predicts larger decay length for a light $\n_{i+3}$,
as already addressed in subsection \ref{lightN}.

A $\n_b$ of the same mass remains experimentally difficult 
for scenario 1 for $\tb \gsim 3.2$. For example,
with $\tb=3.2$ scenario 1 gives $\mgg \approx 0.61$.
All the five measured reduced signal strengths
remain within the $2\sigma$ range for 
$1.6\lsim\tb \lsim 3.2$. Here 
we consider only the CMS limits as given by table \ref{tabcross}. 
On the other hand, a $\n_b$ of the same mass remains
well possible for scenario 2, e.g. one gets 
$\mgg$, $\mWW$, $\mZZ \gsim 1$ for $\tb\lsim 2.9$. Concentrating
on the CMS results, one observes that
all of the five $\mu_{XX}(\S_4)$ remain
within their respective $2\sigma$ ranges
for $2.45\lsim \tb\lsim 3.85$, with the chosen
values of $\RSus$ and $\RSds$.
Clearly, the existence of a light $\n_b$ is more feasible
for $\bla=0.7$ and normally $\Gamma_{\S_4\to \n_b\n_b}$
$>$  $\Gamma_{\S_4\to \n_{i+3}\n_{j+3}}$ up to $\bla \approx 0.6$. 
For $0.6\lsim \bla\leq 0.7$, on the contrary,
$\Gamma_{\S_4\to \n_{i+3}\n_{j+3}}$ $>$ $\Gamma_{\S_4\to \n_b\n_b}$.
Once again, the constraint on the presence of lighter $\S_i,\,\P_i$
states indicates larger decay length as $\bla\to 0.7$.

\vspace*{0.15cm}
In a nutshell, in this region of the parameter space,
depending on the value of $\bla$ and $\tb$, both prompt and displaced
objects are possible from the decays of a singlet-like
$S^0_i,\,P^0_i$ and $\n_{i+3}$. 
Unlike the scalars/pseudoscalars, the neutralino decays,
in the absence of a lighter $\S_i/\P_i$ state,
remain possible for the entire span of $\bla$ values. 
A $\n_b$ is also feasible for larger $\bla$ values. 
Enhanced branching fractions, mainly for 
$S^0_4\to S_i\S_j,\,\P_i\P_j$ decay modes, normally put 
severe constraint on this corner of the parameter space,
especially as $\bla \to 0.7$. This restriction in turn
implies larger decay length for light neutralinos,
this time both for singlino-like $\n_{i+3}$ and $\n_b$.
A detail analysis of $\S_4\to\n_4\n_4$ decay for this region
of the parameter space with $\bla\approx 0.2$ in the context
of the displaced but detectable multi-$\tau$ final state has
already been addressed in ref. \cite{Ghosh:2012pq}.
%

\vspace*{0.25cm}
\noindent
(c) {\bf Dominant $\bla$:} The presence of
large singlet-doublet mixing, as already stated
in section \ref{spsn}, makes it rather hard 
for the light $\S_i,\,\P_i$ states to evade a class of
constraints from colliders. Further, large
$\bla$ values normally predict very large
decay widths for $\S_4\to\S_i\S_j,\,\P_i\P_j$
processes, that are excluded experimentally.  
As an example, with $\bla=1,\, \mu=1000$ GeV one gets 
$\mgg\approx 0.1$, $\mbb,\,\mtata\approx 0.01$,
$\mWW,\,\mZZ\approx 0.05$ and $\GNP\approx 126$ MeV
even when $\tb \to 1$. Here, we have assumed that
$\GNP\approx \Gamma_{\S_4\to \S_M\S_M}+\Gamma_{\S_4\to \P_M\P_M}
+\Gamma_{\S_4\to \n_M\n_M}$, which however, as already mentioned,
gives suppressed contribution compared to the true value
for $\bla \gsim 0.7$.
Nonetheless, even with this suppressed approximation
we observe that all the reduced signal strengths,
excluding $\mbb$, are beyond their respective $3\sigma$
ranges, as shown in table \ref{tabcross}. Moreover,
the estimated decay width is $\gg 22$ MeV and
hence is excluded by the CMS result 
\cite{CMS:2014ala,Khachatryan:2014iha}.
For $\n_{i+3}$, on the contrary, the corresponding
decay widths remain experimentally viable, provided
that the lighter $\S_i,\,\P_i$ states are absent.
For example, $\bla=1,\,\mu=1$ TeV and $\tb\to 1$
give $\Gamma_{\S_4\to \n_M\n_M}\approx 1$ MeV
while $\Gamma_{\S_4\to \S_M\S_M,\,\P_M\P_M}$
is about 62.5 MeV. One must note that in this
corner of the parameter space, similar to 
moderate to large $\bla$ region with $\bla\to 0.7$,
one also needs to consider the other non-negligible contributions 
like $\S_4\to \n_{U_1}\n_{U_1},\,\n_{U_2}\n_{U_2}$,
$\S_{U_1}\S_{U_1},\,\P_{U_2}\P_{U_2}$ etc.
Combining all these modes for $\n_{i+3}$ and
assuming a similar contribution from all of them, the 
reduced signal strengths, apart from $\mbb$ and $\mtata$, remain
beyond their measured $2\sigma$ variations
for $\tb\gsim 1.3$ and, moves to lower values
for larger $\bla$ values. It is thus apparent
that the presence of $\S_i,\P_i$ and $\n_{i+3}$
states, lighter than $m_{\S_4}/2$, is experimentally
unrealistic in this region of the parameter space.

Repeating the same exercise with $\mu=1$ TeV for a $\n_b$ of the 
same mass, in the absence of lighter $\S_i,\,\P_i$ states,
one gets all the reduced signal strengths within
their respective $2\sigma$ CMS ranges for 
$2.4\lsim \tb \lsim 3.8$. For $\mgg,\,\mWW,\,\mZZ$,
it remain possible to get $> 1$ values, within the respective
$2\sigma$ ranges, in the span of $2.4\lsim \tb \lsim 2.85$. 
For larger $\mu$ values, that are anticipated
with larger $\bla$ values keeping $\nu^c$ fixed
at 1 TeV, one observes a slight shifting of the 
aforementioned $\tb$ window towards larger values.
$\Gamma_{S_4\to\n_b\n_b}$ remains $\ll$ compared
to the CMS upper bound of $22$ MeV, unless
one considers $\tb\gg1$ which is already
excluded from the measured reduced signal strengths.

\vspace*{0.15cm}
In summary, a large amount of doublet contamination 
and very high $\Gamma_{\S_4\to \S_i\S_j},\,\Gamma_{\S_4\to \P_i\P_j}$
make it hard for the light $\S_i,\P_i$ states to
survive a class of collider constraints for this region of $\bla$ values.
Consequently, the chance of getting an enhancement
in the number of prompt final states, from $\S_4$ decays, is hardly 
possible for this corner of the parameter space. The situation
is identical for light $\n_{i+3}$, however,
in this case the severe constraints appear through
the measured reduced signal strengths and not from the 
decay width $\Gamma_{\S_4\to \n_{i+3}\n_{j+3}}$.
A light $\n_b$, on the contrary, depending on $\tb$ remains
well possible in this span of $\bla$ values,
given that no lighter $\S_i,\,\P_i$ state
exists in the spectrum. Thus, the associated
decay length would appear longer, especially
for $m_{\n_b}< 40$ GeV \cite{Bartl:2009an}, 
as already mentioned in subsection \ref{lightN}. 
We note in passing, as also stated 
in the context of moderate to large $\bla$ values
as $\bla\to 0.7$, that the absence of light states in the 
spectrum, e.g. $\S_i,\,\P_i$ or $\n_{i+3}$, does not exclude
the possibility of getting $\mgg,\,\mWW,\,\mZZ>1$ 
depending on the scale of $\tb$ and the values
of $\RSds$ and $\RSus$.

%


\section{Conclusions}
\label{Summary-conclusion}

In this work, in the context of the $\mu\nu$SSM, 
we have performed an analytical estimate of all the 
new two-body decays for the SM-like Higgs boson $(\S_4)$, with a mass about
125 GeV, in the presence of light singlet-like scalars $(\S_i)$, 
pseudoscalars $(\P_i)$
and neutralinos $(\n_{i+3}),~i=1,2,3$. We further explored the relative importance
of the different parameters not only in the context of
new $\S_4$ decay modes, but also to accommodate
a 125 GeV doublet-like scalar with properties similar
to that of the SM-like Higgs boson. At the same time we have identified
the singlet-doublet mixing parameters, namely $\la_i$ 
($\epsilon_{ab}\la_i \hat {\nu}^c_i\hat H^a_d\hat 
H^b_u$ term in eq.~(\ref{superpotential})), 
as the key parameters for this analysis
since they are very crucial in determining the 
relative size of those new $\S_4$ decay branching fractions
compared to that of the SM decay modes. The $\la_i$-parameters,
as explored in section \ref{Higgs-sector},
also produce an extra contribution (eq.~(\ref{boundHiggs1})) for 
the tree-level lightest doublet-like scalar mass which
is prominent for low $\tb$. Our discussion is illustrated
over the three different regions of the $\la_i$ $(\equiv \bla/\sqrt{3})$ values, 
namely (a) small to moderate ($0.01 \lsim \bla \leq 0.1$), 
(b) moderate to large ($0.1 < \bla \leq 0.7$) 
and (c) dominant (i.e., $\bla > 0.7$).

In section \ref{spsn} we have presented for the first-time a set of formulas
for the singlet-like scalars, pseudoscalars and neutralinos
mass terms (eq.~(\ref{sps-approx1})) assuming a simplified index 
structure (eq.~(\ref{EWF-param3})) of the relevant
model parameters. We however, did not consider any assumption
regarding the scale of the parameters. Hence, these formulae
are rather generic and can also be applied for the NMSSM, 
with or without multiple singlets along with necessary
changes (e.g., with one singlet $\mu=\bla \nc$ etc.). We explored these mass terms
for the same three different regions of $\la_i$
values. The scale of the other relevant parameters (e.g. $\ka$, $\mu$, etc.)
have also been estimated during our investigation.

Finally, in section \ref{decay} we have discussed the new two-body
decays of $\S_4$ into a pair of $\S_i$, $\P_i$, $\n_{i+3}$ states, 
presented the expressions of the decay widths (eq. (\ref{decay-width})),
and estimated the five well measured reduced signal
strengths, namely $\mgg$, $\mbb$, $\mtata$, $\mWW$
and $\mZZ$ (see eq.~(\ref{reduced-all})), as well as the total decay
width for the SM-like $\S_4$ (eqs.~(\ref{29NPtot}) and 
(\ref{reduced-5})). Our discussion is furthermore extended
to address the feasibility of getting prompt and/or displaced
leptons/taus/jets/photons at colliders. These signatures 
appear through the decays of these new states, 
following the analyses presented in section \ref{lightest},
where issues of the possible leading backgrounds are also discussed.
The final states considered in this article are the different combination
of four prompt/displaced leptons/taus/jets/photons (at least two of
the each type), accompanied by some $\MET$ originating from the light 
neutrinos $(\n_i)$ and/or possible mis-measurements.
Our analysis also addressed the different possible natures (e.g., bino-like, 
right-handed neutrino-like or bino-singlino mixed) of the 
lightest neutralino $\n_4$ and consequently the effect on the length
of the associated displaced vertex.

In this course of analysis we have observed that in terms of the new 
$\S_4$ decays, consistent with experimental observations,
e.g. reduced signal strengths, the upper limit
of total $\S_4$ decay width, etc., small to moderate
$\la_i$ values (i.e., $0.01\lsim \bla \leq 0.1$) 
are the most favoured one, given that one
works in the range of $2.5\lsim\tb\lsim 3.9$.
This range in $\tb$ is estimated with our choice 
of $\RSds=0.1$ and $\RSus=0.9$. We have also shown
the variation in the range of $\tb$ with a different
choice of $\RSds$ and $\RSus$, e.g.
$1.45\lsim\tb\lsim 2.15$ with $\RSus=0.75$ and $\RSds=0.25$.
Further, we have observed that the pure singlet-like $\S_i,\,\P_i$
and $\n_{i+3}$ states appear rather naturally in this corner of the 
parameter space with a negligible to small parameter tuning.
This is also true for their lightness, as discussed in subsection \ref{spsn1}. 
The room for a light $\n_b$, that can survive the collider constraints,
exists for this region of the parameter space
only when $\mu$ takes a large value. 
Note however that the advantage
of getting relevant extra contribution to the tree-level lightest doublet-like
scalar mass is hardly possible for this range of $\la_i$ values.
An interesting observation for this corner of the parameter
space is its capability of accommodating all the five reduced signal
strengths $\approx 1$, the expected SM value, even in the presence
of a non-vanishing new physics contribution.

For moderate to large $\la_i$ (i.e., $0.1 < \bla \leq 0.7$), on the contrary,
a sizable contribution to the tree-level lightest doublet-like scalar mass 
is possible for this region of the parameter space with small $\tb$.
With a larger singlet-doublet mixing (through larger $\la_i$), 
the singlet purity of $\S_i,\,\P_i$ and $\n_{i+3}$ states diminishes
for this corner of the parameter space. Their
lightness can, however, be preserved through a fine cancellation
using a moderate to large parameter tuning. Regarding decays, depending
on the value of $\la_i$, $\S_4\to\S_i\S_j$ and $\S_4\to\P_i\P_j$
processes can produce the experimentally unacceptable 
reduced signal strengths and the total decay width, 
while the displaced objects through $\S_4\to\n_{i+3}\n_{j+3}$ modes still remain
feasible. However, as $\bla\to 0.7$ the light $\n_{i+3}$ states must not be
accompanied by lighter $\S_i,\P_i$ states in order to remain experimentally
viable. A similar conclusion also holds true for a bino-like lightest
neutralino for this entire region of $\bla$ values. The absence of lighter $\S_i,\P_i$ states
leaves its imprints in terms of the associated decay length.
For $\bla$ values closer to $0.2$, once again
a small window of low $\tb$ values,
e.g. about $2.3\lsim \tb\lsim 3.75$, remains consistent
with the experimental observations for the light singlino-like $\n_{i+3}$.
For a light $\n_b$, on the other hand, depending
on the scale of $\bla$, a similar range
lies approximately within 1.6 to 3.85. This range of $\tb$ values, 
however, will change with a \textit{much} different
choice of $\RSds$ and $\RSus$.

Lastly, the dominant $\la_i$ region (i.e. $\bla > 0.7$) mainly serves the purpose
of giving a very large additional contribution to the tree-level lightest
doublet-like scalar mass.  The lightness of the $\S_i,\,\P_i$ and 
$\n_{i+3}$ states may still appear possible with a severe parameter tuning.
With very large $\la_i$, these states appear with a very large
singlet-doublet mixing and hence, are often excluded from the existing
experimental results. Also in this region of the $\la_i$, new $\S_4$ decay 
modes, primarily through
$\S_i,\P_i$, are normally ruled out by the experimentally 
measured reduced signal strengths and the upper limit
of the $\S_4$ decay width.
Regarding $\n_{i+3}$ states, the conclusion remain
the same, especially concerning the measured reduced
signal strengths. A light $\n_b$, on the other hand, remains
experimentally viable in this region of
the parameter space for a small window of low $\tb$ values
around $2.4\lsim \tb\lsim 3.8$, given the absence of any lighter
$\S_i,\,\P_i$ states. 

To probe the model origin of $\S_4$, 
further investigation of these kinds of new Higgs decay
modes with the dedicated experimental analyses (i.e., 
to detect soft, collimated and often displaced objects), would appear
very relevant in the coming years, especially concerning 
the restart of the LHC in 2015 with the enhanced centre-of-mass energy
and increased luminosity. 
Analyses of these kinds can exclude/narrow down the 
hitherto available non-standard decay window for the observed
SM-Higgs like scalar boson. With higher centre-of-mass
energy and enhanced luminosity, the possibilities of detecting these
new decays and hence, indirect evidences of new physics
beyond the SM (e.g., \cite{Ghosh:2014rha}), are also well envisaged. One would, 
however, require a full numerical estimate of these scenarios
with a proper background analysis which we aim to cover with a set
of forthcoming publications \cite{glmmrF2}.

\vspace{1cm}

\noindent {\bf Acknowledgments:} The work  of PG and 
CM is supported in part by the Spanish MINECO  under grant 
FPA2012-34694 and under `Centro de Excelencia 
Severo Ochoa' Programme SEV-2012-0249, and by the Comunidad de Madrid under grant HEPHACOS 
S2009/ESP-1473. The work of DL is supported by the Argentinian CONICET. 
VAM acknowledges support by the Spanish MINECO under the project FPA2012-39055-C02-01, 
by the Generalitat Valenciana through the project PROMETEO~II/2013-017 and by the 
Spanish National Research Council (CSIC) under the JAE-Doc program co-funded by 
the European Social Fund (ESF). The work of RR is  supported 
by the Ram\'on y Cajal program of the Spanish MINECO and also thanks the support of the 
MINECO under grant FPA2011-29678. The authors also acknowledge the support of the 
MINECO's Consolider-Ingenio  2010 Programme under grant MultiDark CSD2009-00064.


\bibliography{glmmr-Hhh-bibV22A}
\end{document}